\documentclass[useAMS,usenatbib]{mnras}
\usepackage{graphicx,verbatim}
\usepackage[normalem]{ulem}
\usepackage{color,ulem}
\usepackage[T1]{fontenc} 
\usepackage{amsmath,amssymb}
\usepackage{lmodern}

\def \be{\begin{equation}}
\def \ee{\end{equation}}
\newcommand       \ba           {\begin{eqnarray}}
\newcommand       \ea           {\end{eqnarray}}
\def \la{\langle}
\def \ra{\rangle}
\def \bea{\begin{eqnarray}}
\def \eea{\end{eqnarray}}

\newcommand{\comments}[1]{}

\definecolor{webgreen}{rgb}{0,.5,0}
\definecolor{webbrown}{rgb}{.6,0,0}
\hypersetup{%
   colorlinks=true,hyperfootnotes=false,%
   breaklinks=true,%
   plainpages=false, bookmarksnumbered, bookmarksopen=true,
   bookmarksopenlevel=1,%
   urlcolor=webbrown, linkcolor=webbrown, citecolor=webgreen,
   }

\title[3D global MHD simulations of RIAFs]{3D global simulations of RIAFs: convergence, effects of azimuthal extent and dynamo}   

\author[P. Dhang \textit{\&} P. Sharma]
{Prasun Dhang,$^{1}$\thanks{E-mail:prasundhang@gmail.com}
Prateek Sharma$^{1}$
\\
$^{1}$Department of Physics and Joint Astronomy Programme, Indian Institute of Science, Bangalore, INDIA 560012
}

\begin{document}
\maketitle
\label{firstpage}
\begin{abstract}
We study the long-term evolution of non-radiative geometrically thick ($H/R\approx 0.5$) accretion flows using 3D global ideal MHD simulations and a pseudo-Newtonian gravity. We find that resolving the scale height with 42 grid
points is adequate to obtain convergence with the product of quality factors $\la \la Q_{\theta} \ra \ra \la \la Q_{\phi} \ra \ra \geq 300$ and magnetic tilt angle $\theta_B \sim 13^{\circ}-14^{\circ}$. Like previous global isothermal thin disk simulations, we find stronger mean magnetic fields for the restricted azimuthal domains. Imposing periodic
boundary conditions with the azimuthal extent smaller than $2\pi$ makes the turbulent field at low $m$ appear as a mean field in the runs with smaller azimuthal extent.
But unlike previous works, we do not find a monotonic trend in turbulence with the azimuthal extent. We conclude that the minimum azimuthal extent should be $\geq \pi/2$ to capture the flow structure, but a full $2 \pi$ extent is necessary to study the dynamo.
We find an intermittent dynamo cycle, with $\alpha$-quenching playing an important role in the nonlinear saturated state. Unlike previous local studies, we find almost similar values of kinetic and magnetic $\alpha$-s, giving rise to an irregular distribution of  dynamo-$\alpha$. 
The effects of dynamical quenching are shown explicitly for the first time in global simulations of accretion flows. 
\end{abstract}


\section{Introduction}
Accretion of matter onto compact objects is the main source power in most energetic sources in the universe; e.g. X-ray binaries (XRBs), active galactic nuclei (AGNs), Gamma ray bursts (GRBs).
\cite{Shakura_sunyaev1973,Novikov_thorne1973} provided a phenomenological model for geometrically thin cold disks with $H/R \ll 1$. Thin disks are supposed to power luminous AGN (\citealt{Koratkar1999}) and XRBs in the high soft state (\cite{Remillard2006}). The accretion rate in the thin disk ranges from few percent of the Eddington rate to the Eddington rate (for a review see \citealt{Yuan2014}). In contrast to the cold thin disk, there are geometrically thick ($H/R\sim 0.5$ ) hot accretion flows with smaller accretion rates that power the under luminous sources; e.g. low luminous AGNs, XRBs in hard state. These hot radiatively inefficient accretion flows (RIAFs) do not radiate efficiently because of a smaller density (e.g. see Fig. 3 in \citealt{Das2013}), and hence have a larger temperature and scale-height (\citealt{Ichimaru1977,Narayan_Yi1994,Narayan_Yi1995,Abramowicz1995}). On the other extreme, the radiation dominated slim disks (\citealt{Abramowicz1988}) with super-Eddington accretion rates, are also geometrically thick. In this paper we study the thick disks of the former kind, referred to as RIAFs.    

Angular momentum transport in accretion flows is a long-standing problem. Microscopic viscosity alone is inadequate to give the required transport in order to explain the observed accretion rates.  \cite{Shakura_sunyaev1973} explained the angular momentum transport in terms of an emergent turbulent viscosity. But the source of turbulence was not identified until \cite{Balbus_hawley1991} rediscovered (considered previously by \citealt{Velikhov1959} and \citealt{Chandrasekhar1960}) and highlighted a local linear MHD instability in Keplerian flows, the magneto-rotational instability (MRI), as a solution to the angular momentum problem (for a review see \cite{Balbus_hawley1998}). Although the MRI ensures outward angular momentum transport in the linear regime, understanding the instability in the nonlinear regime is essential. Till date, the nonlinear regime has been studied extensively in the local shearing box simulations. While MRI is a local instability and in the geometrically thin disks ($H/R \ll1$) the global effects are not that important (but see \citealt{Regev2008, Beckwith2011}) as in hot/non-radiative accretion flows ($H/R \sim 0.5$), the focus of this paper.

For the numerical results to be trusted, it is critical that they are converged. For MRI simulations this means that the statistical properties of the MHD turbulent flow do not depend on the grid resolution. Several local shearing box MRI simulations with and without any explicit dissipation (viscosity or resistivity) and with and without vertical stratification have been performed to 
investigate convergence. Some of these studies reached divergent conclusions. While most of the zero net flux unstratified simulations without explicit dissipation (\citealt{Fromang2007_I, Pessah2007,Guan2009, Bodo2011}) show that the turbulent transport diminishes as the grid resolution is increased, recent study of \cite{Shi2016}  with tall boxes is able to show convergence. In presence of a net flux, all the unstratified (\citealt{Hawley1995, Sano2004, Guan2009}) and stratified (\citealt{Stone1996, Miller2000}) simulations show convergence and non-zero transport in the high resolution limit. Interestingly, \cite{Davis2010} and \cite{Shi2010} claim that vertically stratified zero net flux simulations are converged, however, recent studies of \cite{Bodo2014} and \cite{Ryan2017} challenge this claim.

In contrast to local shearing boxes, global simulations are more realistic as they capture the large scale structure of the accretion flow but cannot resolve the MRI turbulence as nicely. Moreover, the shearing box setup imposes several restrictions/symmetries (e.g., no sense of the direction of mass transport, conservation of magnetic fluxes across various boundaries) on the flow that are not respected by real accretion flows. There are only a handful of works (\citealt{Fromang_nelson2006, Flock2010, Hawley2011, Sorathia2012, Shiokawa2012, Hawley2013, Parkin2013}) that study convergence in global simulations. To quantify convergence in our global simulations, we rely on the  statistical behaviour of the convergence metrics in the quasi-steady state (QSS) of the turbulent accretion flow, that is during the non-linear evolution of the MRI.

Self-consistent MRI simulations have to be carried out in 3D because of the impossibility of the sustenance of magnetic fields (and hence turbulence which is driven by the MRI) for long times in 2D (e.g., see \citealt{ARC1998}). However, 3D simulations are expensive. Therefore it is desirable to reduce the azimuthal domain size to a small fraction of $2\pi$, provided the results (in particular, the level of transport and mean/fluctuating quantities) are similar to the full extent. Till now different studies have found different effects of azimuthal extent. By studying two different sets of simulations (namely {\em initial vertical field and adiabatic} equation of state and {\em initial toroidal field and isothermal} equation of state), \cite{Hawley2001} concluded that azimuthal domain size of $\Phi_0=\pi/2$ produces a quasi-stationary turbulent state qualitatively similar to the full $\Phi_0=2 \pi$ extent. Unstratified global simulations of \cite{Sorathia2012} also did not show any considerable difference between the domain sizes $\Phi_0=\pi/4$ and $\Phi_0 = 2 \pi$. Note that both these works use a cylindrically symmetric (and not the more realistic spherically symmetric) potential. In contrast, by performing explicit comparison of global simulations with different azimuthal extents \cite{Flock2012}, showed that models with smaller azimuthal domain size give rise to a higher accretion stress. They attribute the high accretion stress to the stronger mean fields produced by an $\alpha-\Omega$ dynamo. Here, we systematically investigate the effects of azimuthal domain size for a geometrically thick RIAF ($H/R\sim 0.5$) by looking at its different temporal and spatio-temporal quantities in the turbulent steady-state. 

 Large scale fields  play a crucial role in producing jets (\citealt{Blandford_Znajek1977, Blandford_Payne1982}) and coronal winds in accretion flows.  RIAFs, which are supposed to exist in XRBs in the hard state, are prone to outflows (\citealt{Remillard2006}). The generation of  large scale fields in accretion flows is debatable. The large scale field can be the due to the advection of field (\citealt{Lubow1994, Lovelace2009, Guilet_Ogilvie2012}) or due to {\em in situ} production by a large scale dynamo (\citealt{Brandenburg2005}). Both unstratified (with sufficient vertical extent; \citealt{Lesur2008, Shi2016}) and stratified (\citealt{Brandenburg1995,Davis2010,Shi2010, Gressel2010}) local shearing box, and global thin disk simulations (\citealt{Beckwith2011, Flock2012, Suzuki2014, Hogg2016}) show generation of large scale fields due to a dynamo process. While the early dynamo models were kinematic, with the mean flows and the statistical properties of the flow specified, MRI turbulence cannot be in the kinematic regime because the instability is driven by magnetic tension affecting the velocity perturbations (for a review see \citealt{Balbus_hawley1998}). We investigate the generation of large scale magnetic fields and the self-sustained dynamo process in a geometrically thick RIAF. As  there is no externally imposed net field and we run our simulations for a long time, we expect the advection of field do not play a significant role in generating large scale fields.
 
 In this work, we perform 3D global simulations of radiatively inefficient accretion flows with three aims: i) convergence, that is the minimum number of grid cells required to properly capture the nonlinear saturation of MRI turbulence and to quantify the converged solutions; ii) the effects of azimuthal extent of the computational domain on the different properties of the mean and turbulent field evolution; iii) the saturation process of magnetic energy, i.e. on the dynamo mechanism in a radiatively inefficient geometrically thick hot accretion flow. 
 
 The paper is organized as follows. In section \ref{sect:method_disk} we discuss the physical set-up, solution method and different diagnostics used in the work. In section \ref{sect:flow_evo} we discuss the evolution of the accretion flow for our fiducial run. Convergence is discussed in section \ref{sect:convergence}. We discus the effects of the azimuthal extent on the accretion flow and the dynamo mechanism in  sections \ref{sect:azimuth_effect} and \ref{sect:dynamo} respectively. Finally the key points of results are discussed and summarized in sections \ref{sect:discuss_disk} and \ref{sect:summary}.
\section{Method}
\label{sect:method_disk}
\subsection{Equations solved}
We solve the Newtonian MHD equations in spherical co-ordinates ($r,\theta,\phi$) 
using the {\tt PLUTO} code (\citealt{Mignone2007}). The equations  are
 \ba 
 \label{eq:mass}
&& \frac{\partial \rho}{\partial t} + \nabla .(\rho \textbf{v})= 0, \\
\label{eq:momentum}
&& \frac{\partial }{\partial t}\left(\rho \textbf{v}\right)+ 
\nabla . \left( \rho \textbf{v}\textbf{v} - \textbf{B}\textbf{B} \right)=  -\rho \nabla \Phi - \nabla P^*, \\
\label{eq:energy}
&& \frac{\partial E}{\partial t} + \nabla . \left((E + P^*)\textbf{v} - \textbf{B}(\textbf{B}.\textbf{v})   \right)= -\rho \textbf{v}.\nabla \Phi, \\
\label{eq:induction}
&& \frac{\partial \textbf{B}}{\partial t} + \nabla . \left (\textbf{v} \textbf{B} - \textbf{B}\textbf{v} \right) = 0,
\ea
where $\rho$ is the gas mass density, $\textbf{v}$ and $\textbf{B}$ are the velocity and magnetic fields respectively,
$P^*$ is the total pressure given by $P^* = P + B^2/2$ ($P$ is  gas pressure), $E$ is the total energy density which is related to the internal energy 
density 
$e$ as $E = e + \rho u^2/2 + B^2/2$. We use an {\em ideal} equation of state where internal energy is defined as $e =P/(\gamma -1)$ with $\gamma=5/3$. To mimic the  general relativistic effects close to the black hole,
we use the pseudo-Newtonian potential (\citealt{Paczynsky_wiita1980}) $\Phi = GM/(r-2r_g)$, where $r_g = GM/c^2$ is the gravitational radius,
$M$ and $c$ are the mass of the accreting black hole and the speed of light in vacuum respectively. We work in dimensionless units in which 
$GM=c=1$. Therefore,  in this paper all the length scales and velocities are given  in the units of $GM/c^2$ and $c$ respectively. 
Unless stated otherwise, time scales are expressed in terms of the number of orbits a test particle would do at  the {\em inner most 
stable circular orbit} (ISCO), and is given by
\be
N_{\rm ISCO} = \frac{t_{\rm sim}}{T_{\rm ISCO}},
\ee
where the simulation time is $t_{\rm sim}$ and the orbital period at ISCO $T_{\rm ISCO}= 2 \pi r_{\rm ISCO}^{1/2}(r_{\rm ISCO}-2)$ are 
expressed in the units of $GM/c^3$. For a Schwarzchild black hole the location of ISCO is $r_{\rm ISCO}= 6 r_g$.

{\tt PLUTO} uses a conservative Godunov scheme. All quantities are stored at the cell center, except 
for the magnetic field components that are face centered and staggered. A staggered magnetic field is required for the constrained
transport scheme to evolve the induction equation (Eq. \ref{eq:induction}; \citealt{Evans1988}). This preserved $\nabla \cdot {\bf B}=0$ 
to machine precession.
Out of several possible combinations of algorithms, we follow the recommendations of \cite{Flock2010} to properly capture the MRI modes. 
We use  the HLLD solver (\citealt{Miyoshi2005}) with second-order slope limited 
reconstruction. For time-integration, second order Runge-Kutta (RK2) is used with the CFL number 0.3. For the calculation of the 
electromotive forces (EMFs) for the induction equation, we use the upwind CT `contact' method (\citealt{Gardiner_stone2005}).

In our simulations, we find that the time step is determined by the large Alfv\'en speed $v_A=B/\sqrt{\rho}$. As in Newtonian MHD there is no 
speed limit and $v_A$ becomes unusually large in the low density regions near the poles. This results in impractically short time-steps due to the 
CFL condition. We fix this problem by imposing a density floor such that $v_A < v_{A,{\rm max}}=5$. For most of the runs, $v_A$ seldom touches 
the limit and the total mass added by floors is insignificant. Moreover, mass is added only in the regions near the poles and in the highly 
magnetized `funnels', which we do not incorporate in our analysis.

\subsection{Initial conditions}
We initialize the simulation domain with the equilibrium solution given by \cite{Papaloizou_pringle1984}, which describes a constant angular 
momentum torus embedded in a non-rotating, low density hydrostatic medium. Pressure and density within the torus 
follow a polytropic equation of state $P = K \rho^{\gamma}$ ($K$ is a constant). The initial 
density of the torus is given by
\be
\label{eq:torus_den}
\rho^{\gamma -1} = \frac{1}{(n+1)R_0 K} \left[ \frac{R_0}{r-2} - \frac{1}{2}\frac{R^4 _0}{[(R_0-2)R]^2} - \frac{1}{2d} \right ], 
\ee
where $\gamma$ is the adiabatic index, $n = 1/(\gamma -1)$ is the polytropic index, $R=r {\rm sin} \theta$ is the cylindrical radius, 
$R_0$ is the cylindrical radial distance of the center of the torus from the black hole, $d$ is the distortion parameter which determines 
the shape and size of the torus. The density of the torus is maximum ($\rho = \rho_0$) at $r=R_0$ and $\theta=\pi/2$. This 
information is used to calculate
\be
\label{eq:ken}
K = \frac{1}{(n+1)R_0 \rho^{\gamma-1}_0} \left[ \frac{R_0}{R_0-2} - \frac{1}{2}\frac{R^2 _0}{(R_0-2)^2} - \frac{1}{2d} \right].
\ee
The constant angular momentum $l_0$ associated with the torus is given by its Keplerian value at $R_0$, 
\be
\label{eq:lkep}
l_0 = l_K(R_0)= \frac{R^{\frac{3}{2}}_{0}}{R_0 - 2}
\ee
We choose $\rho_0 = 10^6$ in code units, $R_0 = 20$ and $d=1.15$. The chosen value of $d$ makes the initial torus geometrically thick 
($H/R \sim 0.1$ at $R_0$). We choose the density of ambient the medium to be small enough ($\rho_{\rm amb}=10^{-4}\rho_0$), such that it does not affect our results. For all the runs the initial torus is seeded with white noise of $|\delta v_r| = 10^{-5}$.

We initialize a poloidal magnetic field which threads the initial torus and is parallel to the density contours. This magnetic 
field is defined through a vector potential
\be
\label{eq:vec_poten}
A_{\phi} = C\rho^2,
\ee
which guarantees the divergence free nature of the magnetic field; $C$ is a constant that determines the field strength. The 
initial magnetic field strength is quantified by the ratio of volume averaged (over torus) gas to magnetic pressure
\be
\label{eq:beta_ini}
\beta_{\rm ini} = \frac{P_{V} }{B_V^2/2},
\ee
We choose $\beta_{\rm ini} = 890$ in our simulations. 

\subsection{Numerical set-up}
\label{sect:setup}

\begin{table*}
\centering
\begin{tabular}{ |p{2cm}||p{1.5cm}||p{1.5cm}||p{1.5cm}|p{1.5cm}||p{1.5cm}|p{1.5cm}|p{1.5cm}|  }
 \hline
 \multicolumn{8}{|c|}{Numerical set-up} \\
 \hline
 Name      		& $\Phi_0$ & $N_{rl} $ & $N_{rs}$ & $N_{\theta u}$ & $N_{\theta s}$ & $N_{\phi}$     & $(r \Delta \phi/\Delta r)_{\theta=\pi/2}$       \\
 \hline
 L-P4  			& $\pi/4$         &  144      &  32      &      64        &     16         & 32     & 1.46     \\
 \\
 M-P4			& $\pi/4$	        &  296      &  72      &      128      &     32         & 64      & 1.50      \\
 M-P2  			& $\pi/2$         &  296      &  72      &      128      &     32         & 128    & 1.50        \\
 M-1P			& $\pi$	        &  296      &  72      &      128      &     32         & 256    & 1.50     \\
 M-2P			& $2\pi$	        &  296      &  72      &      128      &     32         & 512    & 1.50     \\

\\
 H-P4			& $\pi/4$	        &  536      &  32      &      232       &     24         & 128   & 1.36      \\
 	
 \hline
\end{tabular}
\caption{In the radial direction, $N_{rl}$ logarithmically spaced grids are used between $r_{\rm in}=4$ to $r_{l}=45$ and $N_{rs}$ stretched grid points are employed in the outer buffer zone extended from $r_l=45$ to $r_{\rm out}=140$. Along the meridional direction, between $\theta_1= 60^{\circ}$ and $\theta_2= 120^{\circ}$ $N_{\theta u}$ number of uniform grid points are used. On both sides of this region we use stretched grids with $N_{\theta s}$ points. Total number of grid points in the radial($N_r$) and meridional ($N_{\theta}$) are given by, $N_r=N_{rl}+N_{rs}$ and $N_{\theta} = N_{\theta u}+2N_{\theta s}$. $N_{\phi}$ is the number of grid points spread uniformly over the azimuthal extent $\Phi_0$.  }
\label{tab:simtab}
\end{table*}

\begin{figure}
\centering
 \includegraphics[scale=0.23,trim=200 100 100 100,clip]{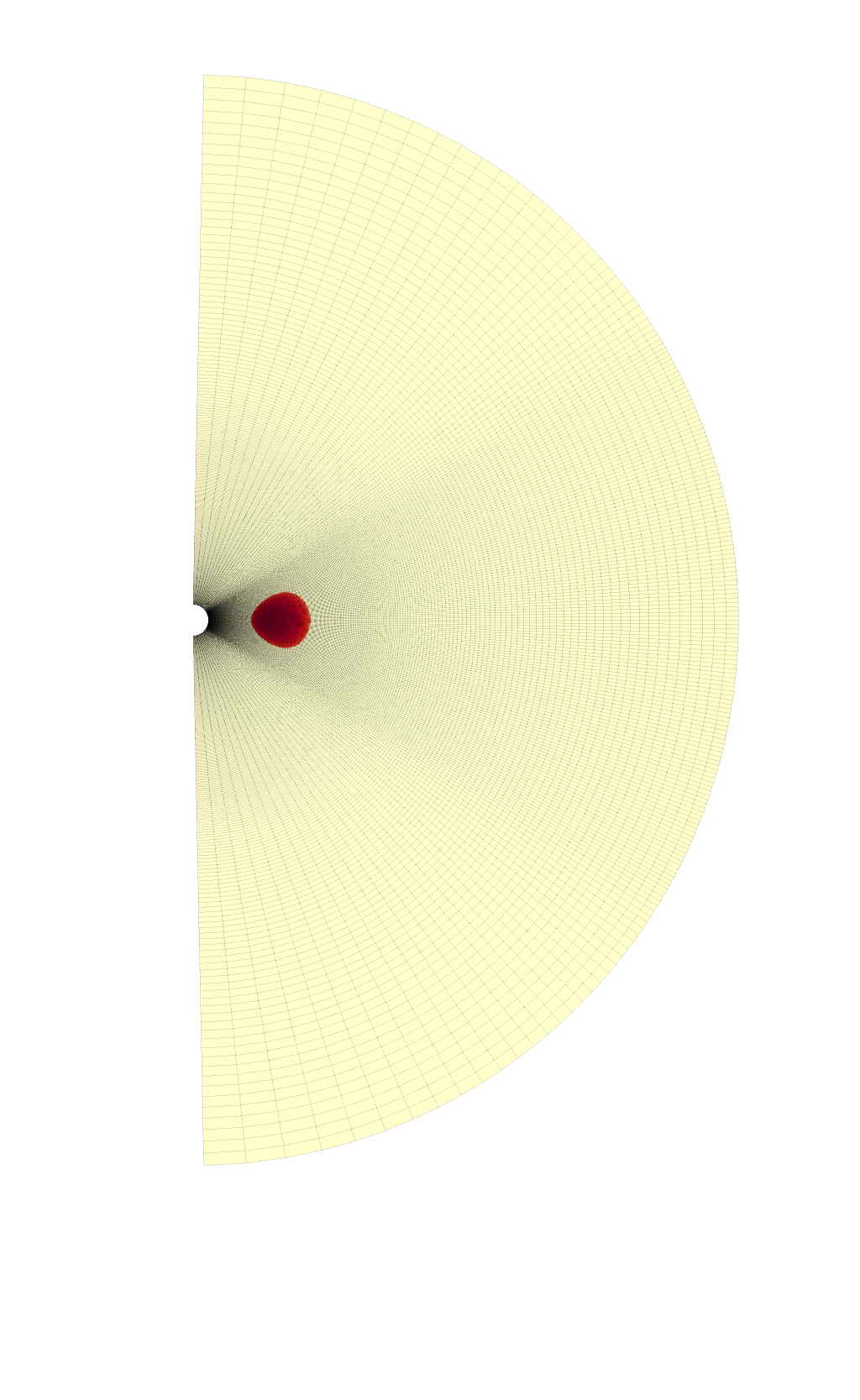}
 \includegraphics[scale=0.16,trim=10 60 30 60,clip]{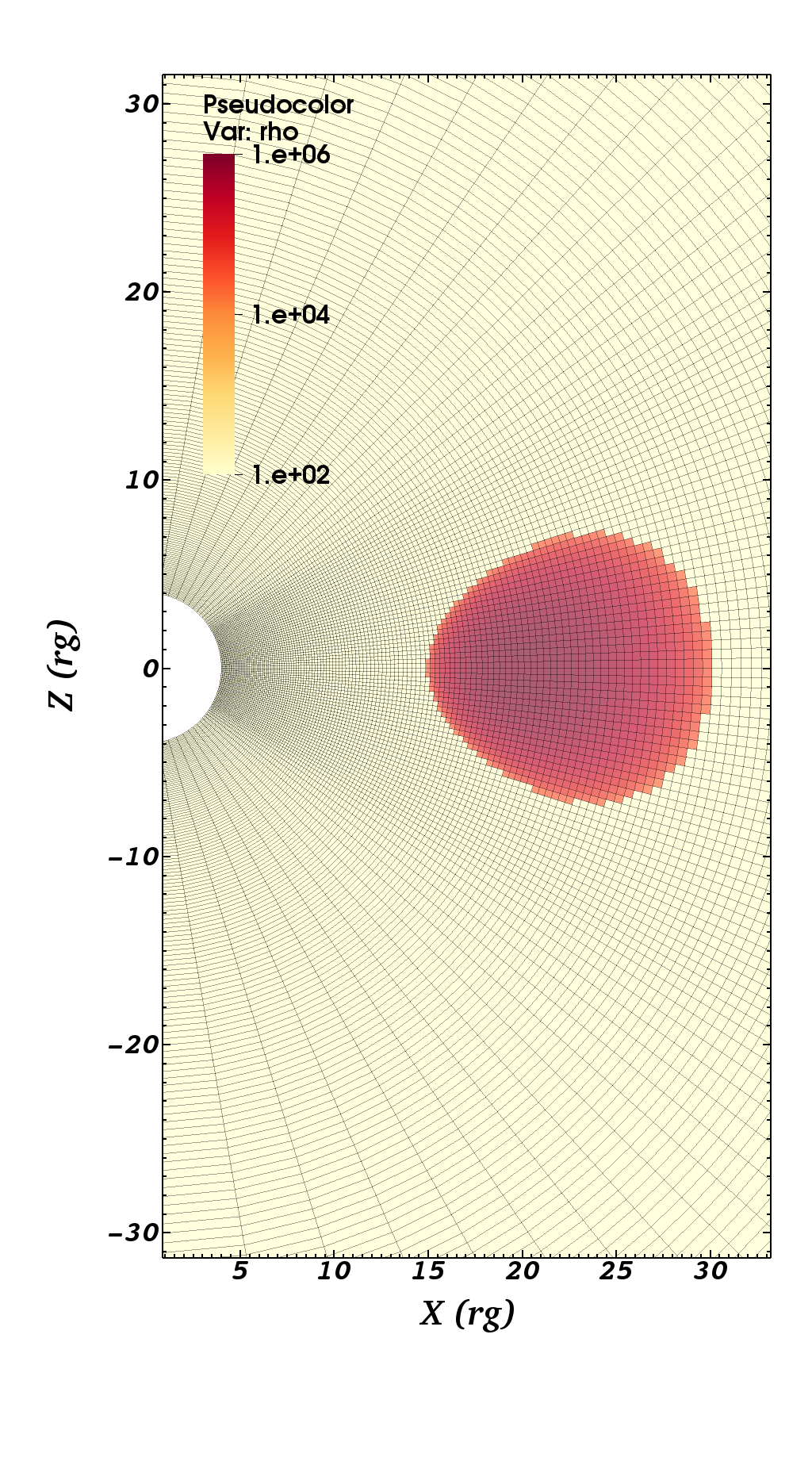}
\caption{Grid structure of the simulation set-up. Top panel shows the full computational domain. Bottom panel shows the zoomed in view of the 
region where the initial torus is embedded. We see that the initial torus is well inside the well-resolved region. The grids are employed in 
such a way that the initial torus as well as the evolved accretion flow are well resolved.}
\label{fig:mesh}
\end{figure}

We carry out six 3D MHD simulations in spherical co-ordinates ($r,\theta,\phi$). The computational domain extends from 
an inner radius $r_{\rm in}=4$ to an outer radius $r_{\rm out}=140$. The advantage of putting the inner boundary within the 
innermost stable circular orbit (ISCO)
is that the accretion velocity $v_r$ becomes supersonic (see the top left panel of Fig. \ref{fig:velocity_r}) before reaching the inner 
boundary. As a result, the flow properties outside sonic radius ($r_c$) become independent of the details of the inner boundary 
conditions (\citealt{McKinney2002}). We also place the radial outer boundary far away from the torus to avoid significant mass loss through it.
We use a logarithmic grid along the radial direction from $r_{\rm in}=4$ to $r_{l}=45$ with $N_{rl}$ grid points. 
In the outer region, from $r_l=45$ to $r_{\rm out}$, we use a stretched grid with $N_{rs}$ grid points. This outer region acts as a buffer zone. Along the 
meridional direction, most of the grid points are concentrated in the equatorial region, with domain extends from $\theta=0.02$ to $\theta=\pi-0.02$.  In the region between $\theta_1= 60^{\circ}$
and $\theta_2= 120^{\circ}$, we use $N_{\theta u}$ grid points with uniform spacing. On both sides of this region we use stretched grids
with $N_{\theta s}$ points. We choose the the value of $\theta_1$
and $\theta_2$ in such a way that initial torus, as well as the evolved accretion flow are inside the well resolved region. 
So the total number of grid points in the radial ($N_r$) and the meridional ($N_{\theta}$) directions are given by $N_r=N_{rl}+N_{rs}$ and 
$N_{\theta} = N_{\theta u}+2N_{\theta s}$. In the azimuthal direction we employ $N_{\phi}$ uniform grid points in the domain of extent
$\Phi_0$. We choose $N_{rl}$, $N_{\theta u}$ in such a way that aspect ratio $\Delta r/ r \Delta \theta \approx 1$ for every cell close to 
mid-plane and in between $r_{\rm in}$ to $r_l$. Previous studies (\citealt{Hawley2011}, \citealt{Sorathia2012}) recommend that  azimuthal 
resolution is very important in order to achieve convergence in the saturated state. Following \cite{Sorathia2012}, we always keep the ratio 
$ (r \Delta \phi/\Delta r)_{\theta=\pi/2} < 2$. Details of the simulation set-ups are tabulated in Table \ref{tab:simtab}. 
A visualization of the grid structure is shown in Fig. \ref{fig:mesh}. The top panel shows the grid structure of the  whole computation domain. 
The bottom panel shows the portion of the computation domain  where the initial torus is embedded. 

We use a pure inflow ($v_r\leq 0$) boundary condition at the radial inner boundary, so that the inner boundary acts as a one-way membrane .
Zero gradient boundary conditions are used for $\rho$, $p$, $v_{\theta}$ and $v_{\phi}$.
At the radial outer boundary, we fix the density and pressure to their initial values, $v_{\theta}$ and $v_{\phi}$ are kept free,
but we set $v_r \geq 0$ so that matter can go out of the computation domain but can not come in. Both at the inner and outer radial boundaries, 
 `force-free' (zero-gradient) conditions are applied  for the tangential components ($B_r$ and $B_{\theta}$) of $\textbf{B}$-field, while the normal 
 component of $\textbf{B}$-field  is determined by CT. Reflective and periodic boundary conditions are used at the meridional and azimuthal boundaries respectively.

\subsection{Diagnostics}
\label{sect:diag}
We discuss temporal, spatial and spatio-temporal behaviour of several quantities. The averaging methods and the common diagnostics are defined below.
\subsubsection{Averaging method}
\label{sect:avg_method}
 For any quantity $q$, we consider following spatial averages, 
 \begin{align}
  \label{eq:vol_avg}
 \begin{split}
 \langle {q(t)} \rangle  &= \frac{1}{V}\int q(r,\theta,\phi)  dV, \\
\langle {q(r,t)} \rangle  &= \frac{1}{S_{r}}\int q(r,\theta,\phi) dS_{r}, \\
\langle {q(\theta,t)} \rangle  &= \frac{1}{S_{\theta}}\int q(r,\theta,\phi) dS_{\theta}, \\
\langle {q(r, \theta, t)} \rangle  &= \frac{1}{l_{\phi}}\int q(r,\theta,\phi) dl_{\phi}, \\
\langle {q(r, \phi, t)} \rangle  &= \frac{1}{l_{\theta}}\int q(r,\theta,\phi) dl_{\theta}, \\
 \end{split}
 \end{align}
where $V =\int dV   = \int r^2 {\rm sin} \theta dr d\theta d\phi, ~S_{r} =\int dS_{r}=  \int {\rm sin} \theta d\theta d\phi, 
~S_{\theta} =\int dS_{\theta}=  \int r^2 dr d\phi,
~ l_{\phi} = \int dl_{\phi} = \int  d\phi, ~l_{\theta}=\int  dl_{\theta}  = \int {\rm sin} \theta d \theta$. The integrations are performed over the
sub-domain $\mathcal{S} = [r_{\rm ISCO}, 40] \times [\pi/2 -\theta_{H}, \pi/2 +\theta_{H}] \times [0,\Phi_0]$, where $\theta_{H}={\rm cos}^{-1}(H/R)$.
$\mathcal{S}$ corresponds to the well-resolved region of the computational domain.
Here, the scale-height $H$ is given by
\be
\label{eq:scale_height}
H = \frac{\langle c_s(r,\theta=\frac{\pi}{2}) \rangle }{\langle \Omega (r,\theta=\frac{\pi}{2}) \rangle},
\ee
where $c_s=\sqrt{\gamma P/\rho}$ and $\Omega=v_\phi/R$ are sound speed  and angular velocity  respectively.

To look at the time averaged radial and meridional variation we compute,
\begin{align}
\label{eq:radial_avg}
\begin{split}
\la \la q(r) \ra \ra = \frac{1}{\Delta T} \int_{T_1}^{T_2} \langle q(r,t)\rangle dt,\\
\la \langle q(\theta) \rangle \ra = \frac{1}{\Delta T} \int_{T_1}^{T_2} \langle q(\theta,t)\rangle dt, \\
\end{split}
\end{align}
respectively; $\Delta T = T_2 - T_1$ is the time interval over which the average is done.

To get a volume and time averaged quantity, we do the following,
\be
\la \la q \ra \ra = \frac{1}{\Delta T} \int_{T_1}^{T_2} \la q(t) \ra dt.
\ee

 \subsubsection{Mass and angular momentum accretion} 
Mass [$\dot{M}(r_{\rm ISCO},t)$] and angular momentum [$\dot{L}_{\rm acc}(r_{\rm ISCO},t)$] accretion  rates across ISCO are given by
\be
\label{eq:mdot_isco}
\dot{M}(r_{\rm ISCO},t) = r^2_{\rm ISCO} \int_{-\theta_H}^{+\theta_H} \int_{0}^{\Phi_0} \rho v_r  dS_{r},
\ee

\be
\label{eq:Ldot_isco}
\dot{L}_{\rm acc}(r_{\rm ISCO},t) = r^3_{\rm ISCO} \int_{-\theta_H}^{+\theta_H} \int_{0}^{\Phi_0} \left (\rho v_r v_{\phi} + W^{\rm Max}_{r \phi } \right )  dS_{r}
\ee
respectively.
Here, the $r-\phi$ components of the total accretion stress $W^{T}_{r \phi}$ is given by
\be
\label{eq:tot_stress}
W^{T}_{r \phi} = W^{\rm Re}_{r \phi} + W^{\rm Max}_{r \phi}, 
\ee
where $W^{\rm Re}_{r \phi}$ and $W^{\rm Max}_{r \phi}$ are the Reynolds and Maxwell accretion stresses respectively.
Following \cite{Hawley2000}, we compute the Reynolds stress $W^{\rm Re}_{r \phi}$ in terms of the difference between instantaneous angular momentum 
flux, and the average mass flux times specific angular momentum,
\be
\label{eq:reynolds_stress}
\la W^{\rm Re}_{r \phi}(r,t) \ra = \langle{\rho v_r v_{\phi}} (r,t) \rangle - \langle{\rho v_r (r,t)} \rangle \langle {v_{\phi}(r,t)} \rangle.
\ee
Since  here MHD turbulence is subsonic, $|v^{\prime}|/c_s \lesssim 0.2$ (see Table \ref{tab:mean_vs_turb}), 
density perturbations are small. Maxwell stress $W^{\rm Max}_{r \phi}$ is defined as,
\be
\label{eq:maxwell_stress}
W^{\rm Max}_{r \phi} = -B_r B_{\phi}.
\ee
Therefore, the normalized stresses  are given by
\be
\label{alpha_t}
\la \alpha^{T} \ra  = \la \alpha^{\rm Re} \ra  + \la  \alpha^{\rm Max} \ra =\frac{\langle W^{\rm Re}_{r \phi} \rangle +\langle W^{\rm Max}_{r \phi} \rangle }{\langle {P} \rangle}.
\ee

\subsubsection{Resolvability}
To check the resolvability of the MRI we look at the two quality factors,
\bea
\label{eq:quality factors}
&& Q_{\theta} = \frac{\lambda^{\rm MRI}_{\theta}}{r \Delta \theta}, \\
\label{eq:qphi}
&& Q_{\phi}  = \frac{\lambda^{C}_{\phi}}{r {\rm sin} \theta \Delta \phi},
\eea
where the quality factor $Q_{\theta}$ measures the number of cells in the $\theta$-direction across a wavelength of the fastest growing mode,
$\lambda^{\rm MRI}_{\theta} = 2 \pi |B_{\theta}|/\Omega \sqrt{\rho} $ and $Q_{\phi}$ measures the number of cells in the $\phi$-direction across a 
wavelength of the fastest growing mode, $\lambda^{C}_{\phi} = 2 \pi |B_{\phi}|/\Omega \sqrt{\rho}$;  $\Delta \theta$ and $\Delta \phi$ are 
the grid sizes in the $\theta$ and $\phi$-directions respectively, $B_{\theta}$ and $B_{\phi}$ are the magnetic field components, 
$\Omega=v_{\phi}/R$ is the fluid's 
angular velocity. Although the quality factors were introduced to check the resolvability of the linear MRI (\citealt{Sano2004}, 
\citealt{Fromang_nelson2006}), they were found to
be an important diagnostic in the fully non-linear turbulent regime too (\citealt{Sorathia2012}, \citealt{Hawley2013}). While quality factors 
give the information of how well the MRI is resolved, they do not reflect the structure of turbulence in the saturated state. Magnetic tilt angle
\be
\label{eq:tilt_angle}
\la \theta_B(t) \ra = \frac{1}{2}{\rm sin}^{-1} \left(\frac{-\langle 2 B_r B_{\phi} (t) \rangle}{\langle B^2 (t) \rangle} \right)
\ee
above a critical value confirms the transition from linear growth of the MRI to the saturated turbulence, 
and the critical value of this metric is $\theta_{B,c} \sim 12^{\circ}$ 
(\citealt{Pessah2010}). It is also a measure of the magnetic field anisotropy which is a key factor behind the angular momentum transport.

\subsubsection{Power spectral density}
\label{sect:PSD}
We also study the azimuthal spectral structure of the turbulent flow in the non-linear regime. We compute the toroidal power spectral density (PSD)
\be
\mathcal{P}_{q}(r,\theta,m,t) = \frac{1}{\Delta m} \left | \int q(r,\theta,\phi,t) e^{i m\phi}d \phi \right | ^2 
\ee
of a physical quantity $q(r,\theta,\phi,t)$, where $\Delta m = 2 \pi/\Phi_0$. The spatio-temporal average of  $\mathcal{P}_{q}(r,\theta,m,t)$ is given by
\be
\label{eq:psd}
\la \langle {\mathcal{P}_q(m)} \rangle \ra = \frac{\int_{T_1}^{T_2} \int_{r=6}^{r=40} \int_{-\theta_H}^{+\theta_H} \mathcal{P}_q(r,\theta,m,t) r^2 {\rm sin} \theta  dr d \theta dt}{\Delta T\int_{r=6}^{r=40} \int_{-\theta_H}^{+\theta_H} r^2 {\rm sin} \theta  dr d \theta }.
\ee

\section{Evolution of the flow}
\label{sect:flow_evo}
\begin{figure*}
    \includegraphics[scale=0.55]{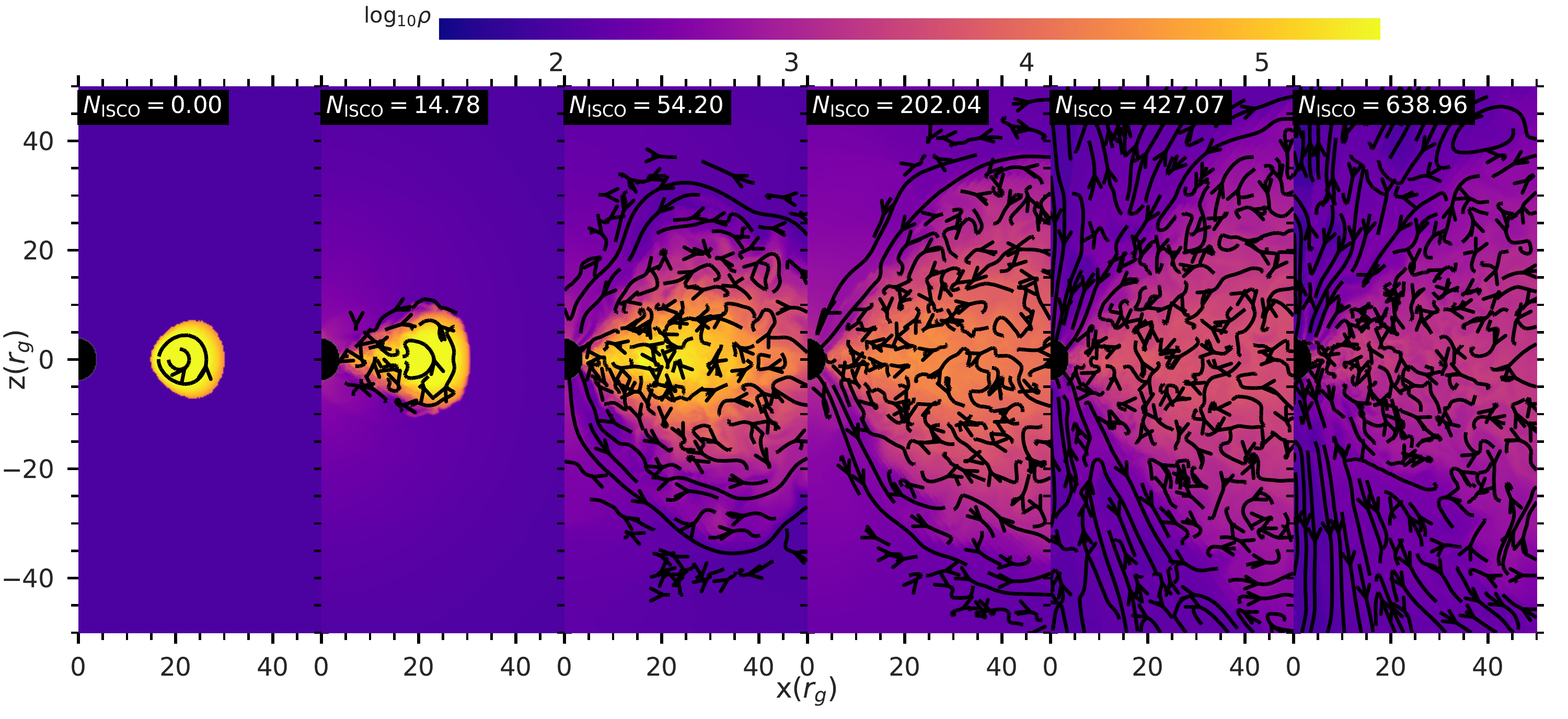}
    \caption{Temporal evolution of the accreting torus for our fiducial model M-2P. The color describes the density, and the instantaneous poloidal magnetic field lines are shown by the streamlines. Starting with the constant angular momentum equilibrium torus threaded by a purely poloidal weak magnetic field (first panel). MRI grows in a dynamical time $t_{\rm dyn}=1/\Omega$ (second panel). Parasitic instabilities take over the MRI when the system enters into the non-linear regime and fully developed MHD turbulence is established  after some time (e.g see the third panel). The last three panels show the accretion flow in the quasi-steady state.      }
   \label{fig:den_strm}
 \end{figure*}

Before discussing the details of our results, we discuss the evolution of the fiducial run M-2P (the medium resolution run with the azimuthal 
extent $\Phi_0=2\pi$), which is converged (see sections \ref{sect:convergence} and \ref{sect:med_quality}). 

\begin{figure}
    \includegraphics[scale=0.22]{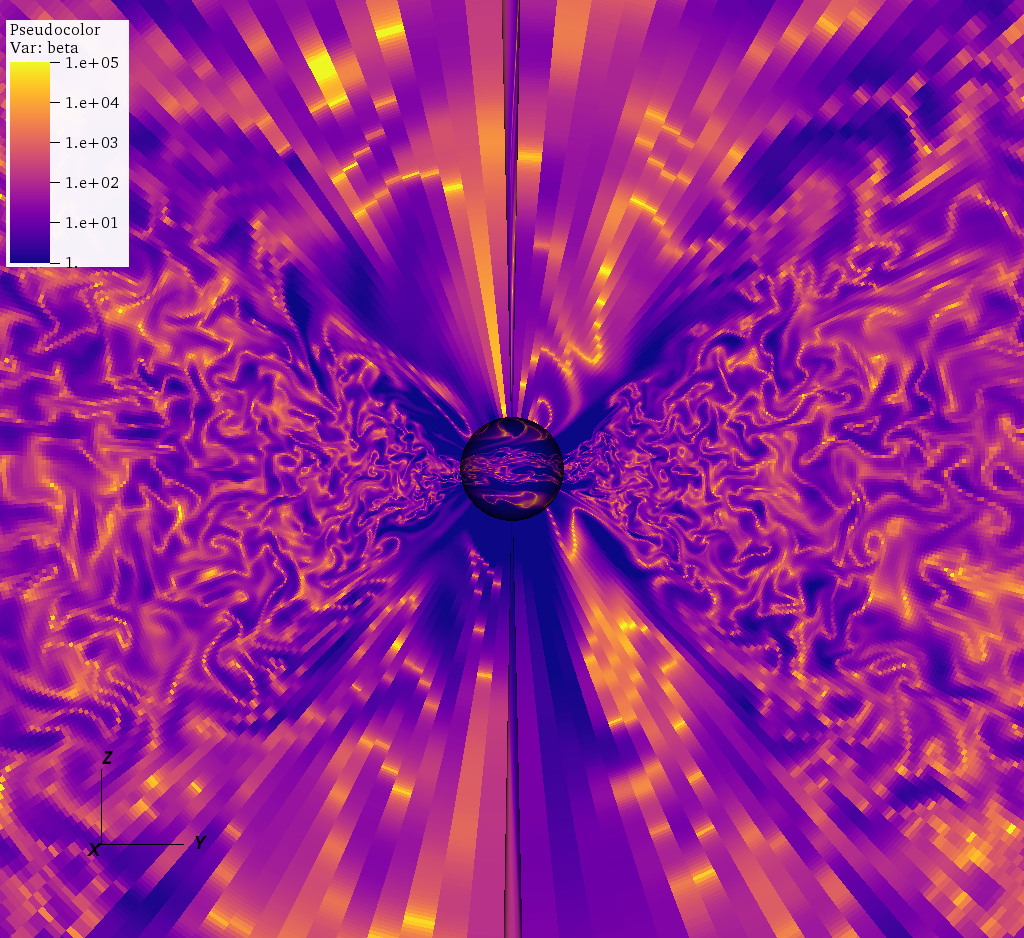}
    \caption{
    A snapshot of $\beta = 2P/B^2$ at time $N_{\rm ISCO}=328.51$ for our fiducial run. The turbulent region on both sides of mid-plane within one scale-height consist of patches of both low and high $\beta$. The laminar region close to the black hole above one scale-height is magnetically dominated ($\beta<1$).}
   \label{fig:beta_iso}
 \end{figure}

Fig. \ref{fig:den_strm} shows the time evolution of density and poloidal magnetic field (shown by streamlines) in the $\phi=0$ plane (x-z plane). The first panel shows the initial condition. Shear produces toroidal field out of the initial poloidal field on a dynamical time scale ($t_{\rm dyn} \approx 1/\Omega$, $\Omega$ is the angular speed, $\Omega \propto 1/R^2$ for the initial constant angular momentum torus). The MRI also grows on a dynamical time. As a result, both poloidal and toroidal fields grow exponentially. Closer to the central black hole, MRI grows faster as the dynamical time $t_{\rm dyn} \propto R^{2}$. As the MRI starts, outward angular momentum transport happens and accretion begins as seen in the second panel of the Fig. \ref{fig:den_strm}. Along with the breaking up of the equilibrium torus due to MRI, torus also expands under the action of magnetic pressure growing due to the background radial shear. As time passes, system enters the non-linear regime and parasitic instabilities take over (\citealt{Goodman1994}, \citealt{Pessah_goodman2009} ), and eventually fully developed MHD turbulence develops throughout the initial torus (see the third panel of Fig. \ref{fig:den_strm}). Last three panels show the snapshots of the accretion flow in the quasi-steady state (QSS). In the QSS, the accretion flow consists of two parts: the highly turbulent region on both sides of the mid-plane within one scale-height and the almost laminar region above that. This can be easily seen in last three panels of Fig. \ref{fig:den_strm} and in Fig. 
\ref{fig:beta_iso}.

 Fig. \ref{fig:beta_iso} shows the three dimensional view of plasma $\beta = 2P/B^2$ in the poloidal plane. Turbulent structures can be seen in the regions around the mid-plane of the accretion flow with both high and low values of $\beta$. As expected, the structures are smaller, closer to the black hole. On the other hand, the laminar region close to the accretor and away from mid-plane is magnetically dominated ($\beta <1$). On average (over time and spatial domain $\mathcal{S}$, see section \ref{sect:avg_method}) $\la \la \beta \ra \ra \approx 20$ (see Table \ref{tab:results_tab}), which shows a stronger magnetization in the QSS compared to the initial magnetization of $\beta_{\rm ini} \approx 890$.
 
 \begin{figure}
    \includegraphics[scale=0.45]{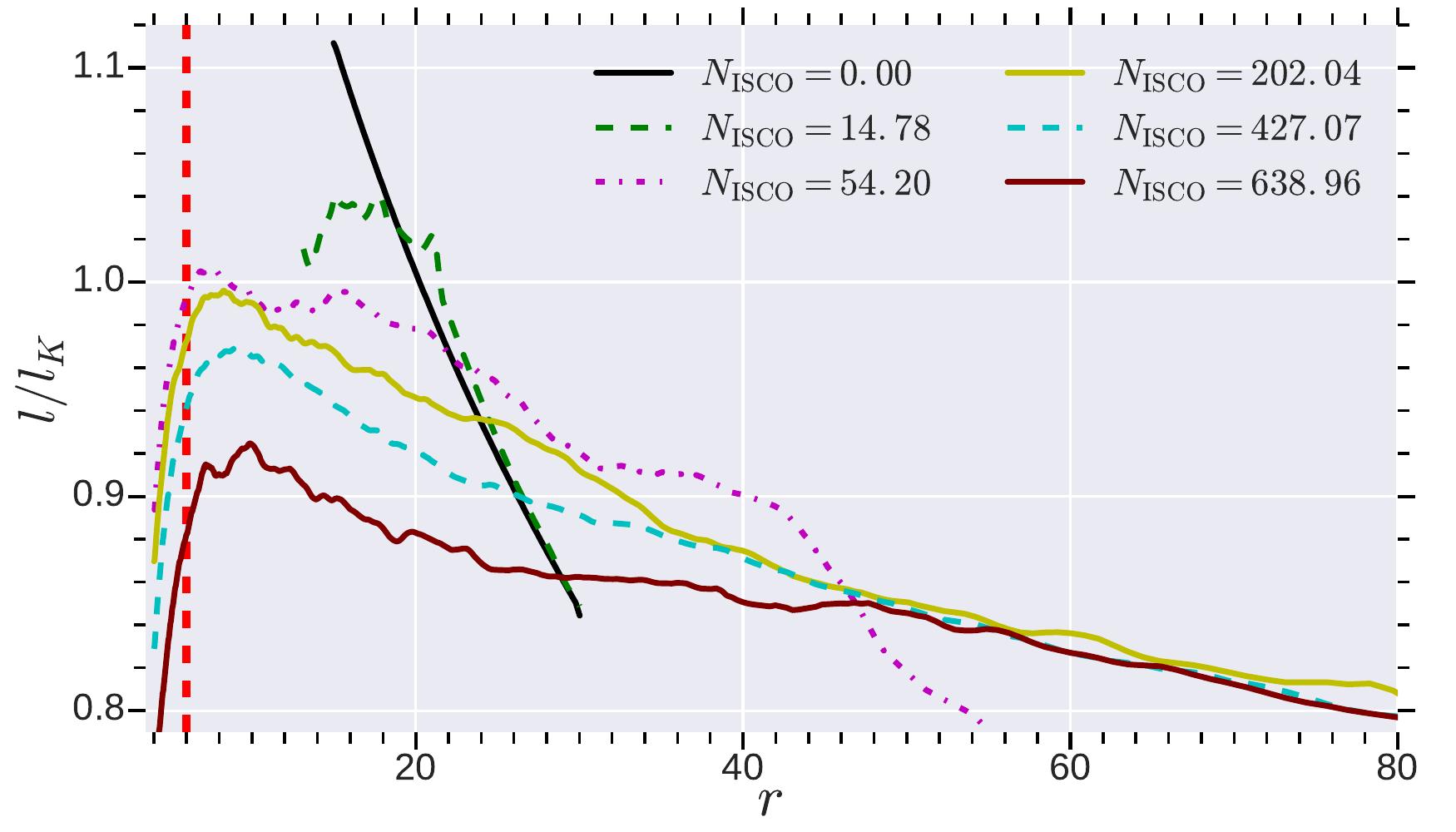}
    \caption{Angular momentum distribution $\la l(r) \ra/l_k(r)$ of accreting matter at different times for our fiducial run M-2P. $l_K=r^{3/2}/(r-2)$ is the Keplerian specific angular momentum at the radius $r$. Red-dashed vertical line shows the location of ISCO. The drop in $l/l_K$ within $6 r_g$ is consistent with an almost constant specific angular momentum of matter plunging within the ISCO.}
   \label{fig:lambda}
 \end{figure}

 Fig. \ref{fig:lambda} shows the radial distribution of average angular momentum $\la l(r) \ra$ at different times. Spatial average is done over the whole azimuthal domain and within one scale-height in the meridional direction. Vertical red line denotes the location of the ISCO. We begin with a constant angular momentum torus with $l_0=l_K(r=R_0=20)$. Initially, for $r<R_0$, the angular momentum distribution $\la l(r<R_0) \ra$ is super-Keplerian, whereas for $r>R_0$, $\la l(r>R_0) \ra$ is sub-Keplerian. As the MRI grows, inner part of the torus loses angular momentum first and subsequently the whole torus becomes unstable against MRI.  As the fully developed MHD turbulence is established, a sub-Keplerian angular momentum profile ($\la l(r)\ra < l_K(r)$) is developed throughout the accretion flow. The flow is more sub-Keplerian further from the accretor. Within the ISCO, as accreting matter plunges in with an almost constant specific angular momentum, $\la l(r)\ra/l_K$ drops rapidly. It is interesting to see that with time, $\la l(r)\ra$ becomes more and more sub-Keplerian. As a result, the amount of angular momentum carried by the matter accreted onto the central black hole also decreases (see the decreasing trend of the $\la l(r)\ra$ at ISCO  with time in Fig. \ref{fig:lambda}; also see the time evolution of $j_{\rm acc}=\dot{L}(r_{\rm ISCO},t)/\dot{M}(r_{\rm ISCO},t)$ in  Figs. \ref{fig:jnet_conv_p4} and \ref{fig:med_mdot_jnet}). The decrease in the angular momentum reaching the black hole at late times is due to two reasons, i) the mass with initial sub-Keplerian angular momentum distribution for $r > R_0$ starts contributing to accretion onto the black hole, ii) the relative increase in total accretion stress $\alpha^T$ at late times (see Fig. \ref{fig:med_alpha_R_M_t}). From Figs. \ref{fig:med_mdot_jnet} and \ref{fig:med_alpha_R_M_t} it looks like that the second factor is dominant over the first one because the runs (M-P2, M-1P, M-2P) that display a higher increase in $\alpha^{T}$, show larger decrease in $j_{\rm acc}$ at late times.

 \section{Convergence}
\label{sect:convergence}
\begin{table*}
\centering
\begin{tabular}{ |p{1cm}||p{2.1cm}||p{2.1cm}||p{2.0cm}|p{1.5cm}||p{1.5cm}|p{1.5cm}|p{1.5cm}|  }
 \hline
 \multicolumn{8}{|c|}{Simulation results} \\
 \hline
 Name      		& $\la \la \alpha^{\rm Re}_{\rm sat} \ra \ra  \times 10^{-2}$ & $\la \la \alpha^{\rm Max}_{\rm sat}\ra \ra \times 10^{-2} $ & $\la \la \alpha^{T}_{\rm sat} \ra \ra \times 10^{-2}$ & $\la \la \beta_{\rm sat}\ra \ra $ & $\la \la Q_{\theta,{\rm sat}} \ra \ra$ & $\la \la Q_{\phi,{\rm sat}} \ra \rangle $     & $\la \la \theta_{B,{\rm sat}}\ra \ra $       \\
 \hline
 L-P4  		& $0.08 \pm 0.04$          &  $0.24 \pm 0.07$        &  $0.3 \pm 0.1$      &      $73 \pm 19$        &     $1 \pm 2$         & $9.7 \pm 0.8$     & $4.6\pm0.6$     \\
 \\
 M-P4		& $0.3 \pm 0.1$	        &  $1.5 \pm 0.3$      & $1.9 \pm0.4$      &      $31 \pm 5$      &     $10.5 \pm 0.9$         & $33 \pm 3$      & $13.4 \pm 0.7$      \\
 M-P2  		& $0.8 \pm 0.3$             &  $3.2 \pm 0.7$        & $4.0 \pm 0.9$      &     $15 \pm 3$             &   $19\pm 4$         & $54 \pm 11$      & $14 \pm1$        \\
 M-1P		& $0.8 \pm 0.2$	        & $3.0 \pm 0.5 $     &  $3.8 \pm 0.6$      &      $16 \pm 2$          &   $18 \pm 3$        & $52 \pm 7$    & $14.3 \pm 0.6$     \\
 M-2P		& $0.7 \pm 0.1$	        &  $2.4 \pm 0.3$      & $3.1 \pm 0.3$          &      $20 \pm 2$   	     & $16 \pm 3$             & $44 \pm 7$    & $14.1 \pm 0.4$    \\ 
\\
 H-P4		& $0.4 \pm 0.1$	        &  $1.4 \pm 0.3$      &  $1.8 \pm 0.4$      &    $34 \pm 6$       &    $20 \pm 1 $        & $58 \pm 4 $  & $13.8 \pm 0.5$      \\
 	
 \hline
\end{tabular}
\caption{Value of different volume and time averaged quantities for all the runs. Volume average is within the sub-domain $\mathcal{S}$ and time average is done in the QSS between $N_{\rm ISCO}=200-600$. See section \ref{sect:avg_method} for the details of averaging.   }
\label{tab:results_tab}
\end{table*}

In most practical situations there are no exact analytical solutions of the fluid equations. Therefore, we discretize  the continuous 
partial differential equations and solve them using finite difference/element/volume methods. The approximate
solutions rely on discretization of the computational domain. The credibility of the resulting solution is often described in terms of its convergence. 
While true numerical convergence implies that the truncation errors in the solution vanish as the resolution is increased sufficiently. In case of 
ideal magnetohydrodynamic (MHD) simulations, or more precisely for turbulent systems in the absence of explicit dissipation, the concept of convergence is 
ill-defined. This is because, in ideal MHD there are no fixed viscous and resistive length scales, and the dissipation scales are proportional to the grid size. Therefore, as the
resolution is increased, new structures are created.
We do not include explicit viscosity and resistivity to be able to simulate as large as possible Reynolds and magnetic Reynolds numbers. 
Since point-wise convergence is not expected in this case, by convergence we mean that the physically important, globally averaged 
observables (Maxwell stress, magnetic energy, etc.) do not change significantly with the change in numerical resolution.

To study convergence, we consider three runs L-P4, M-P4, H-P4 with three different resolutions (low, medium and high for the domain extending $\pi/4$ in  the azimuthal direction). We do a convergence study for an azimuthal extent of $\Phi_0=\pi/4$, because doing a higher resolution run for a larger $\Phi_0$ is computationally expensive. Following \cite{Sorathia2012}, we classify the convergence metrics into three categories: physical, numerical and spectral. 

\subsection{Physical metrics}
\label{sect:metric_phy}
 \begin{figure}
    \includegraphics[scale=0.43]{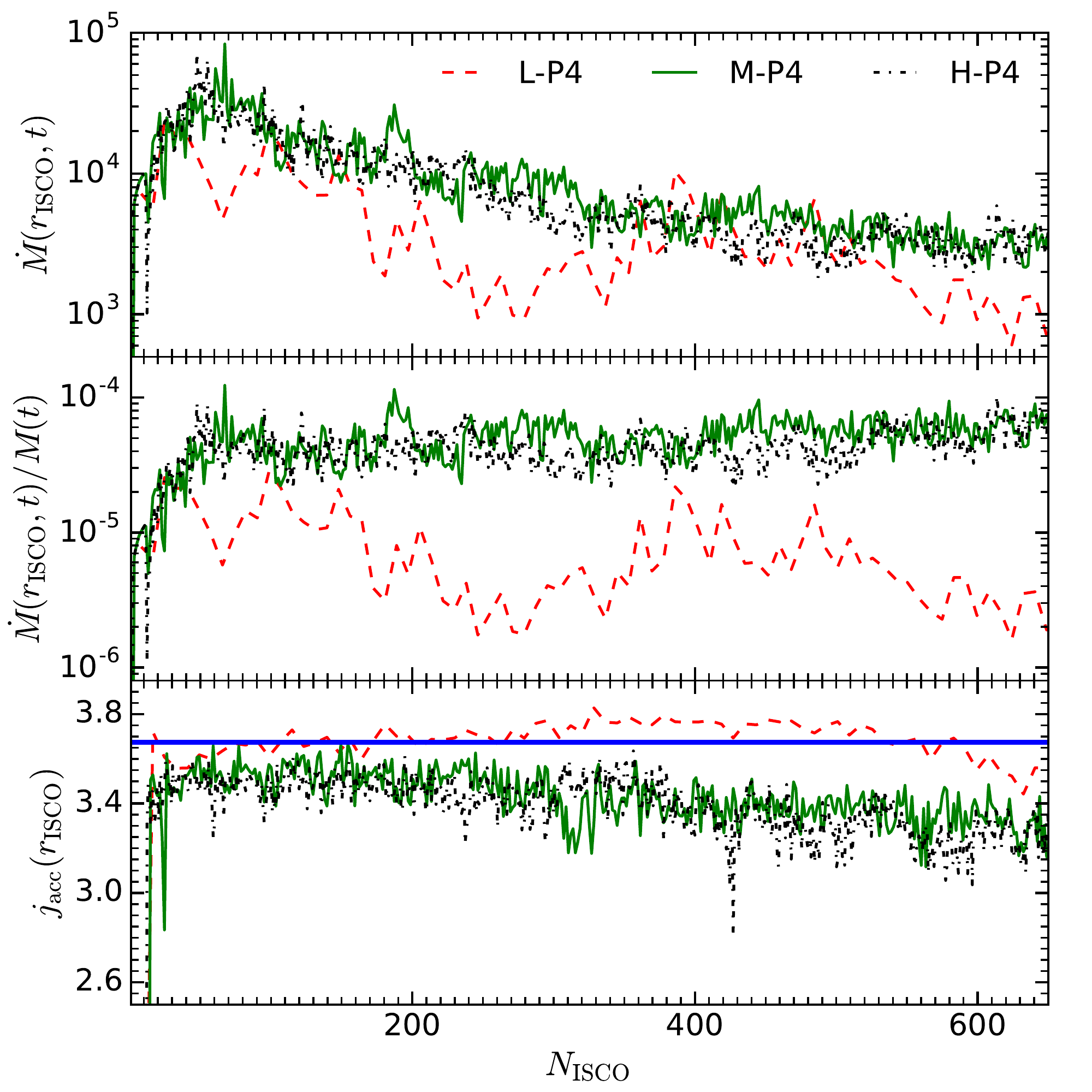}
    \caption{ Time variation of $\dot{M}(r_{\rm ISCO},t)$, $\dot{M}(r_{\rm ISCO},t)/M(t)$ and $j_{\rm acc}=\dot{L}(r_{\rm ISCO},t)/\dot{M}(r_{\rm ISCO},t)$ 
    for three different resolution runs with $\Phi_0=\pi/4$. Blue solid line in the bottom panel depicts the Keplerian (see equation \ref{eq:lkep}) angular momentum at ISCO.  While the two high resolution runs (M-P4 and H-P4) show similar trends, the low resolution run stands out.}
   \label{fig:jnet_conv_p4}
 \end{figure}
 
 \begin{figure}
    \includegraphics[scale=0.48]{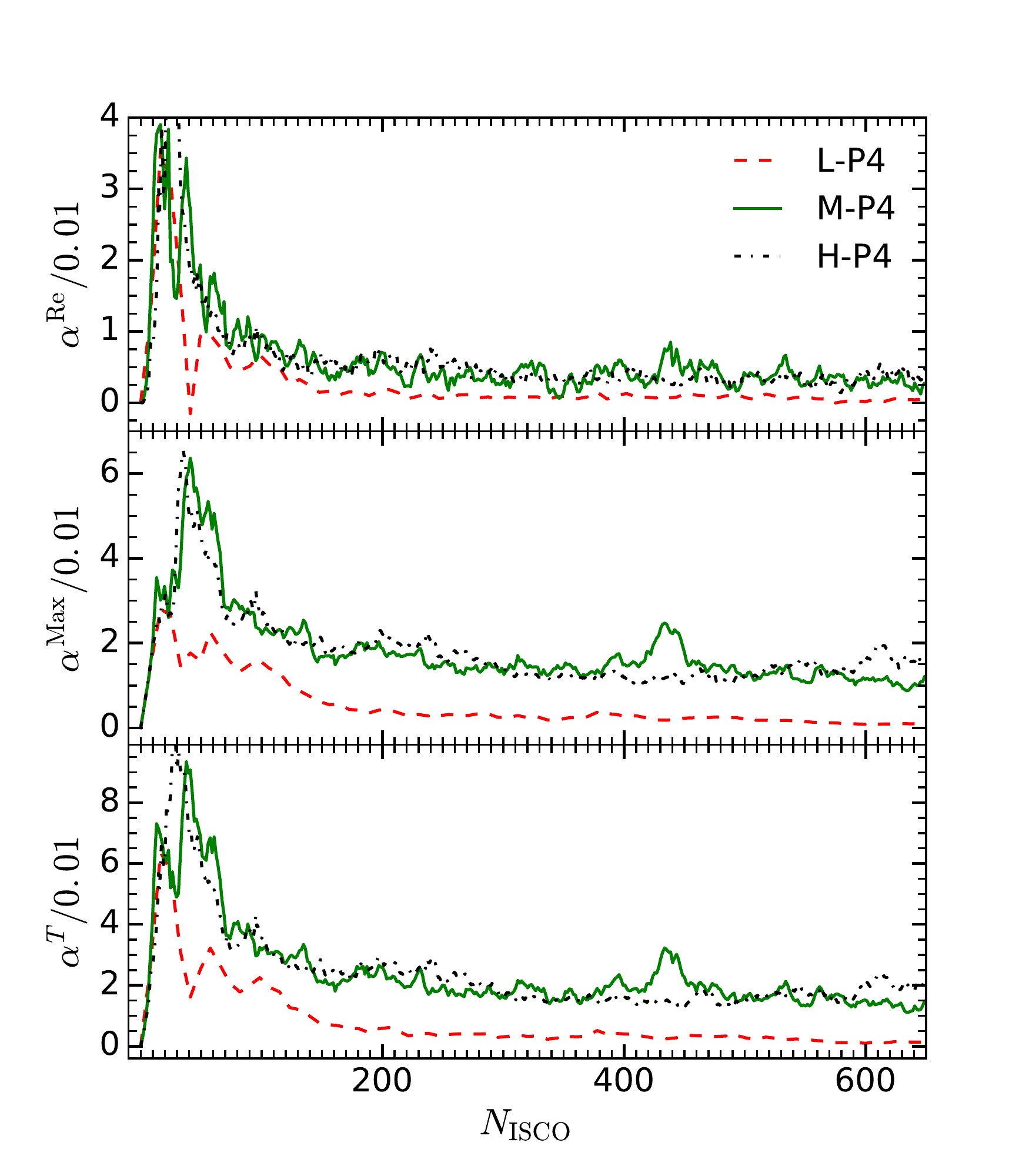}
    \caption{ The variation of the normalized Reynolds stress  $\la \alpha^{\rm Re}(t) \ra$ (top panel) , Maxwell stress $\la \alpha^{\rm Max}(t) \ra$ (middle panel) and total  
    stress (Reynolds+Maxwell) $\la \alpha^{T}(t) \ra$ (bottom panel) with time for three different resolutions. The match between accretion stresses of M-P4 and H-P4 indicates the convergence of medium and high resolution runs. }
   \label{fig:alpha_conv_p4}
 \end{figure}
 
\begin{figure}
    \includegraphics[scale=0.42]{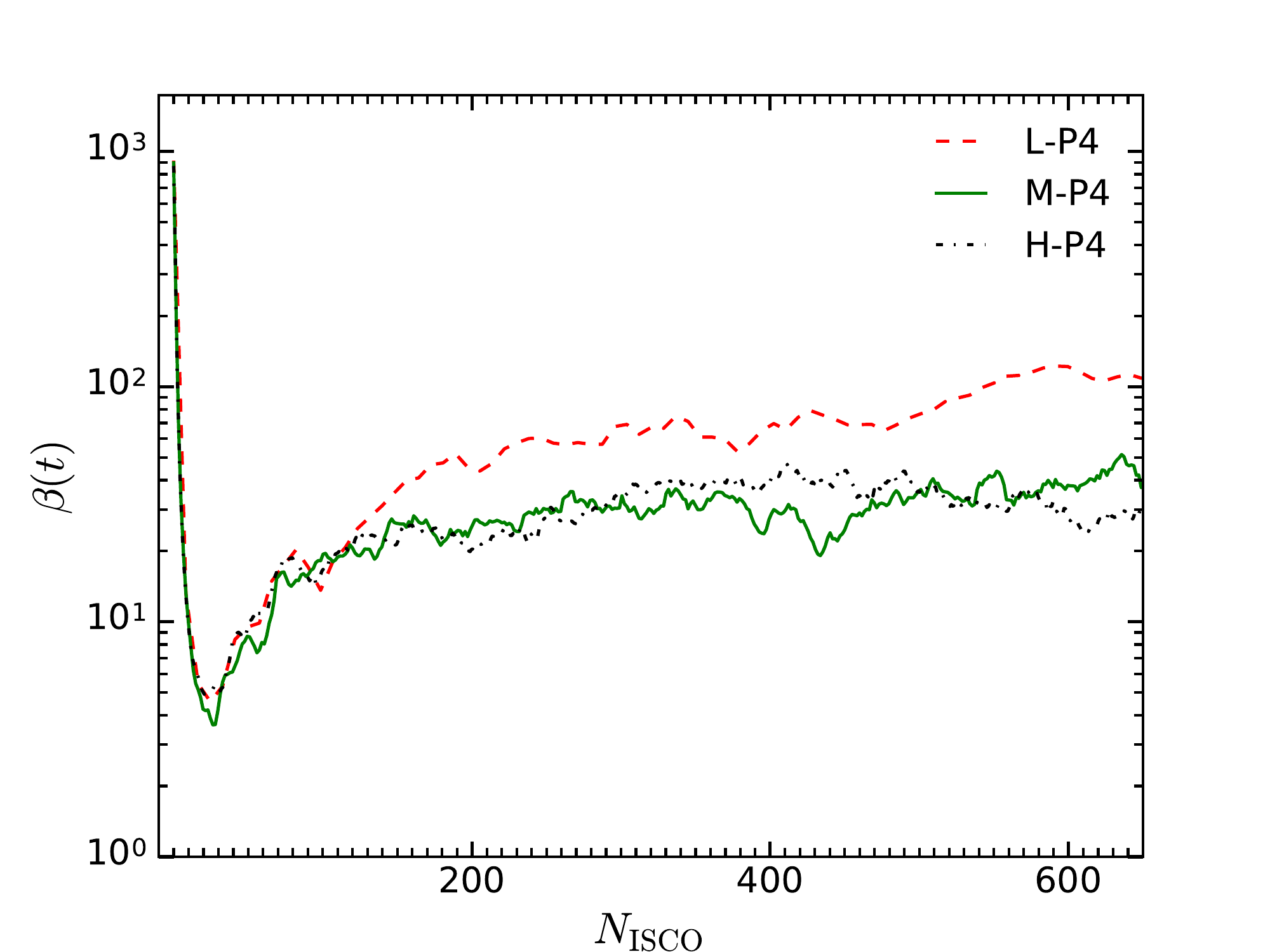}
    \caption{ Time variation of plasma $\la \beta (t) \ra = \langle P(t) \rangle/ \langle B^2(t)/2\rangle$ for different grid resolutions. Magnetization of the accretion flow is similar for runs M-P4 and H-P4; the low resolution run shows weaker magnetization. }
   \label{fig:beta_conv_p4}
 \end{figure}

 To check whether MRI remains well resolved, we examine the following physical metrics: mass and angular momentum accretion rate through the ISCO;
 normalized accretion stresses; and plasma $\beta$.
 
 Top panel of Fig. \ref{fig:jnet_conv_p4} shows the temporal evolution of the mass accretion rate $\dot{M}(r_{\rm ISCO},t)$ through the
 $r=r_{\rm ISCO}$ surface. During the linear phase of the MRI, $\dot{M}(r_{\rm ISCO},t)$ attains a maximum for all three runs. After that,
 while for higher resolution runs (M-P4, H-P4), $\dot{M}(r_{\rm ISCO},t)$ shows a similar declining trend with time, 
 for the lowest resolution run (L-P4) it shows
 a non-monotonic behavior. Around $N_{\rm ISCO}=300$, $\dot{M}(r_{\rm ISCO},t)$ for the low resolution run 
 shows an increase and equals that of the higher resolutions 
 runs for some time. The reason is that the mass density in the computation domain is higher in L-P4 due to a lower accretion rate 
 during the initial phase of evolution. To remove the effects of secular decrease of mass from the computational domain, we 
 normalize $\dot{M}(r_{\rm ISCO},t)$ by the total mass $M(t) = \int_{\mathcal{S}} \rho dV$ within the sub-domain $\mathcal{S}$ 
 ($\mathcal{S}$ is defined in section \ref{sect:avg_method}). The middle panel of Fig. \ref{fig:jnet_conv_p4} shows the time 
 variation of $\dot{M}(r_{\rm ISCO},t)/M(t)$. For the higher resolution runs $\dot{M}(r_{\rm ISCO},t)/M(t)$ is almost constant, 
 but for the low resolution run it shows a different erratic behavior. 

 The bottom panel of  Fig. \ref{fig:jnet_conv_p4} shows the time variation of the specific angular momentum accreted across 
 ISCO, $j_{\rm acc} = \dot{L}(r_{\rm ISCO},t)/\dot{M}(r_{\rm ISCO},t)$. \cite{Noble2010} found that $j_{\rm acc }$ increases secularly 
 when the MRI becomes under-resolved. In our simulations, for higher resolution runs (M-P4 and H-P4) $j_{\rm acc}$ remains almost 
 constant with sub-Keplerian values throughout the evolution. There is a slight decrease in $j_{\rm acc}$ at late times for these two 
 higher resolution runs due to the two facts discussed 
 at the end of section \ref{sect:flow_evo}. For the low resolution run, initially $j_{\rm acc}$ shows an increasing trend, indicating an under-resolved MRI (\citealt{Noble2010}).
 
A direct measure of the efficiency of angular momentum transport is provided by the accretion stresses. 
FIG. \ref{fig:alpha_conv_p4} shows the variation of the
normalized Reynolds stress $\alpha^{\rm Re}(t)$ (top panel), Maxwell stress $\la \alpha^{\rm Max}(t) \ra$ (middle panel) and the total stress 
$\la \alpha^{T}(t) \ra = \la \alpha^{\rm Re}(t)\ra + \la \alpha^{\rm Max}(t)\ra $ (bottom panel) with time. During the initial linear growth of the MRI, all three stresses attain a maximum and
then saturate for the non-linear evolution. We can clearly see that the normalized accretion stresses are about a factor of $\approx 6$ smaller for the low resolution run (also see Table \ref{tab:results_tab}).

 Another physical metric we look at is the plasma $ \la \beta(t) \ra  = \langle P(t) \rangle/ \langle B^2(t)/2\rangle$, which is a measure of magnetization of the accretion flow
 (a smaller value of $\beta$ implies greater magnetization). FIG. \ref{fig:beta_conv_p4} shows the temporal variation of $\beta$ for three
 different resolutions. We start with a moderately high $\beta_{\rm ini} \approx 890$. During the linearly growing phase of MRI, $\la \beta(t) \ra$ decreases (i.e magnetization of 
 the medium increases) due to the immense field amplification; after $N_{\rm ISCO}=200$, $\la \beta(t) \ra$ saturates. While for the low resolution run L-P4, the time-averaged
 (between $N_{\rm ISCO}=200$ and $N_{\rm ISCO}=600$ ) $\beta$ in the saturated state is $\la \la \beta_{\rm sat}\ra \ra \approx 73 \pm 19$, for the higher resolution runs M-P4 and H-P4, $\la \la \beta_{\rm sat}\ra \ra \approx 31 \pm 5$ and $\la \la \beta_{\rm sat} \ra \ra \approx 34 \pm 6$ respectively.
 
 So from the above discussion of physical metrics we see that the runs M-P4 and H-P4 show convergence, but the low resolution run 
 L-P4 does not. These results quantify the minimum resolution required for converged simulations. We expect this resolution criterion
 (per unit scale height) to be valid for  the runs with a larger azimuthal extent.
 
 \subsection{Numerical metrics and the magnetic tilt angle}
 \label{sect:metric_num}
 \begin{figure}
    \includegraphics[scale=0.45]{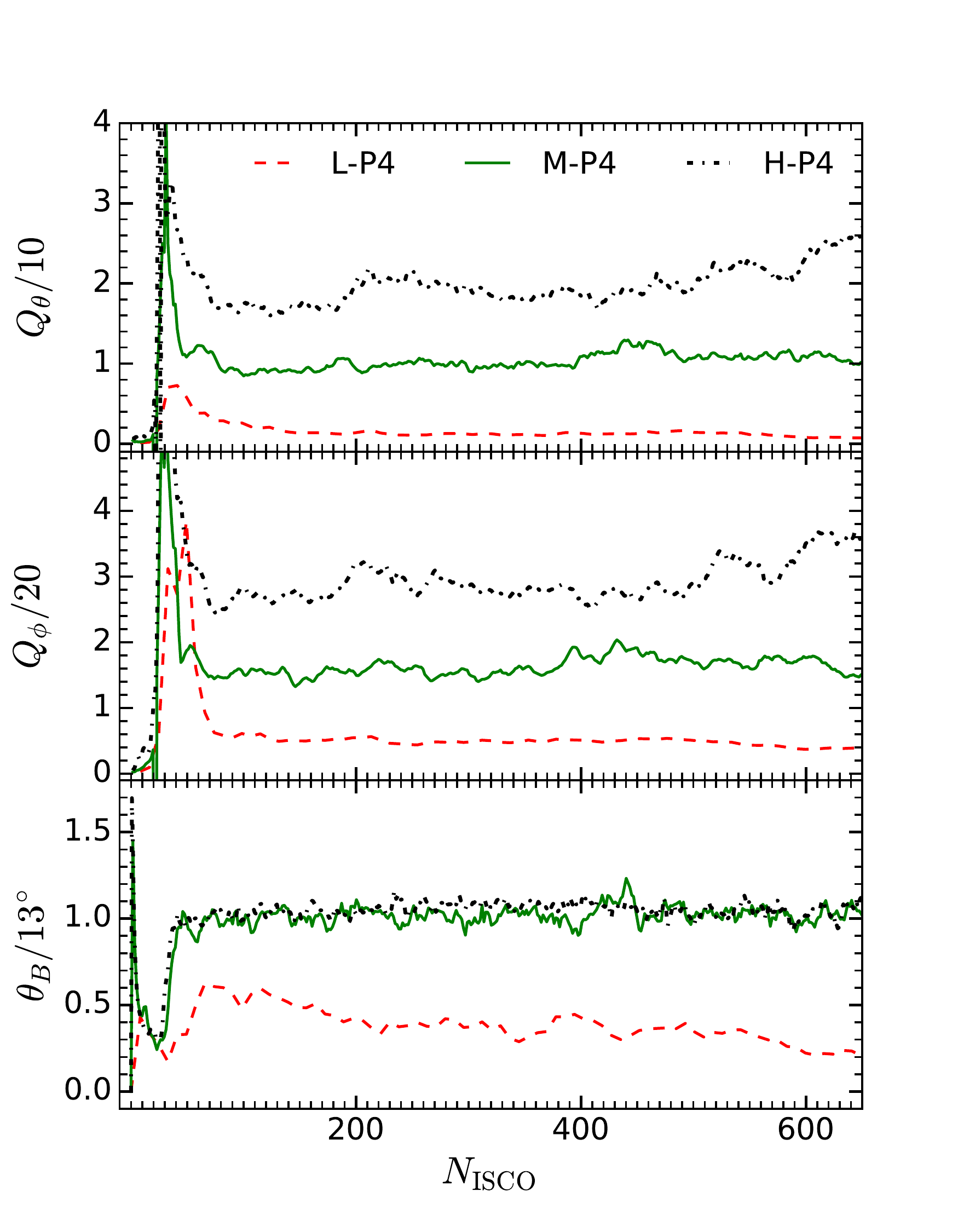}
    \caption{ Top and middle panels show the temporal evolution of average numerical metrics: the poloidal quality factor $\langle Q_{\theta}(t)\rangle$ 
    (Eq. \ref{eq:quality factors}) and 
   the toroidal quality factor $\langle Q_{\phi}(t) \rangle$ (Eq. \ref{eq:qphi}). Time variation of the magnetic tilt angle, $\la \theta_B(t) \ra$ (Eq. \ref{eq:tilt_angle}), is shown in the bottom panel.}
   \label{fig:num_conv_p4}
 \end{figure}
 
 To check how well the MRI is resolved, we study the poloidal ($Q_{\theta}$) and the toroidal ($Q_{\phi}$) quality factors (equations (\ref{eq:quality factors}) and (\ref{eq:qphi})). Along with them, we also look at the degree of correlation between $B_r $ and $B_{\phi}$, measured by the magnetic tilt angle $\theta_B$ (equation (\ref{eq:tilt_angle})) to study the structure of turbulence. Fig. \ref{fig:num_conv_p4} shows the variation of $\langle Q_{\theta}(t) \rangle$, $\langle Q_{\phi}(t) \rangle$ and $\la \theta_B(t) \ra$ with time for different resolutions. For all three resolution runs, $\langle Q_{\theta}(t) \rangle$ and $\langle Q_{\phi}(t) \rangle$ show similar trends; after initial increase in the linear phase they saturate in the non-linear regime of evolution. The time averaged (between $N_{\rm ISCO}=200$ and $N_{\rm ISCO}=600$) saturated value of $\la \la Q_\theta \ra \ra$ and $\la \la Q_\phi \ra \ra$ for the
 converged resolutions scale inversely with the grid resolution; these parameters are much smaller than this scaling for the lowest resolution run (see also
 Table \ref{tab:results_tab}). The magnetic tilt angle $\la \theta_B(t) \ra$ is a more physical diagnostic that is independent of resolution. It attains a quasi-steady value for 
 the medium and high resolution runs with $\la \la \theta_B \ra \ra \approx 13^\circ-14^\circ$. 
 The lower resolution run L-P4 stands out with $\la \la \theta_{B,{\rm sat}}\ra \ra=(4.6\pm0.6)^{\circ}$; i.e., with much smaller magnetic field anisotropy.
 
\cite{Hawley2013} compared the dependence of meridional and azimuthal resolutions, and suggested a resolvability condition $\la Q_{\theta}(t)\ra \la Q_{\phi}(t)\ra \gtrsim 250$. On the other hand, the criteria for convergence with respect to the magnetic tilt angle is $\theta_B>12^{\circ}$ (\citealt{Pessah2010}). Previous global studies (\citealt{Sorathia2012}, \citealt{Hawley2013}) found $\theta_B \approx 13^{\circ}$ for the converged runs. Both the runs M-P4 and H-P4 satisfy all the criteria required for the quality factors and the magnetic tilt angle.

 \subsection{Spectral metrics}
\label{sect:metric_spec}
 \begin{figure}
    \includegraphics[scale=0.35]{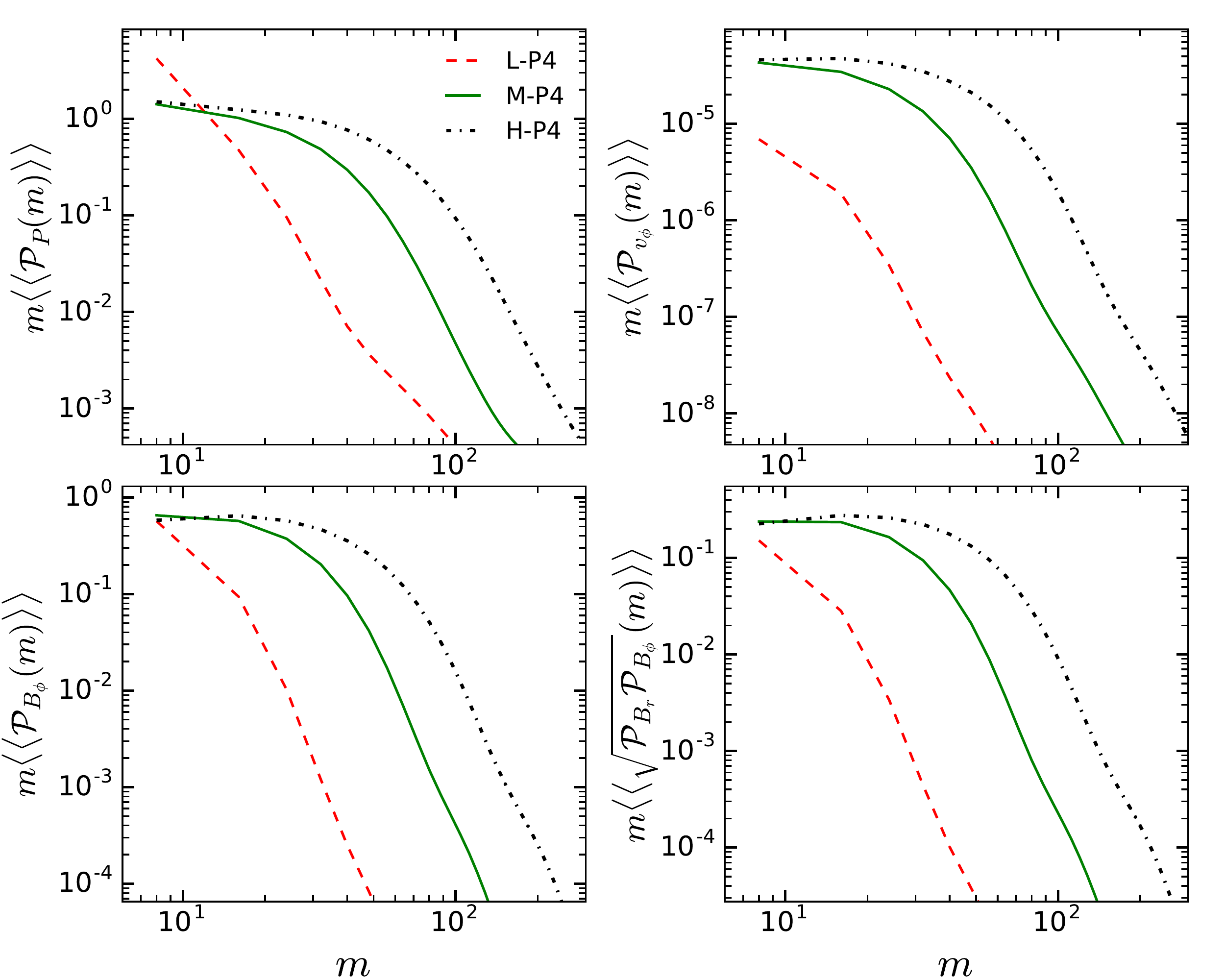}
    \caption{Compensated azimuthal power spectra (see equation \ref{eq:psd}) of thermal pressure $P$ (top left), $v_{\phi}$ (top right), $B_{\phi}$ (bottom left) and $B_r B_{\phi}$
    (bottom right). Time average is done between $N_{\rm ISCO}= 450$ and $600$.}
   \label{fig:spect_conv_p4}
 \end{figure}

 Power spectrum is an important tool to investigate the structure of turbulence. Fig. \ref{fig:spect_conv_p4} shows the toroidal power spectral density (in the azimuthal direction)
 for thermal pressure $P$ (top left), $v_{\phi}$ (top right), $B_{\phi}$ (bottom left) and $B_r B_{\phi}$ (bottom right).  While calculating $\langle \la \mathcal{P}_{P}(m) \ra \rangle$, 
 $\langle \la \mathcal{P}_{v_{\phi}}(m) \ra \rangle$ and $\langle \la \mathcal{P}_{B_{\phi}}(m)  \ra \rangle$, we follow the procedure described in Sec. \ref{sect:PSD}, 
 for $B_r B_{\phi}$, we calculate $\langle \la \sqrt{\mathcal{P}_{B_r} \mathcal{P}_{B_\phi}} (m) \ra  \rangle$. All the spectra display the typical features 
 of MRI turbulence; compensated spectra are flat at
 large scales and most of the power is concentrated at large scales. For higher resolution runs (M-P4 and H-P4), we also see the build up of an inertial range. As expected, with
 the increase in resolution, the inertial range extends.
 
The azimuthal power spectra we obtained agree with those obtained by \cite{Sorathia2012} for their zero-net-flux converged global models but not with those obtained for the 
zero-net-flux non-converged local simulations of \cite{Fromang2007_I}.  

\section{Effects of azimuthal domain size}
\label{sect:azimuth_effect}
In the previous section we see that the medium resolution is adequate to attain convergence.
In this section we compare the results obtained for four medium resolution runs with different azimuthal extents $\Phi_0$: M-P4 ($\Phi_0=\pi/4$), M-P2  ($\Phi_0=\pi/2$), M-1P  ($\Phi_0=\pi$) and M-2P  ($\Phi_0=2 \pi$). We keep the same resolution for per unit azimuthal extent for all these runs. For details of the runs, see Table \ref{tab:simtab}. 

\subsection{Mass and angular momentum accretion through ISCO}
\begin{figure}
    \includegraphics[scale=0.4]{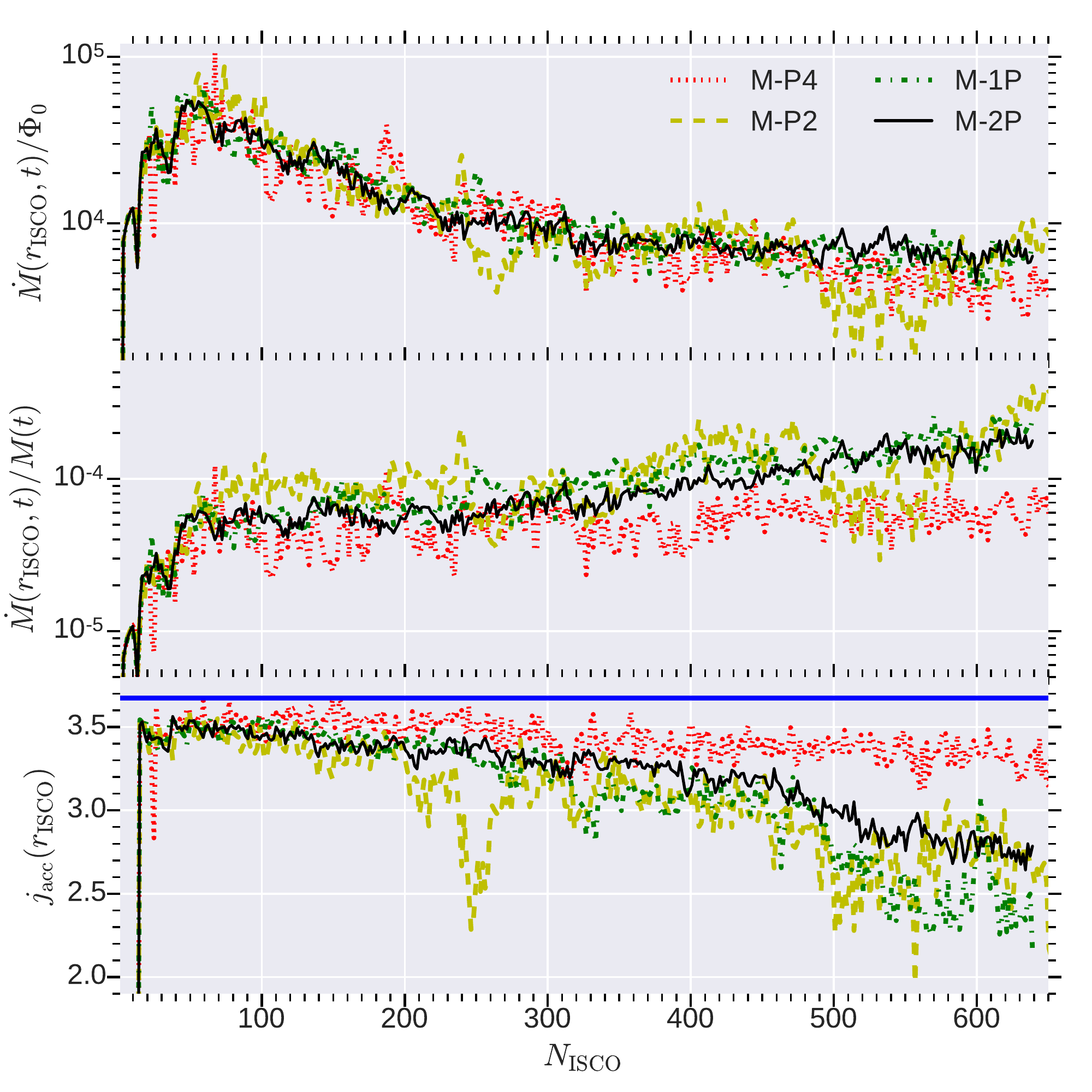}
    \caption{Temporal variation of mass and angular momentum accretion through ISCO for runs with different azimuthal extents $\Phi_0$. Top panel: mass accretion rate thorough the ISCO per unit azimuthal angle $\dot{M}(r_{\rm ISCO}, t)/\Phi_0$; middle panel:  $\dot{M}_(r_{\rm ISCO}, t)/M(t)$; bottom panel: specific angular momentum accreted through the ISCO ($j_{\rm acc}$).  }
   \label{fig:med_mdot_jnet}
 \end{figure}
 
The three panels of Fig. \ref{fig:med_mdot_jnet} show the time variation of the mass accretion rate per unit azimuthal angle $\dot{M}(r_{\rm ISCO}, t)/\Phi_0$, $\dot{M}(r_{\rm ISCO}, t)/M(t)$ and the specific angular momentum accreted through the ISCO ($j_{\rm acc}$) for runs with different azimuthal extents. We compute $\dot{M}(r_{\rm ISCO}, t)/\Phi_0$ instead of $\dot{M}(r_{\rm ISCO}, t)$ to compare the models with different azimuthal extents ($\Phi_0$) on equal footing. While $\dot{M}(r_{\rm ISCO}, t)/\Phi_0$ for all models are almost similar, run M-P4 shows a smaller $\dot{M}(r_{\rm ISCO}, t)/M(t)$ after $N_{\rm ISCO}=300$ . For model M-P2 we see some rapid fluctuations both in $\dot{M}(r_{\rm ISCO}, t)/\Phi_0$ and $\dot{M}(r_{\rm ISCO}, t)/M(t)$ at certain times. A similar trend is followed by $j_{\rm acc}$; $j_{\rm acc}$ for the run M-P4 stands out from that of the other three runs. For both M-P2 and M-1P, we observe depressions in the value of $j_{\rm acc}$ at certain times. This is an indication of higher angular momentum transport at those times which will be discussed in the next subsections.
 
 \subsection{Accretion stresses}
  \begin{figure}
    \includegraphics[scale=0.45]{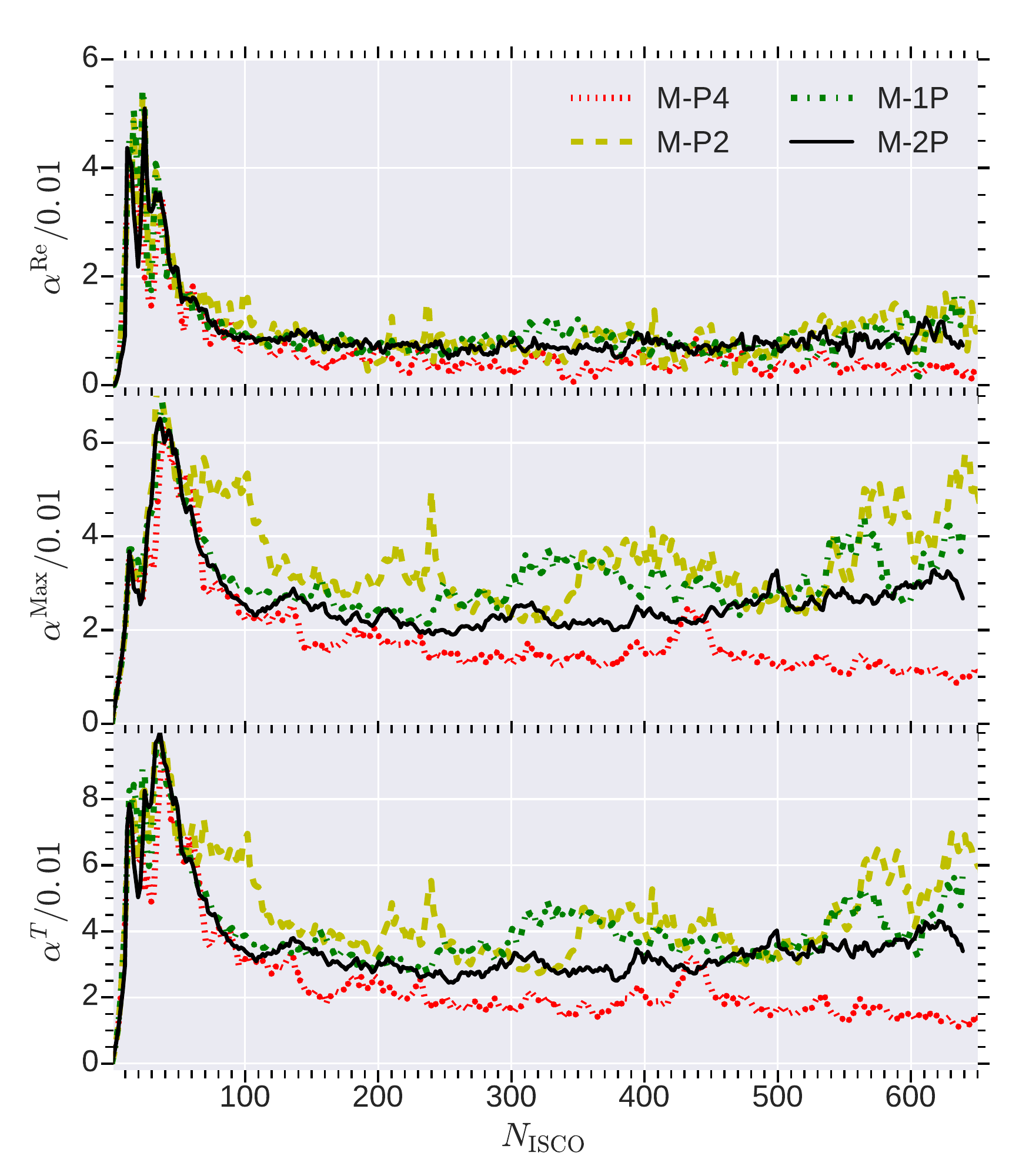}
    \caption{Time evolution of normalized Reynolds stress $\la \alpha^{\rm Re}(t)\ra$ (top panel), Maxwell stress $\la \alpha^{\rm Max}(t) \ra$ (middle panel) and total stress $\la \alpha^{T}(t) \ra$ (bottom panel) for four different runs M-P4, M-P2, M-1P and M-2P.  The runs M-P2 and M-1P show larger variability as well as larger stresses compared to the runs M-P4 and M-2P. }
   \label{fig:med_alpha_R_M_t}
 \end{figure}
 
  \begin{figure}
    \includegraphics[scale=0.345]{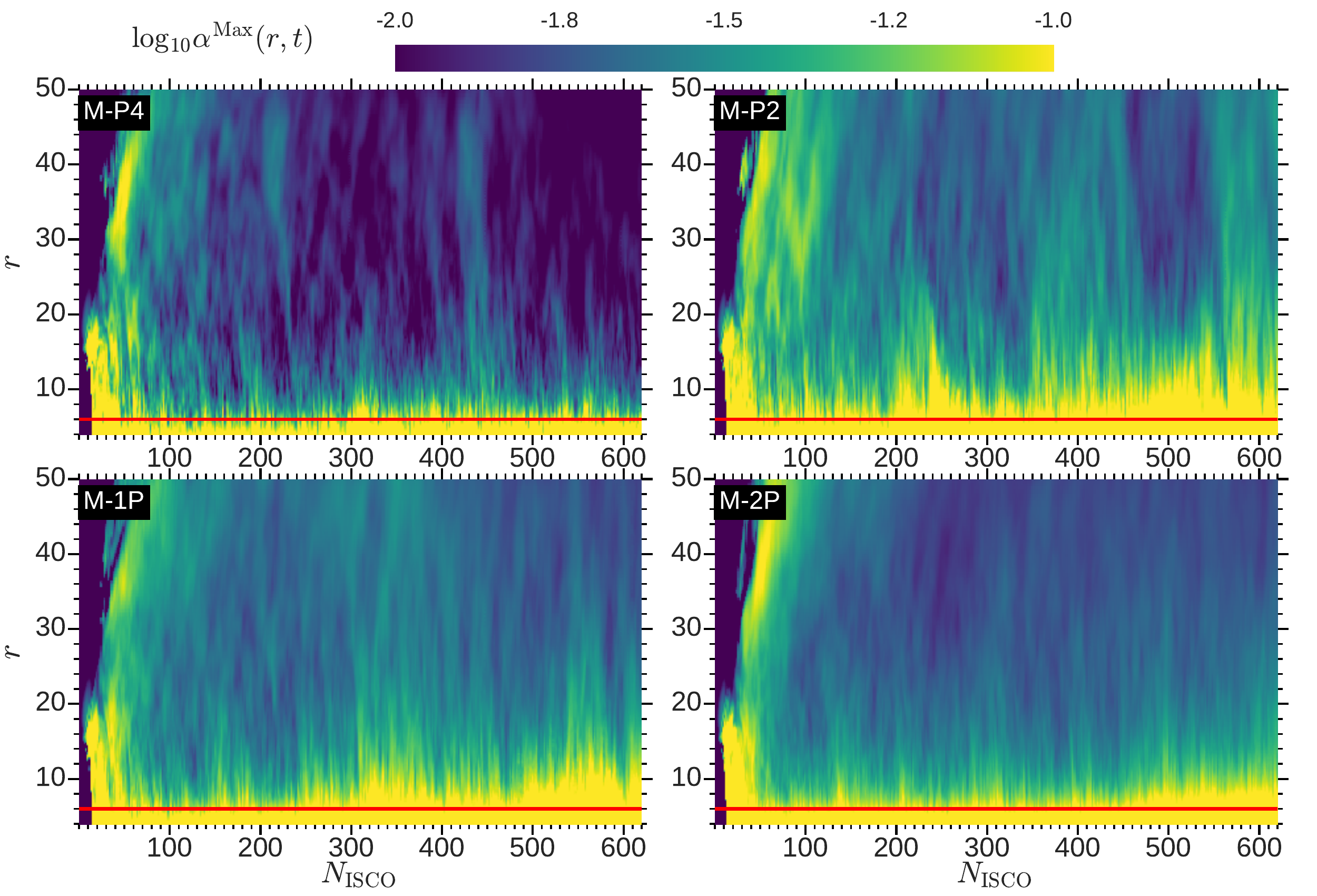}
    \caption{Spatio-temporal distribution of the normalized Maxwell stress $\la \alpha^{\rm Max}(r,t) \ra=-\la B_r B_{\phi}(r,t) \ra/ \la P (r,t)\ra$ for the four different runs M-P4, M-P2, M-1P, M-2P. The horizontal red line represents the location of the ISCO. The runs with restricted azimuthal domain (M-P4, M-P2, M-1P) show larger variability in $\la \alpha^{\rm Max}(r,t) \ra$ (especially at large radii) compared to the run M-2P with the natural azimuthal extent.  }
   \label{fig:alpha_max_rt}
 \end{figure}
 
 Fig. \ref{fig:med_alpha_R_M_t} shows the time variation of the normalized accretion stresses ($\la \alpha^{\rm Re}(t) \ra$, $\la \alpha^{\rm Max}(t) \ra$ and  $\la \alpha^{T}(t) \ra$ respectively) for runs with different azimuthal extents. While Reynolds stresses $\la \alpha^{\rm Re}(t) \ra$ are similar for runs M-P2, M-1P and M-2P, that for M-P4 is smaller (top panel). Maxwell stress  $\la \alpha^{\rm Max}(t) \ra$ (middle panel) for runs M-P2 and M-1P show strong time variability and is larger than that for the runs M-P4 and M-2P. As a result, for the runs M-P2 and M-1P, the total accretion stress $\la \alpha^{T}(t) \ra$ (bottom panel) shows outbursts at certain times which is an indication of higher angular momentum transport at those instances. Sometimes this gives rise to variability in $j_{\rm acc}(t)$ and $\dot{M}(r_{\rm ISCO},t)/\Phi_0$ during outbursts in $\la \alpha^{T}(t) \ra$ (see Fig. \ref{fig:med_mdot_jnet}). 
 
 It is to be mentioned that the volume averaged value of the accretion stress does not show the full picture, and its spatio-temporal behavior is a useful diagnostic. For instance, for run M-P2, we see three outbursts in $\la \alpha^{T} \ra$, around $N_{\rm ISCO}= 250, 400, 600$. While during the outburst around $N_{\rm ISCO}= 250$, we see a decrease in $\dot{M}(r_{\rm ISCO},t)/\Phi_0$ and a dip in $j_{\rm acc}$; during outbursts at $N_{\rm ISCO}= 400, 600$, we observe a slight rise in $\dot{M}(r_{\rm ISCO},t)/\Phi_0$ and a steady $j_{\rm acc}$, indicating an efficient and steady angular momentum transport. On the other hand, we see a decrease both in  $\dot{M}(r_{\rm ISCO},t)$ and in $j_{\rm acc}$ around time $N_{\rm ISCO}=500$, even though $\la \alpha^{T} \ra$ does not show any outburst. This apparent uncorrelated behaviour will be clear if we  see Fig. \ref{fig:alpha_max_rt}. 
 
 We show the spatio-temporal distributions of dominant accretion stress $\la \alpha^{\rm Max}(r,t) \ra = -\la B_r B_{\phi}(r,t) \ra/ \la P (r,t)\ra$ for different runs in Fig. \ref{fig:alpha_max_rt}. While spatio-temporal distribution of $\alpha^{\rm Max}$ is very smooth for run M-2P, it shows strong fluctuations for runs with restricted azimuthal domains, specially for the runs M-P2 and M-1P, which show outbursts in accretion stresses at certain times. Let us correlate the fluctuations in accretion stress with that in mass accretion rate and the angular momentum flux at ISCO as described before (in Fig. \ref{fig:med_mdot_jnet}) and to do that we consider run M-P2. For run M-P2, around $N_{\rm ISCO} = 400$ and $600$, increase in $\alpha^{\rm Max}$ happens across a large radial range, giving rise to a smooth mass accretion rate at ISCO. While around $N_{\rm ISCO}=250$, the outburst in volume averaged $\alpha^{\rm Max}$ is mostly due to the  increase in its value near the central accreting black hole. The large radial gradient in $\alpha^{\rm Max}$ gives rise to a higher angular momentum transport at small radii ($r<20$), and a less efficient angular momentum transport at larger radii. As a result, matter close to black hole is drained quickly due to efficient angular momentum transport, while the outer region can not supply enough matter to the inner region due to inefficient transport. This results in a reduction in $j_{\rm acc}$ and in $\dot{M}(r_{\rm ISCO},t)/\Phi_0$. For the same reason, we see a dip both in $\dot{M}(r_{\rm ISCO},t)$ and $j_{\rm acc}$ around $N_{\rm ISCO}=500$, although the volume average value of $\alpha^{\rm Max}$ is close to the average value. Also, the long coherent structures of $\la \alpha^{\rm Max} (r,t) \ra$ at certain times for models with restricted azimuthal domains hint at the sudden increase in the mean field at those instances. 
 
\subsection{Quality factors and magnetic tilt angle}
\label{sect:med_quality}
   \begin{figure}
    \includegraphics[scale=0.45]{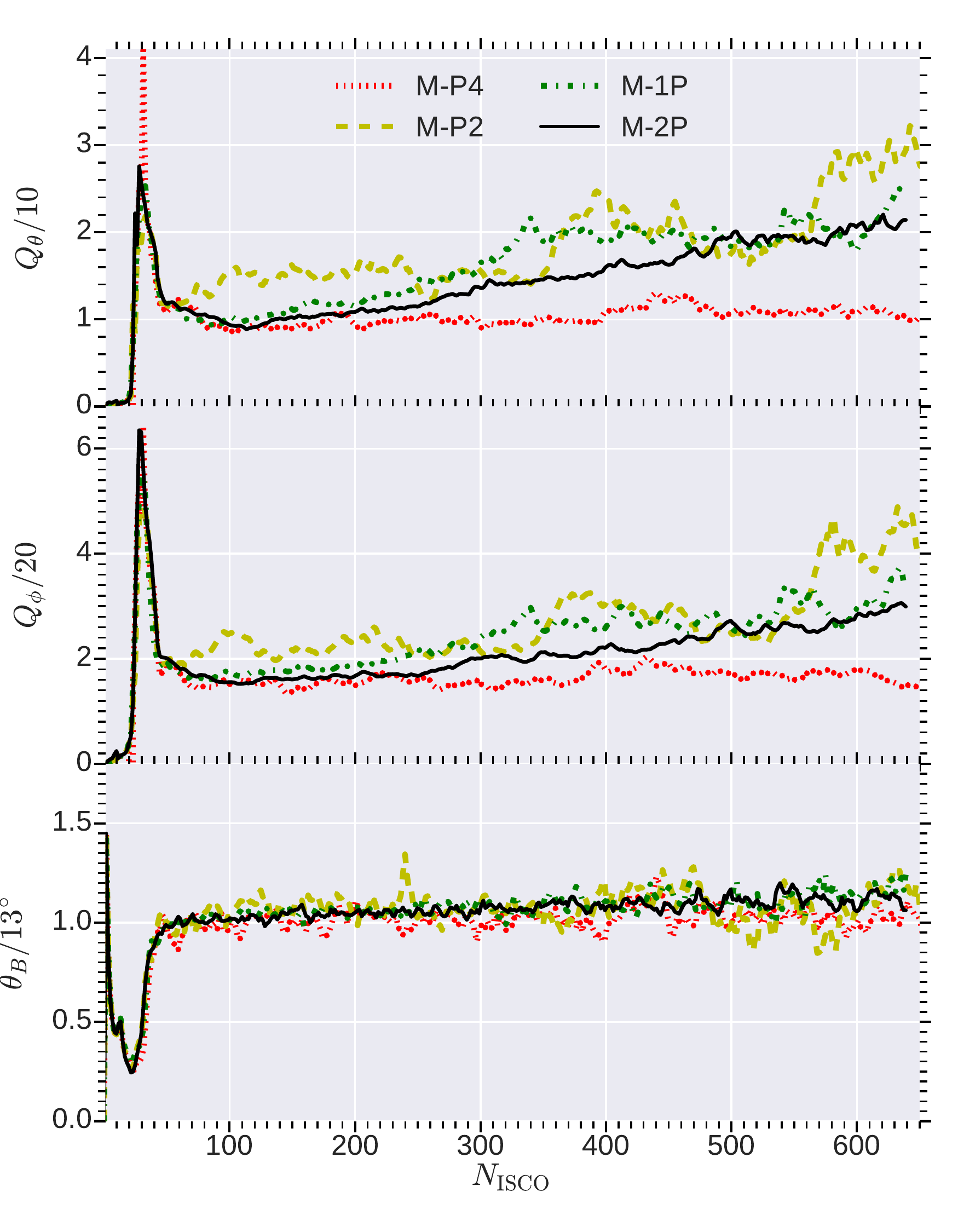}
    \caption{Time evolution of  $\la Q_{\theta}(t)  \ra$, $\la Q_{\phi}(t)  \ra$ and $\la \theta_B(t) \ra$ for our four runs with different azimuthal extents. All runs satisfy the convergence criteria, $\la  Q_{\theta, \rm sat}(t) Q_{\phi}(t) \ra \geq 250$ and $\la \theta_B (t) \ra \geq 12^{\circ}$. It is interesting to note that apart from the run M-P4, all the other runs become better resolved with time. Also note the similarity of $\la \theta_B (t)\ra$ for all the four runs.}
   \label{fig:med_qfactor}
 \end{figure}
 
 Fig. \ref{fig:med_qfactor} shows the temporal evolution of the poloidal ($\la Q_{\theta}(t)  \ra$; top panel) and toroidal ($\la Q_{\phi}(t) \ra$; 
 middle panel) quality factors along with the evolution of magnetic tilt angle $\la \theta_B(t) \ra$ (bottom panel). 
 All the runs satisfy the convergence criteria -- $\la Q_{\theta}(t) Q_{\phi}(t) \ra \geq 250$ and $\la \theta_B (t) \ra \geq 12^{\circ}$. 
 Other than for the run M-P4 with azimuthal extent $\pi/4$, for all other runs $\la Q_{\theta}(t)$ and $\la Q_{\phi}(t) \ra$ show an 
 increasing trend, indicating better resolvability at late times. The most interesting part is the evolution of $\la \theta_B (t)\ra$; 
 for all runs its value is clustered around $13^{\circ}-14^{\circ}$ (also see Table \ref{tab:results_tab}). This strengthens the claim 
 (see section \ref{sect:metric_num} and \citealt{Sorathia2012}) that $\theta_B$ is the most useful metric of convergence for MRI turbulence.      

 \subsection{Evolution of the magnetic field}
   \begin{figure}
    \includegraphics[scale=0.45]{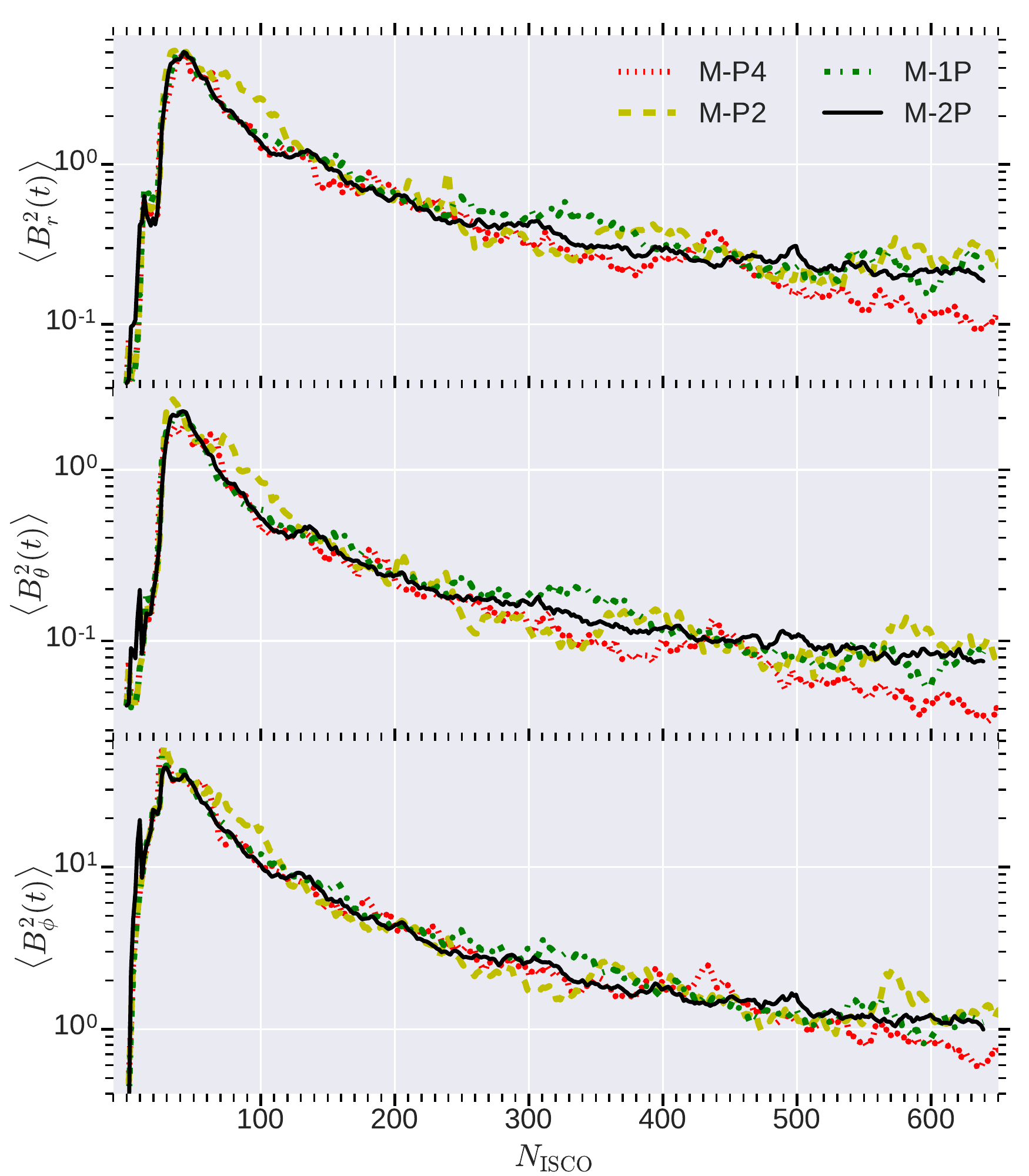}
    \caption{Temporal evolution of magnetic energies associated with radial (top), meridional (middle) and azimuthal (bottom) components of magnetic field for the runs with different azimuthal domain sizes. While the runs with restricted azimuthal domains (M-P4, M-P2, M-1P) show larger variability, the run with the natural azimuthal domain (M-2P) shows a smoother variation. Also notice the rapid decrease in magnetic energy at late times for the run M-P4.}
   \label{fig:med_EB}
 \end{figure}

   \begin{figure}
    \includegraphics[scale=0.4]{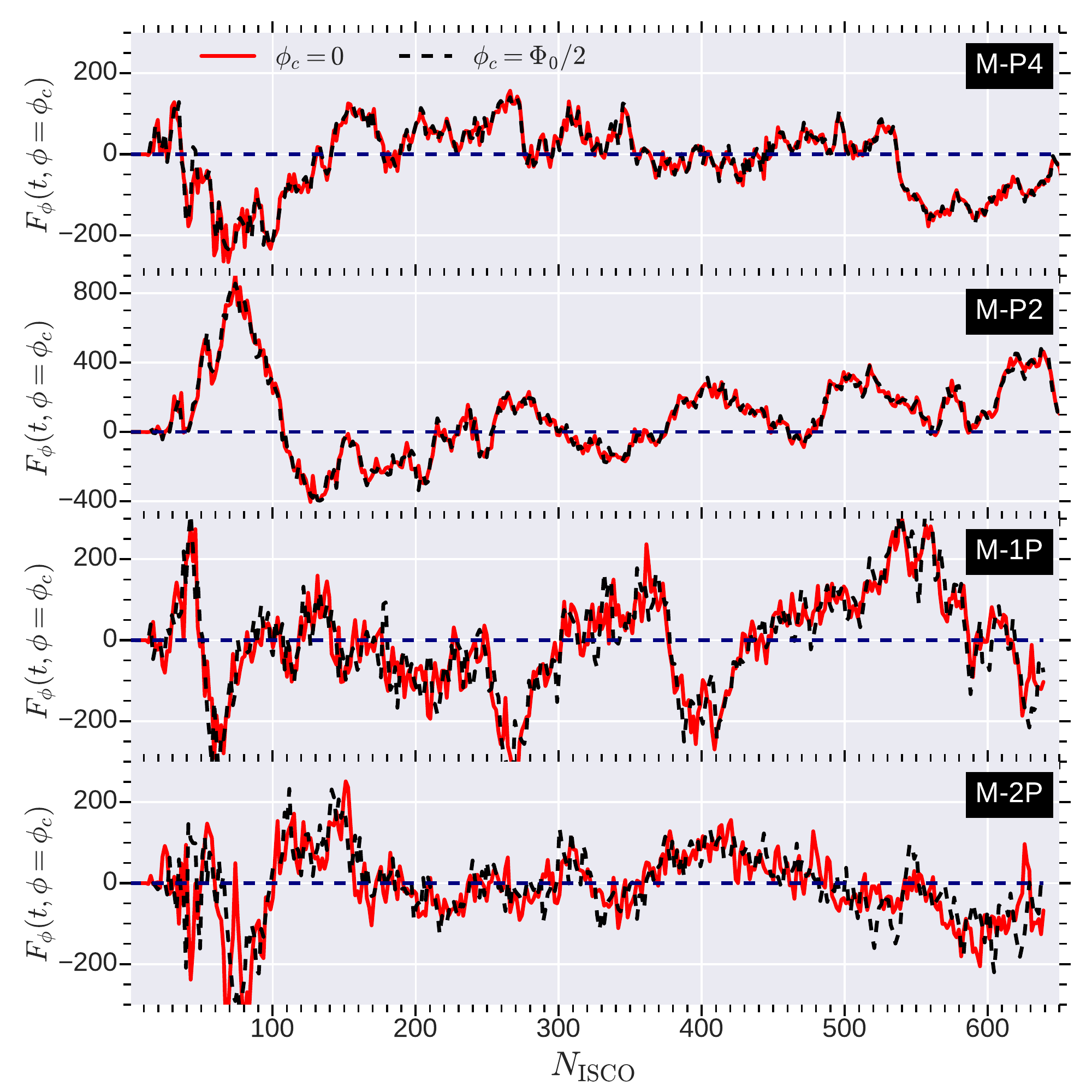}
    \caption{Toroidal flux $F_{\phi}(t)$ through the $\phi=\Phi_c$ (constant) surfaces for the four runs with different azimuthal extents. 
    The change in sign with time in the net toroidal flux is due to the expulsion of the toroidal field (due to magnetic buoyancy) from the 
    mid-plane to the tenuous sub-Keplerian region and the generation of field due to a dynamo. The closeness of the $F_{\phi}(t)$ 
    calculated at two different $\Phi_c~(=0,\Phi_0/2$) indicates the almost axisymmetric nature of $B_{\phi}$. }
   \label{fig:med_fphi}
 \end{figure}

Fig. \ref{fig:med_EB} shows the time variation of magnetic energies associated with radial (top), meridional (middle), and azimuthal (bottom) components of the magnetic field for the four runs with different azimuthal extents. All the runs show similar magnetic energy in the quasi-steady state, albeit runs with restricted azimuthal extents (M-P4, M-P2, M-1P) show larger time variability. Also, for the run M-P4, magnetic energy decreases rapidly at late times (for $N_{\rm ISCO} > 500$). On the other hand, the run M-2P with the natural azimuthal extent displays a smoother magnetic energy evolution.

Fig. \ref{fig:med_fphi} shows the time evolution of the net toroidal flux $F_{\phi}(t)$ defined as, 
\be
F_{\phi} (t) = \int_{r=6}^{40}\int_{\theta=-\theta_H}^{+\theta_H} B_{\phi}(r,\theta, \phi=\Phi_c) r dr d\theta.
\ee
We calculate $F_{\phi}(t)$ at two locations, at $\Phi_c=0$ and at $\Phi_c=\Phi_0/2$. Since we start with a poloidal field, $F_{\phi}=0$ at the beginning.
Toroidal fields build up due to shear, with different signs in the two hemispheres. Due to the underlying symmetry, net toroidal flux is zero until MRI induced MHD turbulence generates asymmetry  and the disk attains a net toroidal flux. With time $F_{\phi}(t)$ changes sign  aperiodically unlike in the previous global simulations of geometrically thin disks (\citealt{Fromang_nelson2006,Oneill2011, Beckwith2011,Flock2011}), which show a more periodic behavior. The time reversal of net toroidal flux hints at gradual expulsion of mean $B_{\phi}$ from the mid-plane towards the tenuous sub-Keplerian regions and replacement with the field of opposite sign due to a dynamo process (\citealt{Davis2010}). The dynamo process is discussed in detail in later sections. Although the toroidal flux evolution is qualitatively similar for all the runs, runs M-P2 and M-1P show noticeably higher amplitudes of the toroidal flux. Also, for M-P2 $F_{\phi}(t)$ remains positive for a longer time ($\approx 65$ percent of the total run time) compared to the other three runs, where the ratio of time spent in positive and negative state of  $F_{\phi}(t)$ is roughly $1:1$. In addition, the coherence between the fluxes calculated at different $\phi=\Phi_c$ surfaces (for all runs) reflects the large scale structure of $B_{\phi}$. 
  
 \subsection{Mean versus turbulent fields}

 \begin{figure}
    \includegraphics[scale=0.45]{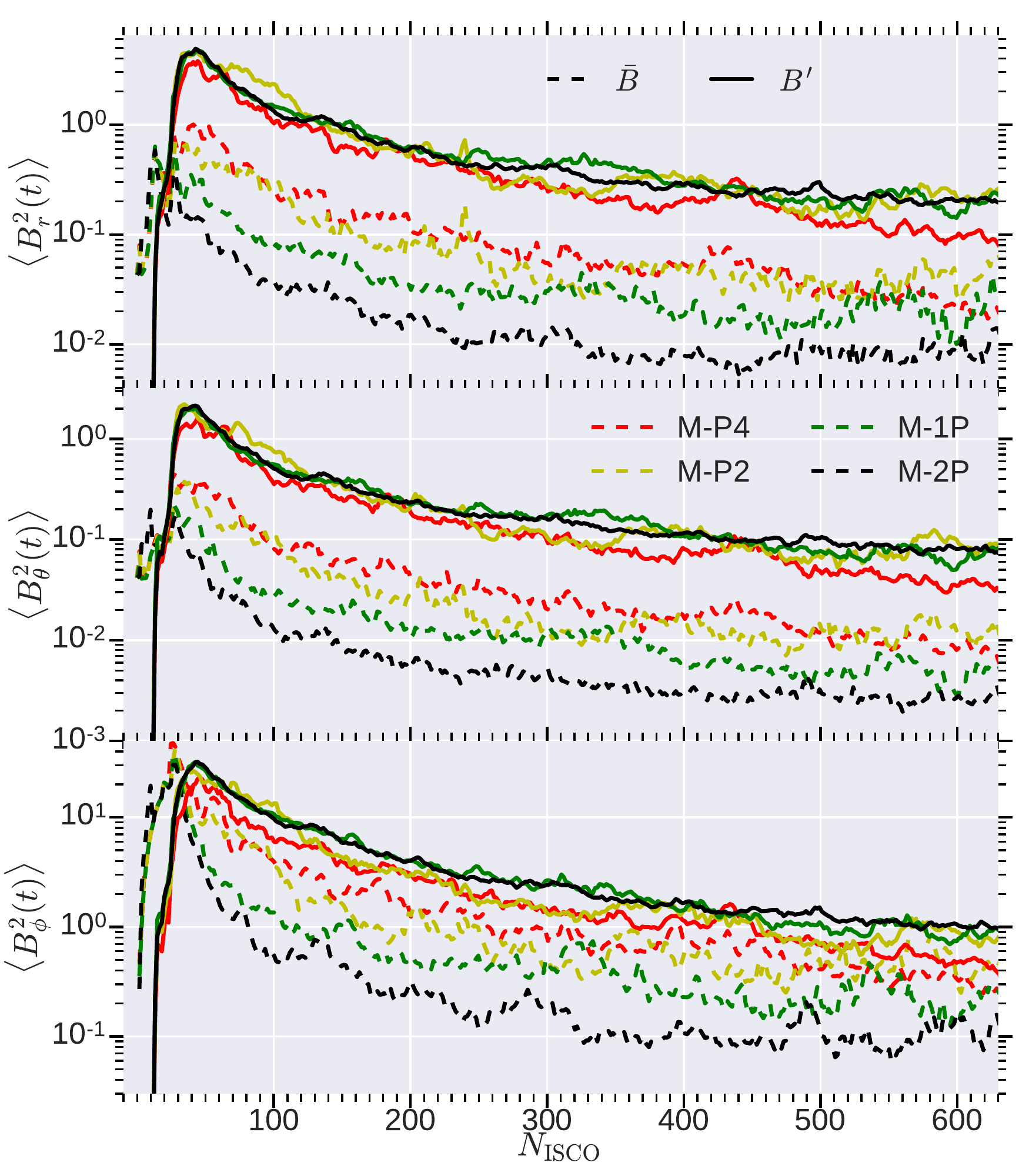}
    \caption{Temporal evolution of the mean (dashed lines) and the turbulent (solid lines) energies associated with the radial (top panel), meridional (middle panel) and azimuthal components (bottom panel) of the magnetic field for the four runs M-P4, M-P2, M-1P and M-2P. For all models and all components of magnetic fields, the contribution from turbulent fields is larger compared to the mean fields. Mean fields are larger for runs with restricted azimuthal domains.  }
   \label{fig:med_turb_mean_B}
 \end{figure}

  \begin{figure}
    \includegraphics[scale=0.4]{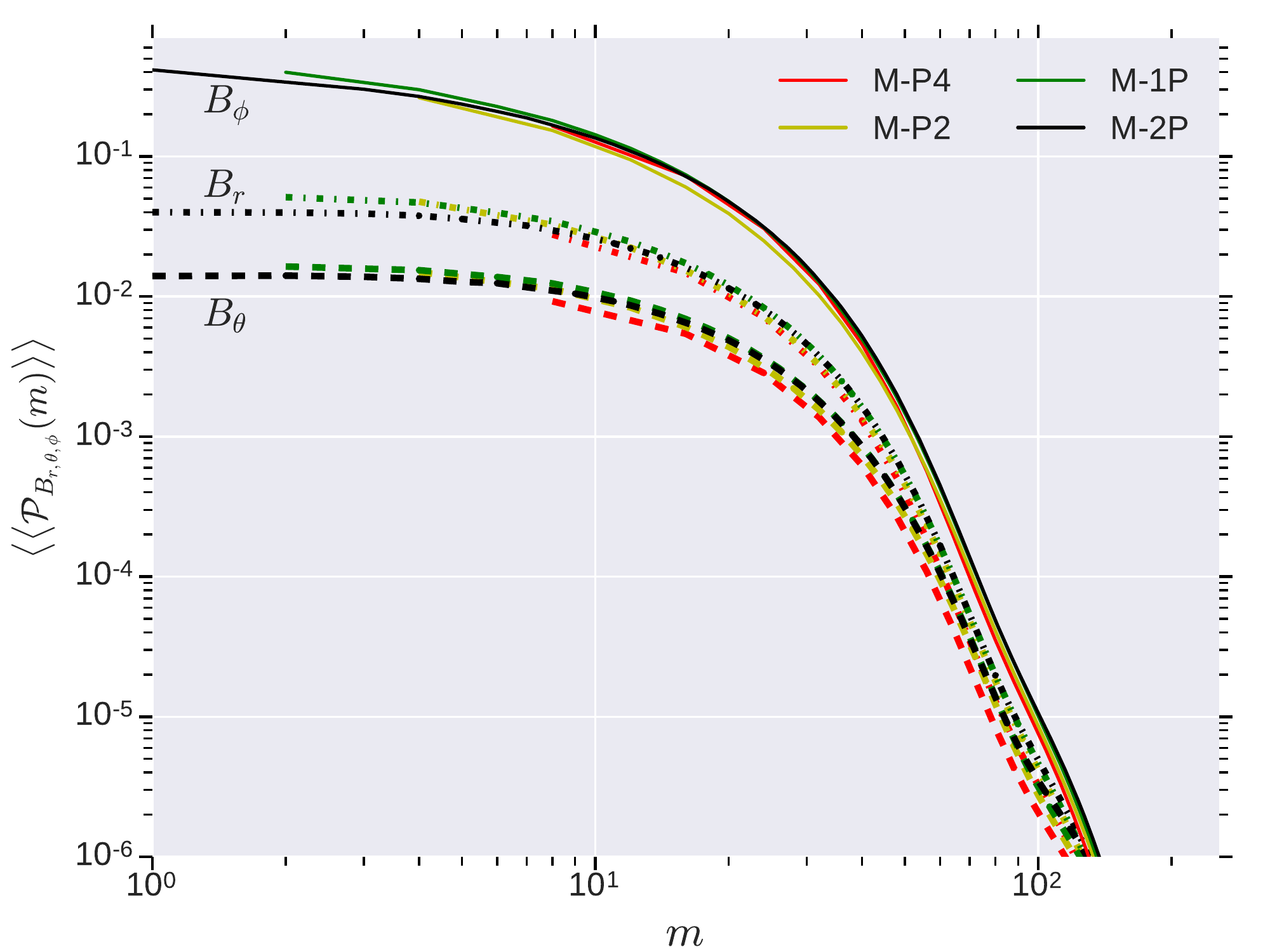}
    \caption{Power spectral density $\la \la {\mathcal{P}_{B_{r,\theta,\phi}}(m)} \ra \ra $ (Eq. \ref{eq:psd}) of different magnetic field components for the runs M-P4, M-P2, M-1P and M-2P. Time average is done between $N_{\rm ISCO}=200-600$. Most of the power is concentrated at small wave numbers. While power for $B_r$ and $B_{\theta}$ start saturating at $m=8$, $\la \la {\mathcal{P}_{B_{\phi}}(m)} \ra \ra$ keeps increasing for lower $m$. For all three components, run M-1P shows a slightly  greater power at large scales.}
   \label{fig:med_psd}
 \end{figure}
 
   \begin{figure}
    \includegraphics[scale=0.45]{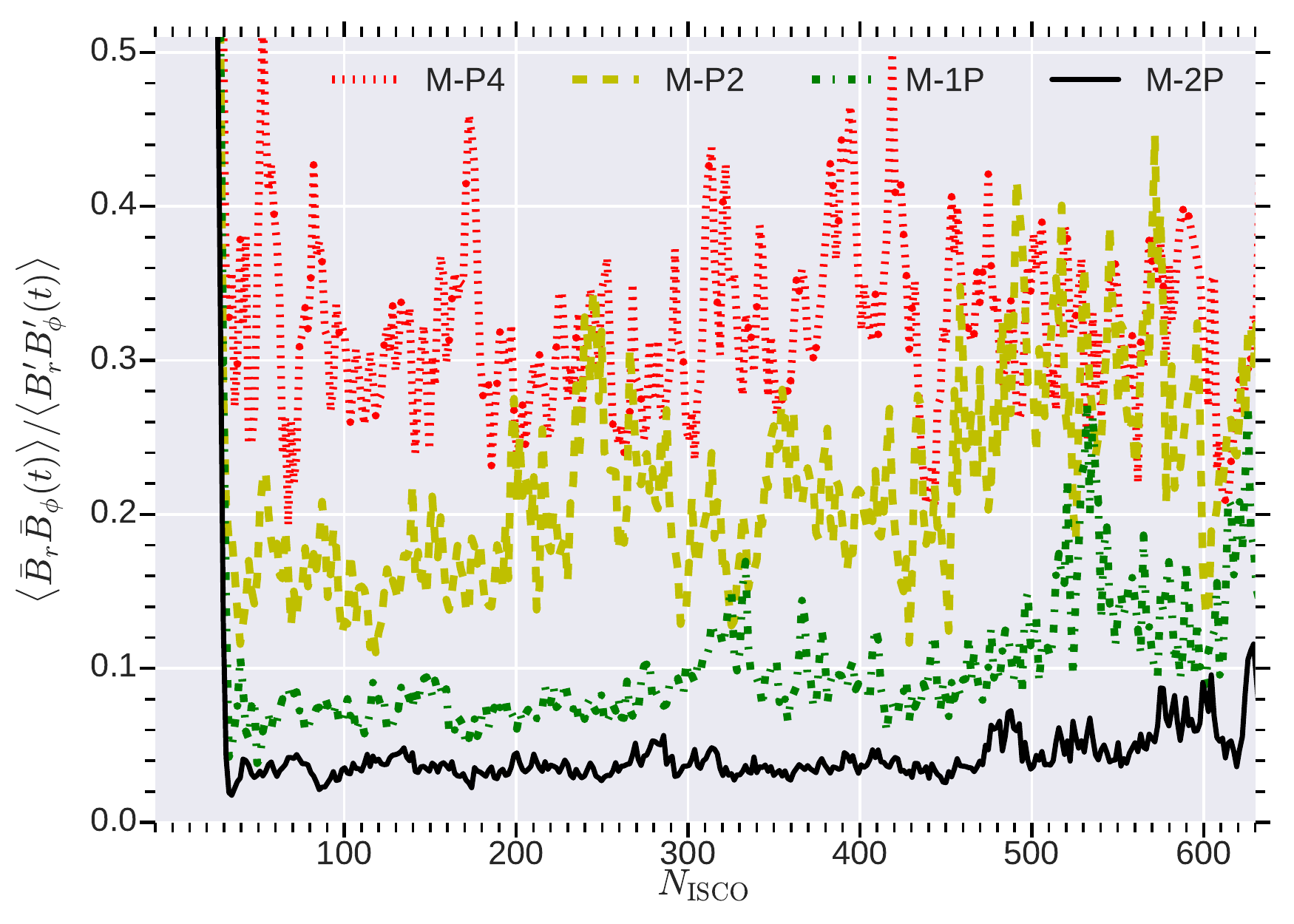}
    \caption{The ratio of mean to turbulent Maxwell stress for the runs with different azimuthal domain sizes. The relative contribution of mean Maxwell stress $\bar{B}_r \bar{B}_{\phi}$ relative to the turbulent Maxwell stress $\bar{B^{\prime}_{r} B^{\prime}_{\phi}}$ decreases as the azimuthal extent of the simulation domain increases.}
   \label{fig:med_turb_mean_Max}
 \end{figure}
 
   \begin{figure}
    \includegraphics[scale=0.45]{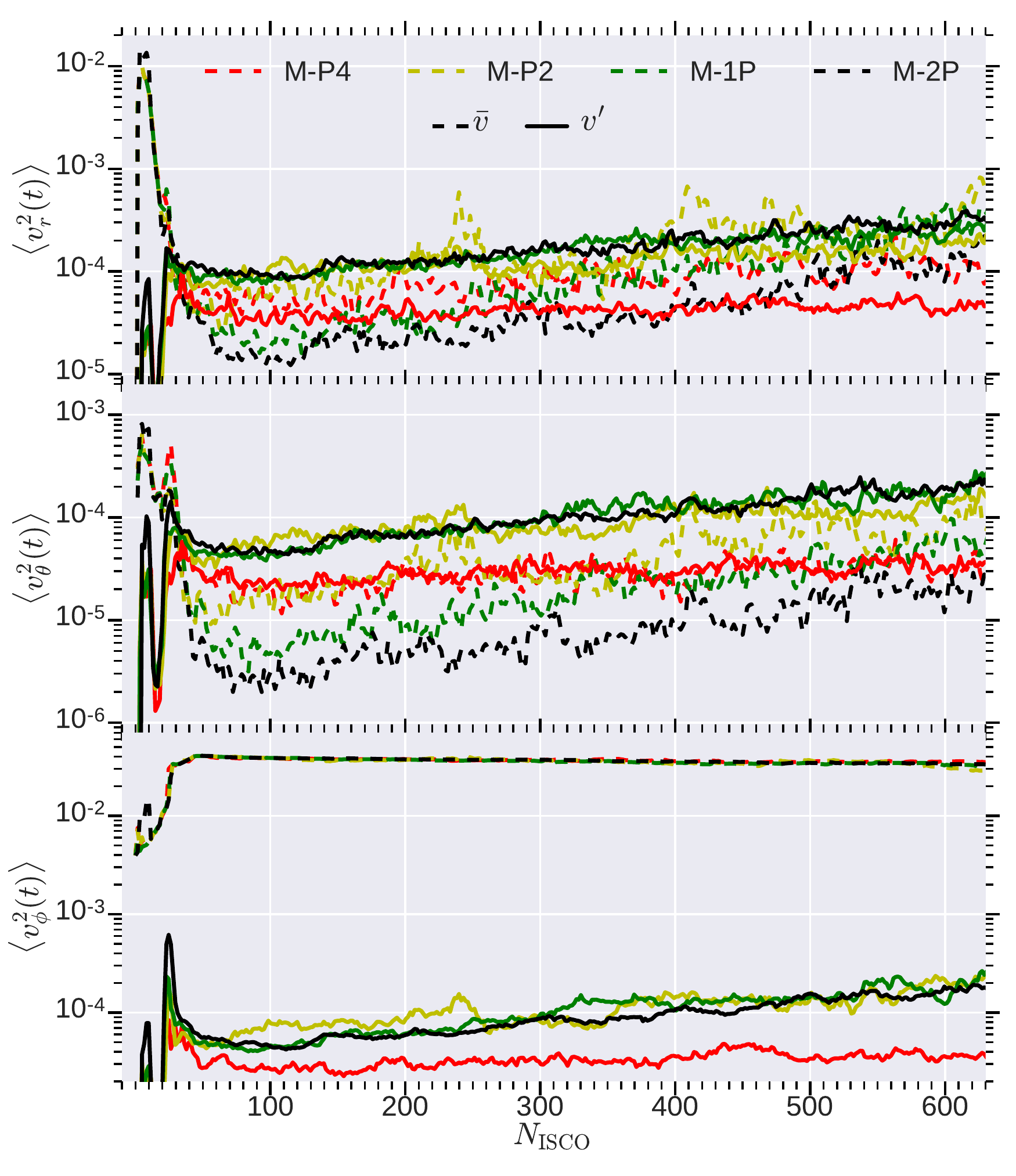}
    \caption{Temporal evolution of the mean (dashed lines) and the turbulent (solid lines) velocity fields for the runs M-P4, M-P2, M-1P and M-2P with different azimuthal extents. Unlike magnetic energy density, which scales with the mass density, $v^2$ is an intensive quantity. So it does not depend on the mass available in simulation domain and hence it does not show any secular decrease in time as a result of mass accretion through the inner boundary. The details are discussed in the main text. }
   \label{fig:med_turb_mean_v}
 \end{figure}

In Fig. \ref{fig:alpha_max_rt}, the coherence in $\alpha^{\rm Max}$ across a large radius for the runs with restricted azimuthal domains indicates a significant role of the mean fields in those instances. 
In this section we compare the relative contribution from mean and turbulent parts of the flow. We decompose some variable $q$ into a mean $\bar{q}$ (where the average is done over $\phi$) and a turbulent $q^{\prime}$
\be
q = \bar{q} + q^{\prime},
\label{eq:mean_turb}
\ee
where the decomposition follows the Reynolds rules, $\bar{\bar{q}} = \bar{q}$, $\bar{q^{\prime}} = 0$. 

Fig. \ref{fig:med_turb_mean_B} shows the time variation of the mean (dashed line) and the turbulent (solid line) energies associated with the radial (top panel), meridional (middle panel) and azimuthal components (bottom panel) of the magnetic field for the four runs with different azimuthal extents. For all the runs and all the components of magnetic fields, the contribution from the mean fields is smaller compared to that of turbulent fields. In addition, mean field evolution shows a coherent trend -- larger the azimuthal extent of the simulation domain, smaller the mean magnetic field energy (see Table \ref{tab:mean_vs_turb}). On the other hand, turbulent magnetic energy evolution does not show any definite trend. These results are different from \cite{Flock2012}, who found a higher strength both for mean and turbulent fields for the restricted domain sizes. While the turbulent magnetic energy evolution for run M-2P shows a smooth behavior, that for the runs with restricted azimuthal extents show larger variability. On average, turbulent energies for the runs M-1P and M-2P are higher than the other two runs (see Table \ref{tab:mean_vs_turb}). 

Fig. \ref{fig:med_psd} shows the azimuthal power spectral density (PSD) $\la \la {\mathcal{P}_{B_{r,\theta,\phi}}(m)}  \ra \ra$ (equation \ref{eq:psd}) of different magnetic field components for the runs M-P4, M-P2, M-1P and M-2P. The time average is done over $N_{\rm ISCO}=200-600$. All the runs show almost similar PSD. For all the models and magnetic field components, most of the power is concentrated at small wavenumbers (i.e., at large scales). While PSDs for $B_r$ and $B_{\theta}$ start saturating at $m=8$, $\la \la {\mathcal{P}_{B_{\phi}}(m)} \ra \ra$ does not show any saturation. For the most dominant field component $B_{\phi}$, run M-1P has a slightly larger power at small $m$ compared to the other three runs. This is reflected in the time averaged value of $B^{\prime 2}$ in Table \ref{tab:mean_vs_turb} (in accordance with the Parseval's theorem).

Total accretion stress is comprised of the Reynolds stress and the dominant Maxwell stress. While Reynolds stress is solely due to turbulence, Maxwell stress can arise due to the statistical correlation between the turbulent fields ($B^{\prime}_ {r} B^{\prime}_{\phi}$) as well as between mean fields ($\bar{B}_r \bar{B}_{\phi}$). Fig. \ref{fig:med_turb_mean_Max} shows the ratio of mean Maxwell stress $\bar{B}_r \bar{B}_{\phi}$ to turbulent Maxwell stress $B^{\prime}_ {r} B^{\prime}_{\phi}$ for the runs with different azimuthal extent. For all the four runs, the relative contribution from the turbulent field is higher than the mean fields. Although, as expected, the relative contribution of the mean fields for restricted azimuthal domains is higher. Also notice the increase in the mean to turbulent stress ratio at times when the total stress and quality factors show fluctuations (compare Fig. \ref{fig:med_turb_mean_Max} with Figs. \ref{fig:med_alpha_R_M_t} and \ref{fig:med_qfactor}). This coincidence indicates a better correlation between mean fields in those instances. It is interesting to see that $\la \la B_r B_{\phi} \ra \ra = \la \la \bar{B}_r \bar{B}_{\phi} \ra \ra + \la \la B^{\prime}_r B^{\prime}_{\phi} \ra \ra$ for run M-P2 is smaller compared to that for runs M-1P and M-2P (see Table \ref{tab:mean_vs_turb}). Still run M-P2 shows highest value of $\la \la \alpha^{\rm Max}_{\rm sat}\ra \ra$ among all the runs (see Table \ref{tab:results_tab}). The reason is that due to a higher mass loss, run M-P2 has comparatively lower $\la P(t)\ra$ and hence a larger normalized stresses.
  
Fig. \ref{fig:med_turb_mean_v} shows the variation in kinetic energy per unit mass $v^2_i$ ($i$ stands for $r,\theta,\phi$) associated with the mean and turbulent flows for the four runs. Unlike magnetic energy, which is an extensive variable, $v^2_i$ is an intensive quantity. Therefore, its evolution is independent of mass available in the simulation domain. As a result, $v^2_i$ does not show any secular decrease in time due to the mass loss through the inner/outer boundaries. Instead, we see a slow secular increase in $v^2$ (both for mean and turbulent flows) with time, which implies that the resolvability gets better as the time passes (also see Fig. \ref{fig:med_qfactor}). The increasing trend is less prominent for the run M-P4 compared to other three runs. One marked difference between the evolutions of $v^2$ and $B^2$ is that while the turbulent magnetic field amplitudes are always larger than the mean magnetic field amplitudes in the quasi-steady state, velocity fields do not follow any definite trend. While for the radial component of velocity, for runs M-1P and M-2P, $v^{\prime 2}_r > \bar{v}^2_r$ and for run M-P2, $v^{\prime 2}_r \approx \bar{v}^2_r$, $v^{\prime 2}_r < \bar{v}^2_r$ for run M-P4; for meridional component, $v^{\prime 2}_{\theta} > \bar{v}^2_{\theta}$ for the runs M-P2, M-1P and M-P2, and $v^{\prime 2}_{\theta} \approx \bar{v}^2_{\theta}$ for the run M-P4. Due to the dominant rotation, as expected, for all runs $v^{\prime 2}_{\phi} \ll \bar{v}^2_{\phi}$. It is interesting to see that for all the velocity components, turbulent velocity amplitude is smallest for the run M-P4, compared to the other three runs which have a comparable amplitude. This is a signature of a higher level of turbulence in the runs M-P2, M-1P, M-2P.

 \begin{table*}
\centering
\begin{tabular}{ |p{1.0cm}||p{1.0cm}||p{1.5cm}||p{1.5cm}|p{1.8cm}||p{1.5cm}||p{2.0cm}||p{1.8cm}||p{1.0cm}| }
 \hline
 \multicolumn{9}{|c|}{Mean versus turbulent and parity of mean fields} \\
 \hline
 run      		& $ \la \la B^{\prime2 }\ra \ra$ & $\la \la \bar{B}^2 \ra \ra$  & $\la \la B^{\prime}_{r} B^{\prime}_{\phi} \ra \ra $ & $\la \la \bar{B}_r \bar{B}_{\phi} \ra \ra$ & $ \la \la v^{\prime2 }\ra \ra/10^{-4}$ &  $\la \la \bar{v}^2 \ra \ra$ & $\la \la c_s \ra \ra$  & $Pa_{\rm av}$ \\
 \hline
 M-P4		& $1.5 \pm 0.7$      & $0.8 \pm 0.4$      &   $0.4 \pm 0.2$        & $0.12 \pm 0.06$  & $1.1 \pm 0.1$ & $0.0363 \pm 0.0008 $ & $0.089 \pm 0.007$ & $0.2 \pm 0.7$  \\
  M-P2  	  	&  $1.7 \pm 0.7$        & $0.7 \pm 0.3$      &    $0.5 \pm 0.2$        & $0.10 \pm 0.04$ & $3.6 \pm 0.8$ & $0.036 \pm 0.002$ & $0.098 \pm 0.009 $ & $0.2 \pm 0.6$ \\
  M-1P		& $2 \pm 1$     &  $0.4 \pm 0.1$      &     $ 0.6 \pm 0.3$          & $0.06 \pm 0.02$  & $4 \pm 1$ & $0.035 \pm 0.001$ & $0.099 \pm 0.009$ & $0.2 \pm 0.6$  \\
  M-2P		&  $2.3 \pm 0.9$      & $0.14 \pm 0.05$          & $ 0.6 \pm 0.2$     & $ 0.022 \pm 0.008$  & $4 \pm 1$&$0.036 \pm 0.001$ &$0.095 \pm 0.009 $ &$0.1 \pm 0.7$ \\ 

 \hline
\end{tabular}
\caption{Average (both over space and time) mean and turbulent quantities. $c_s$ and $Pa_{\rm av}$ are the average sound speed and parity respectively. Time average is done between $N_{\rm ISCO}=200-600$. }
\label{tab:mean_vs_turb}
\end{table*}

 \subsection{Spatio-temporal evolution of mean magnetic field}
  In previous sub-sections, we see that accretion stress attains saturation and mean fields play a significant role in the time evolution of the flow 
  (by causing sudden rise in $\alpha^{\rm Max}$, for example). To get a clearer picture of the scenario, in this sub-section we look into the spatio-temporal (both radial and meridional) behavior of the mean magnetic field.
 
 \subsubsection{Meridional variation}
  \begin{figure*}
      \includegraphics[scale=0.4]{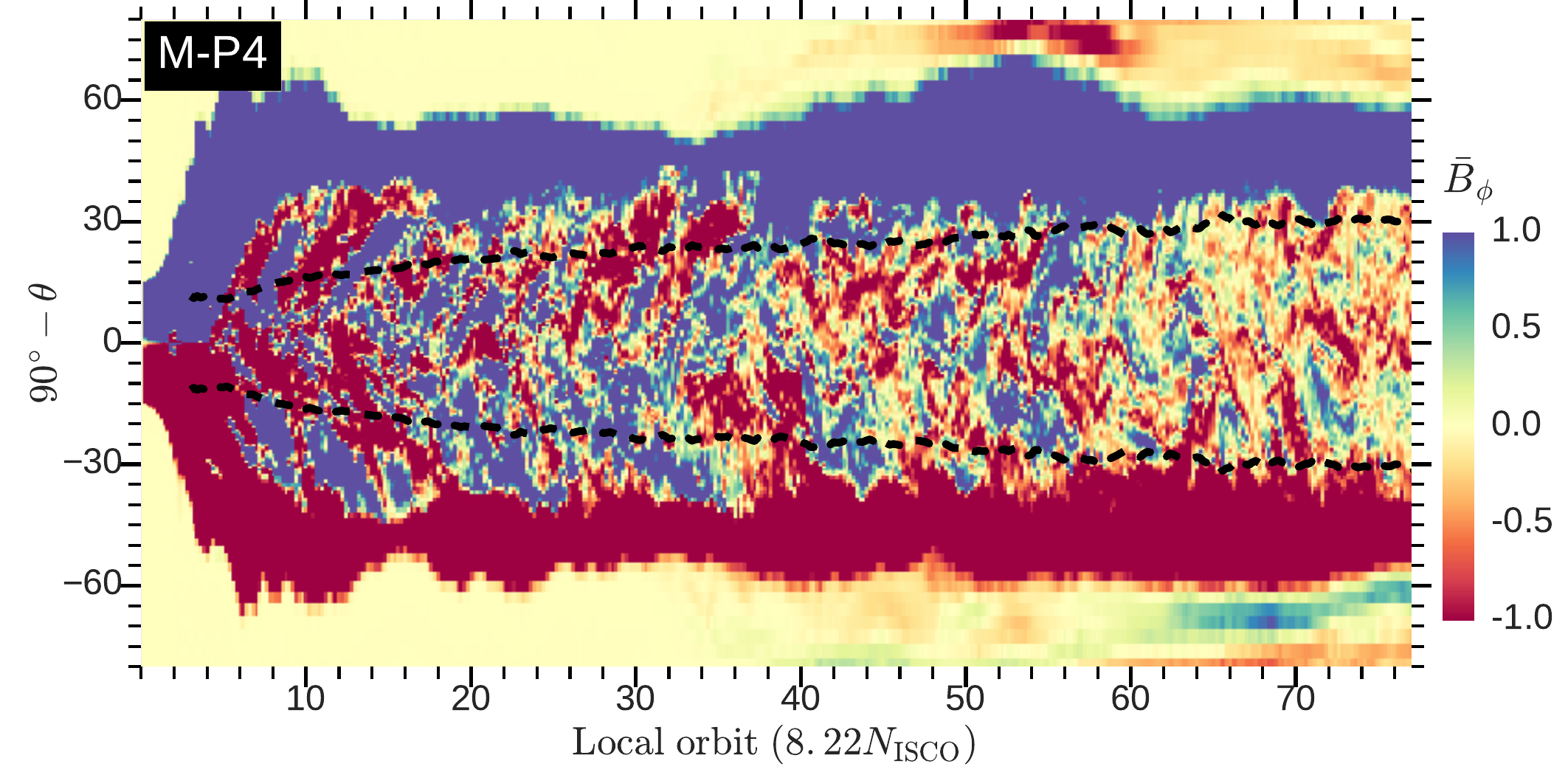}
    \includegraphics[scale=0.4]{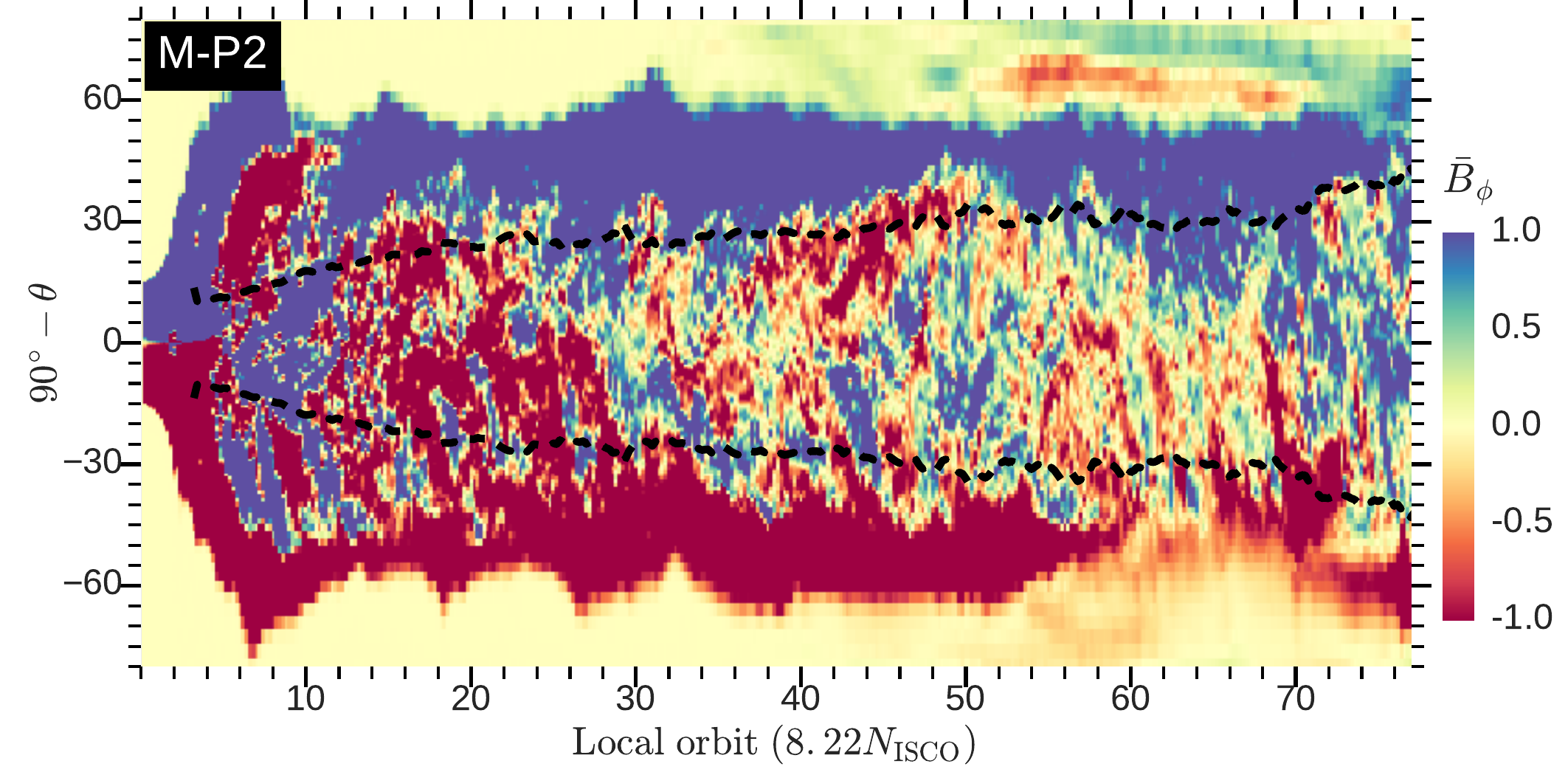}
    \includegraphics[scale=0.4]{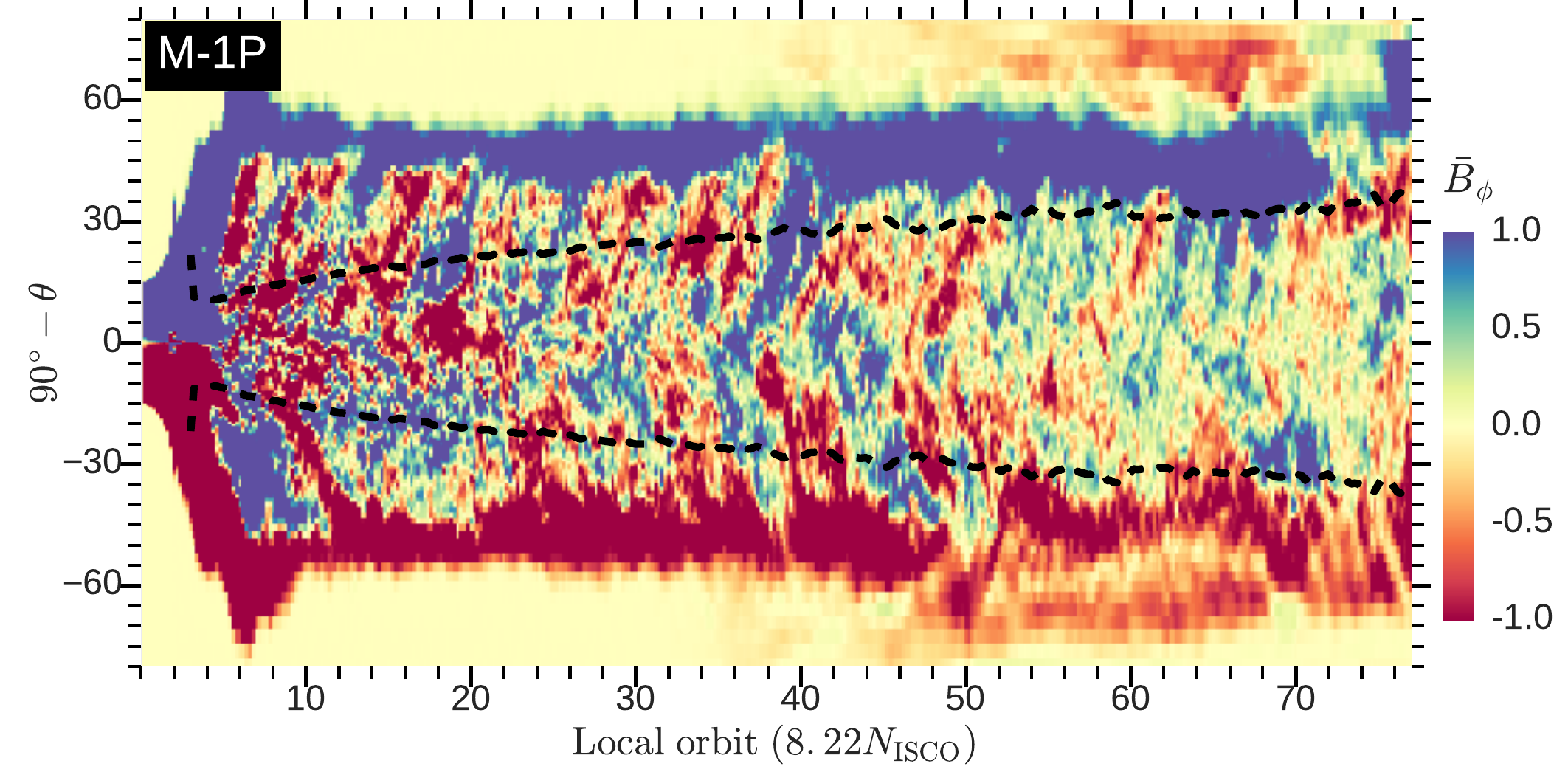}
    \includegraphics[scale=0.4]{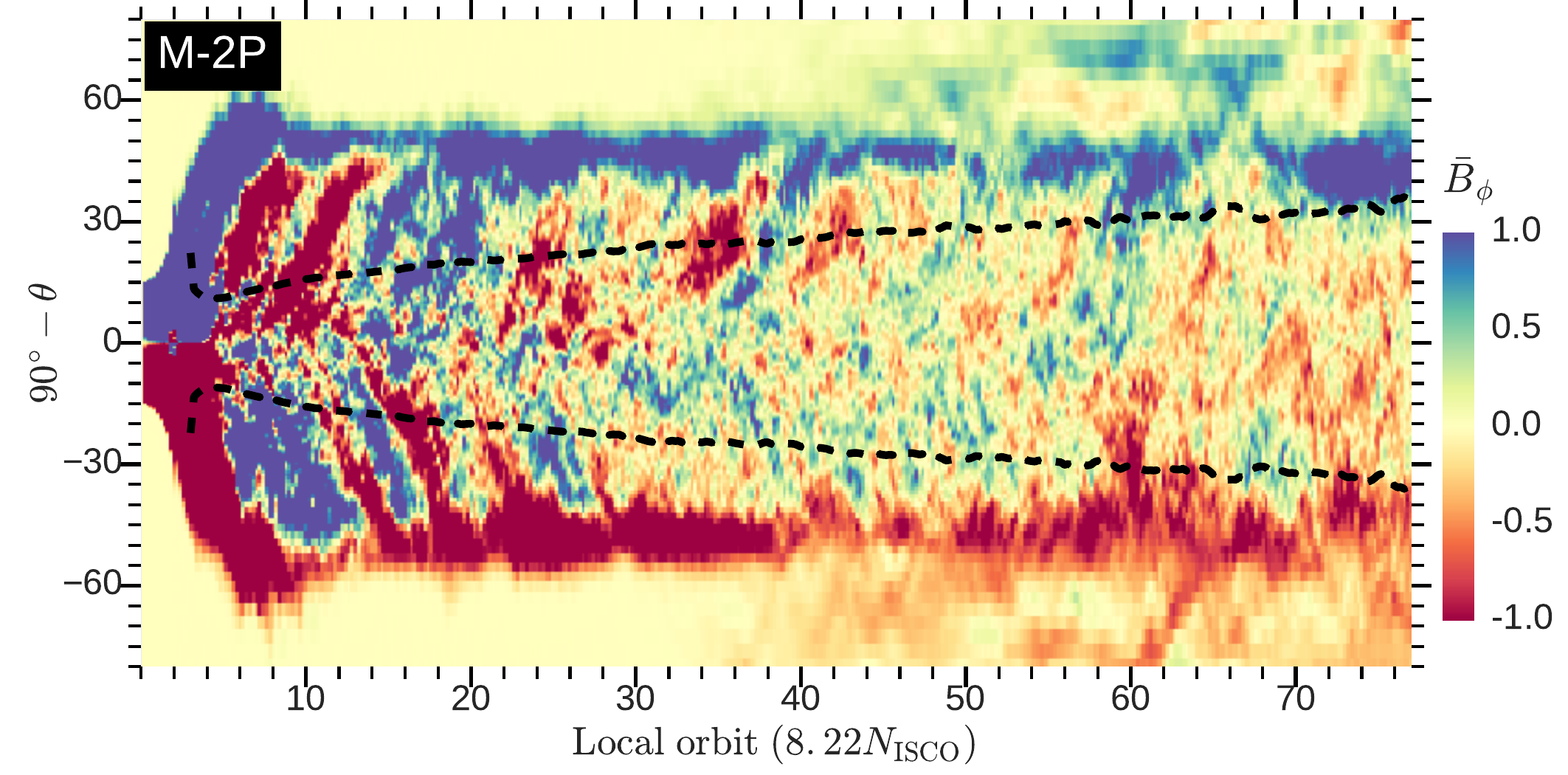}
    \caption{ Butterfly diagram: variation of the mean toroidal field $\bar{B}_{\phi}(R_0=20,\theta,t)$ with latitude ($\pi/2 - \theta$) and time. Black dashed lines indicates the one scale height above and below the mid-plane. Time is expressed in units of local orbit at $R_0=20$ ($=8.22 N_{\rm ISCO}$).}
   \label{fig:med_butterfly}
 \end{figure*}
 
A well-demonstrated phenomenon in stratified accretion disks is an oscillating mean toroidal magnetic field due to buoyant rise 
of the toroidal magnetic field from the mid-plane to the upper coronal regions. This famous feature known as the `butterfly diagram' is seen in  local (\citealt{Brandenburg1995}, \citealt{Hawley1996}, \citealt{Stone1996}, \citealt{Johansen2009}, \citealt{Davis2010}, \citealt{Gressel2010}, \citealt{Simon2012}, \citealt{Bodo2012} )  
 as well as  in global (\citealt{Oneill2011}, \citealt{Flock2012}, \citealt{Parkin2013}, \citealt{Parkin2014}, \citealt{Hogg2016}) MHD simulations  with the 
  oscillation timescale $\sim 10$ local orbits. Fig. \ref{fig:med_butterfly} shows the variation of mean toroidal field  $\bar{B}_{\phi}(r=20,\theta,t)$ with 
  latitude ($\pi/2 - \theta$) and time for our geometrically thick ($H/R \sim 0.5$) radiatively inefficient accretion flow. Here time is expressed in local orbit at $r=R_0=20$, which equals $8.22$ $N_{\rm ISCO}$. Unlike previous studies mentioned above, we do not see a regular cyclic behaviour of $\bar{B}_{\phi}$, rather we observe an intermittent sign reversal in the mean toroidal field. Recently, \cite{Hogg2018} also observed an intermittent cycle of the mean toroidal field with $H/R \sim 0.4$.  The runs with restricted azimuthal domain (M-P4, M-P2, M-1P) show stronger mean fields variations compared to the run M-2P. Also notice the ubiquity of positive $\bar{B}_{\phi}$ at late times for the run M-P2; that is the reason why toroidal flux $F_{\phi}(t)$ remains positive for longer time compared to the other runs in Fig. \ref{fig:med_fphi}.      

 \subsubsection{Radial variation}  
 \begin{figure}
    \includegraphics[scale=0.35]{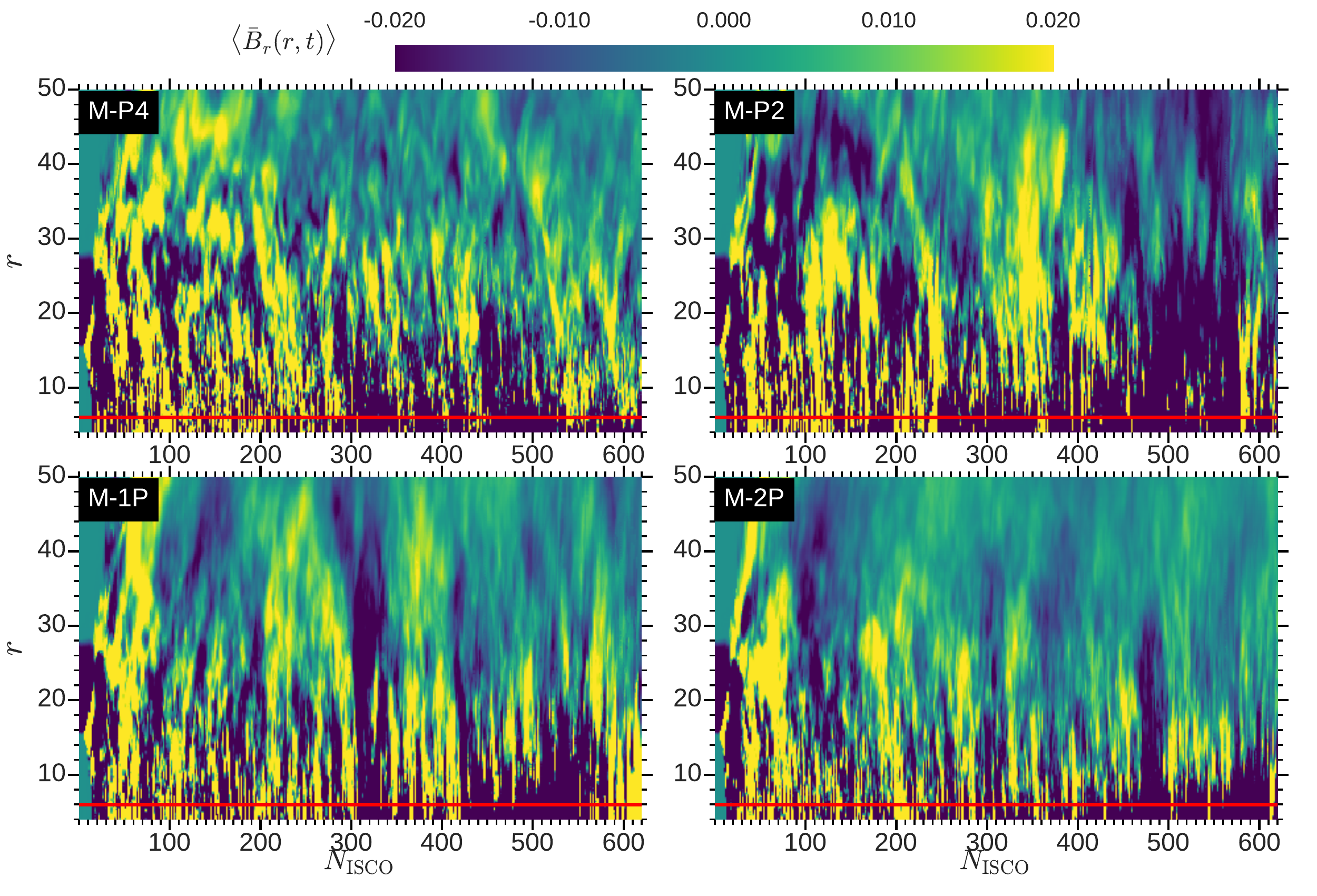}
     \includegraphics[scale=0.35]{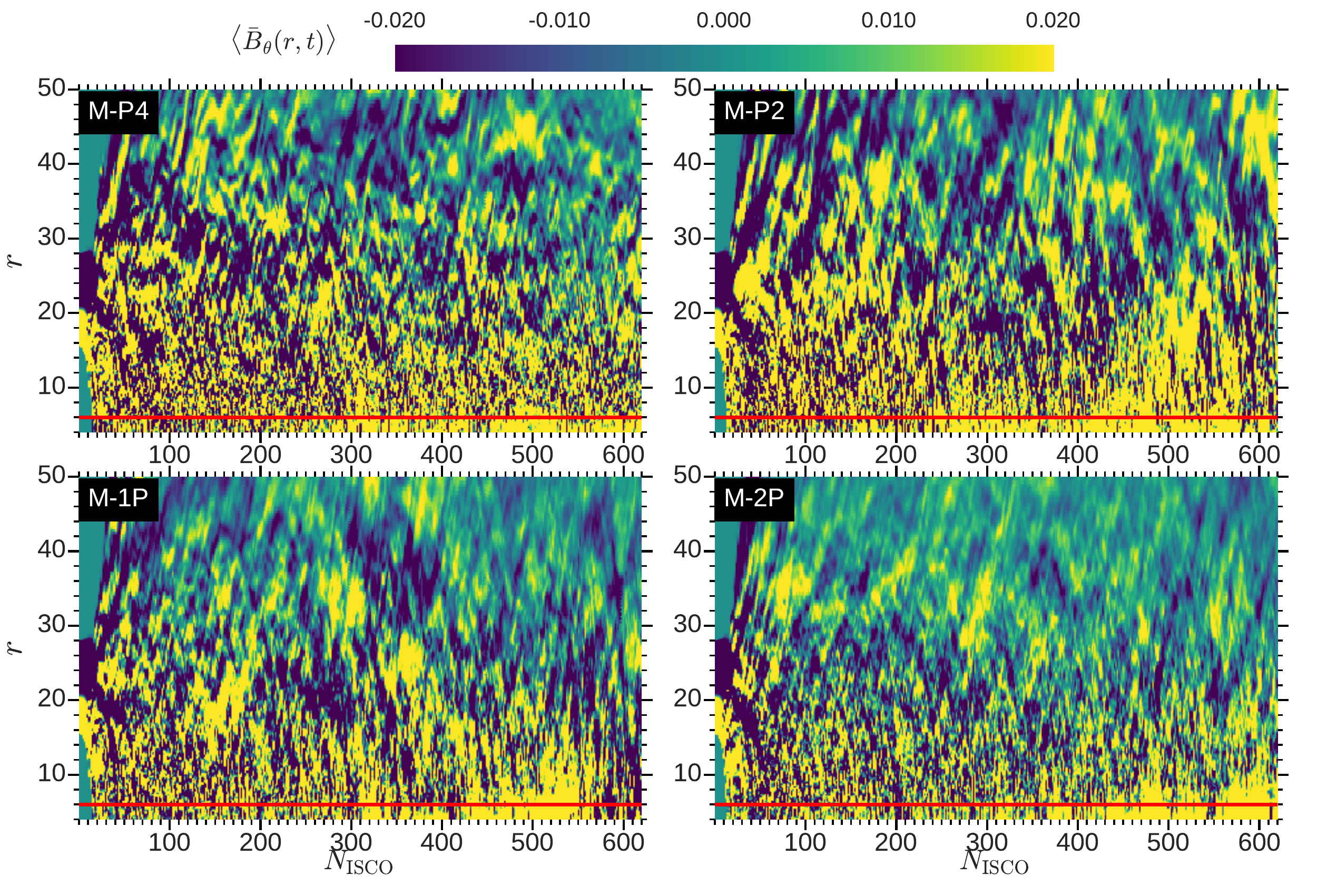}
     \includegraphics[scale=0.35]{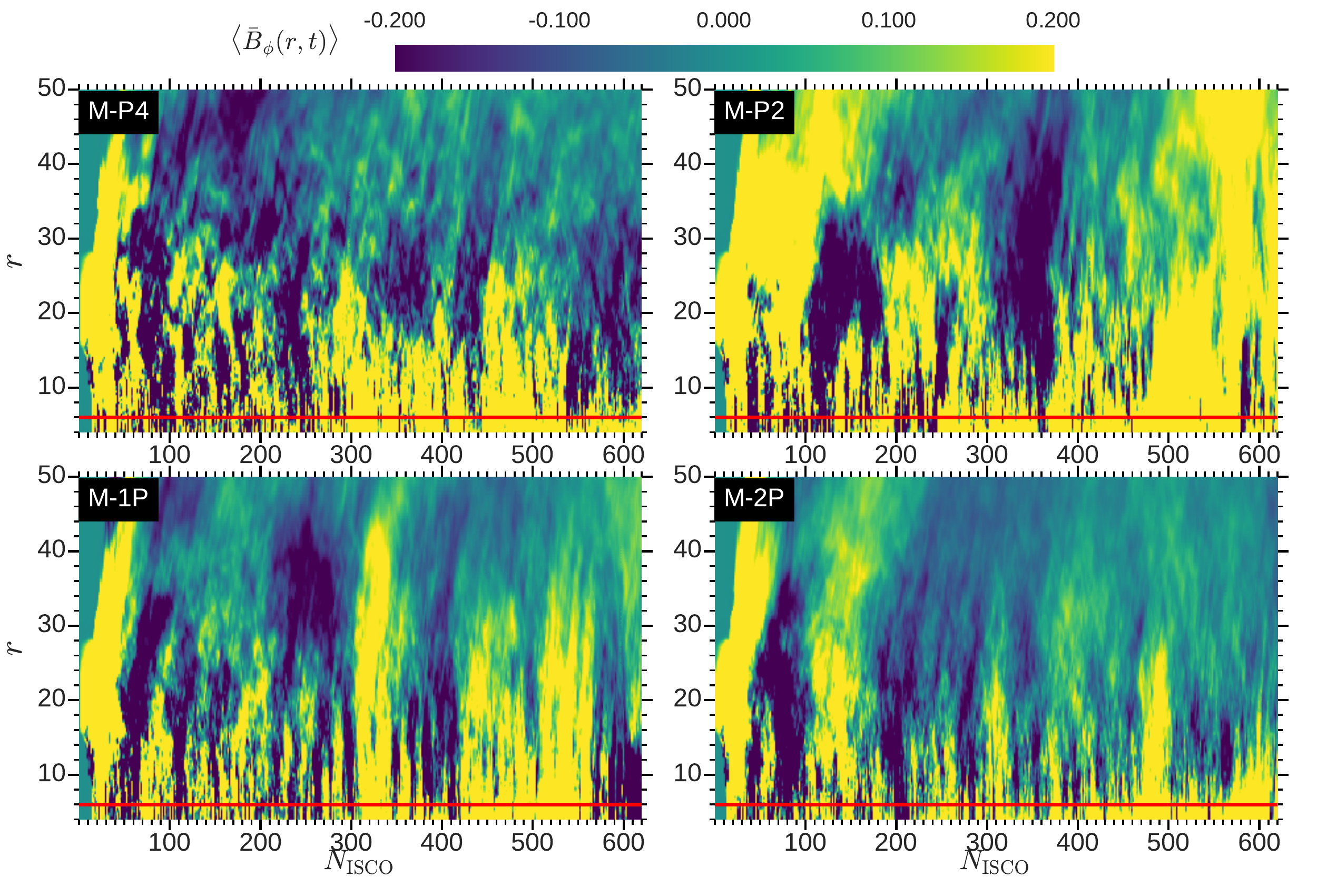}
    \caption{Spatio (radial)-temporal evolution of the mean radial ($\la \bar{B}_r (r,t) \ra$, top panel), meridional ($\la \bar{B}_{\theta}(r,t) \ra$, middle panel) and toroidal ($\la \bar{B}_{\phi}(r,t) \ra$, bottom panel) magnetic fields for the the runs with different azimuthal extents. Meridional average is done over one scale-height in the northern hemisphere. The horizontal red line denotes the location of the ISCO.}
   \label{fig:med_mean_br}
 \end{figure}
 
In this section we study the time evolution of the mean fields in the radial direction. In the meridional direction we average from mid-plane ($\theta=\pi/2$) to $\theta_H$ in the northern hemisphere (NH), as mean fields often have opposite signs in both the hemispheres (as seen in Fig. \ref{fig:med_butterfly}).
 
 Fig. \ref{fig:med_mean_br} shows the time evolution of the mean radial (top panel), meridional (middle panel) and toroidal (bottom panel) magnetic fields  along the radius for the four runs. For all the runs, we see an intermittent sign reversal of $\la \bar{B}_r(r,t) \ra$ with  time. Long coherent radial structures are seen, which are most prominent for runs with restricted domain sizes. Unlike the mean radial magnetic field, mean meridional magnetic field $\la \bar{B}_{\theta}(r,t) \ra $ shows a patchy distribution. Beyond $r=30$, patchy patterns are observed (specially for the run M-2P) to  translate in time, which is a signature of the presence of  an outflow at $r \gtrsim 30$. Like $\la \bar{B}_r(r,t) \ra$, the mean toroidal magnetic field $\la \bar{B}_{\phi}(r,t)\ra$ shows long coherent radial structures, which are anti-correlated  with $\la \bar{B}_r(r,t) \ra$. The strong anti-correlation between the mean radial and toroidal fields at certain times give rise to enhanced accretion stresses (see Fig. \ref{fig:med_alpha_R_M_t} and Fig. \ref{fig:alpha_max_rt}), which results in a higher mass accretion rate through the ISCO. As expected, all the runs with restricted azimuthal domains show stronger radial filaments ($\bar{B}_r$ and $\bar{B}_{\phi}$)/ patches ($\bar{B}_{\theta}$) of mean fields compared to the  run M-2P.
 
 \subsection{Structure of the flow}
 In the previous sub-sections we discuss the effects of azimuthal extent of the computational domain on the temporal evolution of different quantities. In this section, we study how time averaged properties of the accretion flow differ with different azimuthal domain sizes. As we are studying the radiatively inefficient accretion flows (RIAFs), which are geometrically thick ($H/R \sim 0.5$), we concentrate only on the radial profiles, defined by equation \ref{eq:radial_avg}. Time average is done over $N_{\rm ISCO}=200-600$.
 
 \subsubsection{Velocities}
  \begin{figure}
    \includegraphics[scale=0.34]{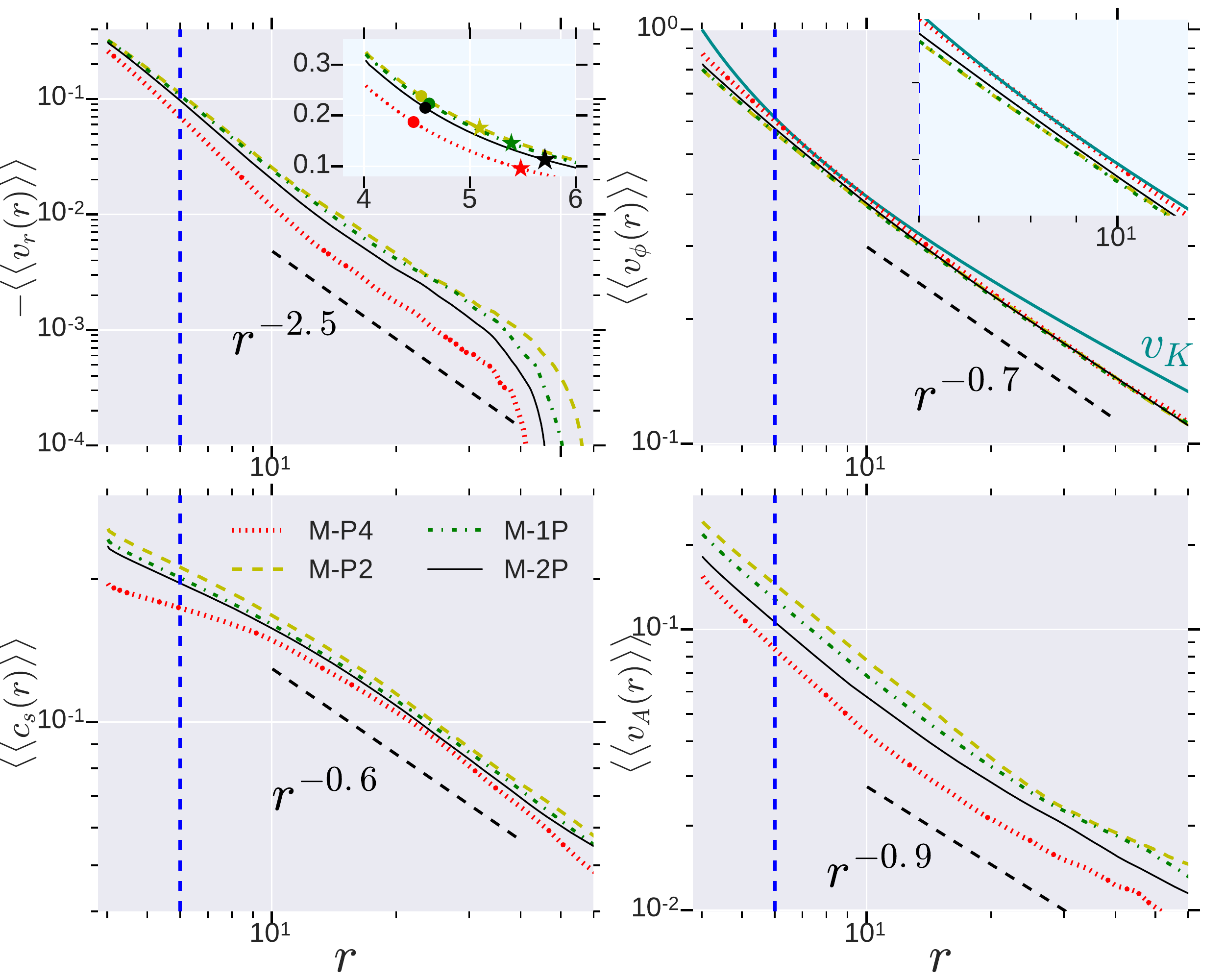}
    \caption{Comparison of radial variation of characteristic velocities for runs M-P4, M-P2, M-1P and M-2P. Left and right panels on top describe average radial ($\la \la v_{r}(r) \ra\ra$) and azimuthal ($\la \la v_{\phi}(r)\ra \ra$) velocities, respectively.  Profiles of sound ($\la \la c_s(r) \ra \ra$) and Alfv\'en ($\la \la v_A(r) \ra \ra$) speeds are shown in the bottom-left and bottom-right panels respectively. The radii pointed by filled circles and $\star$s (in the inset of the top-left panel) denote the locations where the flow becomes super-sonic and super-Alfv\'enic, respectively. The location of ISCO is shown by the vertical blue line.     }
   \label{fig:velocity_r}
 \end{figure}
 Four panels of Fig. \ref{fig:velocity_r} show the radial profiles of different flow velocities for our four runs. The dashed blue vertical line shows the location of the ISCO. Radial 
 \be
 \la \la v_{r}(r) \ra \ra=\frac{\la \la \rho v_r(r)\ra\ra}{\la \la \rho(r) \ra\ra}
 \ee
 and azimuthal
 \be
 \la \la v_{\phi}(r)\ra \ra=\frac{\la \la \rho v_{\phi}(r)\ra\ra}{\la \la \rho(r) \ra\ra}
 \ee
 velocities are compared in the top panels. The bottom panels show the radial profiles of sound 
 \be
 \la \la c_s(r) \ra \ra= \sqrt{\frac{\gamma \la \la P(r)\ra \ra}{ \la \la \rho(r) \ra \ra}}
 \ee
 and Alfv\'en 
 \be
 \la \la v_A(r) \ra \ra = \sqrt{\frac{\la \la B^2(r) \ra \ra }{ \la \la \rho(r) \ra \ra}}
 \ee
 speeds respectively.  
 
 At $r \sim 40$, $\la \la v_{r}(r) \ra \ra$ shows a sharp gradient indicating the region of `inflow equilibrium', the radius inside which the accretion flow is in  a statistically stationary state. This scale is given by $r_{\rm eq} \sim t_{\rm visc} |v_r |$, where $t_{\rm visc}$ is the viscous time. We see almost the same  values of $r_{\rm eq}\approx 40$ for all the runs. In the inset, the radii pointed by filled circles and $\star$s denote the locations where flow becomes super-sonic ($|v_r| > c_s$) and super-Alfv\'enic ($|v_r| > v_A$) respectively. As subsonic and sub-Alfv\'enic flow moves inward, it becomes both super-sonic and super-Alfv\'enic within the ISCO. As a result, accretion flow outside the outermost critical point (here Alfv\'en point) is not affected by any disturbances created around the inner boundary. Although, the nature of the radial profiles $\la \la v_{r}(r) \ra \ra$ are approximately similar in all four runs, its absolute value for the run M-P4 is smaller by a factor of 2 compared to that for the other three runs. 
 
 While the radial velocity shows a transonic nature, azimuthal velocity $\la \la v_{\phi}(r)\ra \ra$ is highly super-sonic and super-Alfv\'enic at all radii. Apart from the run M-P4, all other runs
 maintain a similar sub-Keplerian profile of $\la \la v_{\phi}(r)\ra \ra$ over all radii (see also Fig. \ref{fig:lambda} for the time evolution of angular momentum for the run M-2P). For the run M-P4, the region between $r=6-10$, $\la \la v_{\phi}(r)\ra \ra$ has an almost Keplerian profile (for clarity see the inset), which again indicates a less efficient angular momentum transport compared to the other three runs. Alfv\'en speed always remains sub-thermal ($\beta = 2c^2_s/\gamma v^2_A > 1$) all the way up to the inner boundary for all four runs, in spite of the fact that  the Alfv\'en speed rises more steeply with decreasing radius compared to the sound speed.
 
 Therefore, the study of the time-averaged (in QSS) radial profiles of characteristic velocities suggest a somewhat different nature of the 
 run M-P4 compared to the other three runs.
 
 \subsubsection{Magnetic fields}
  \begin{figure}
    \includegraphics[scale=0.4]{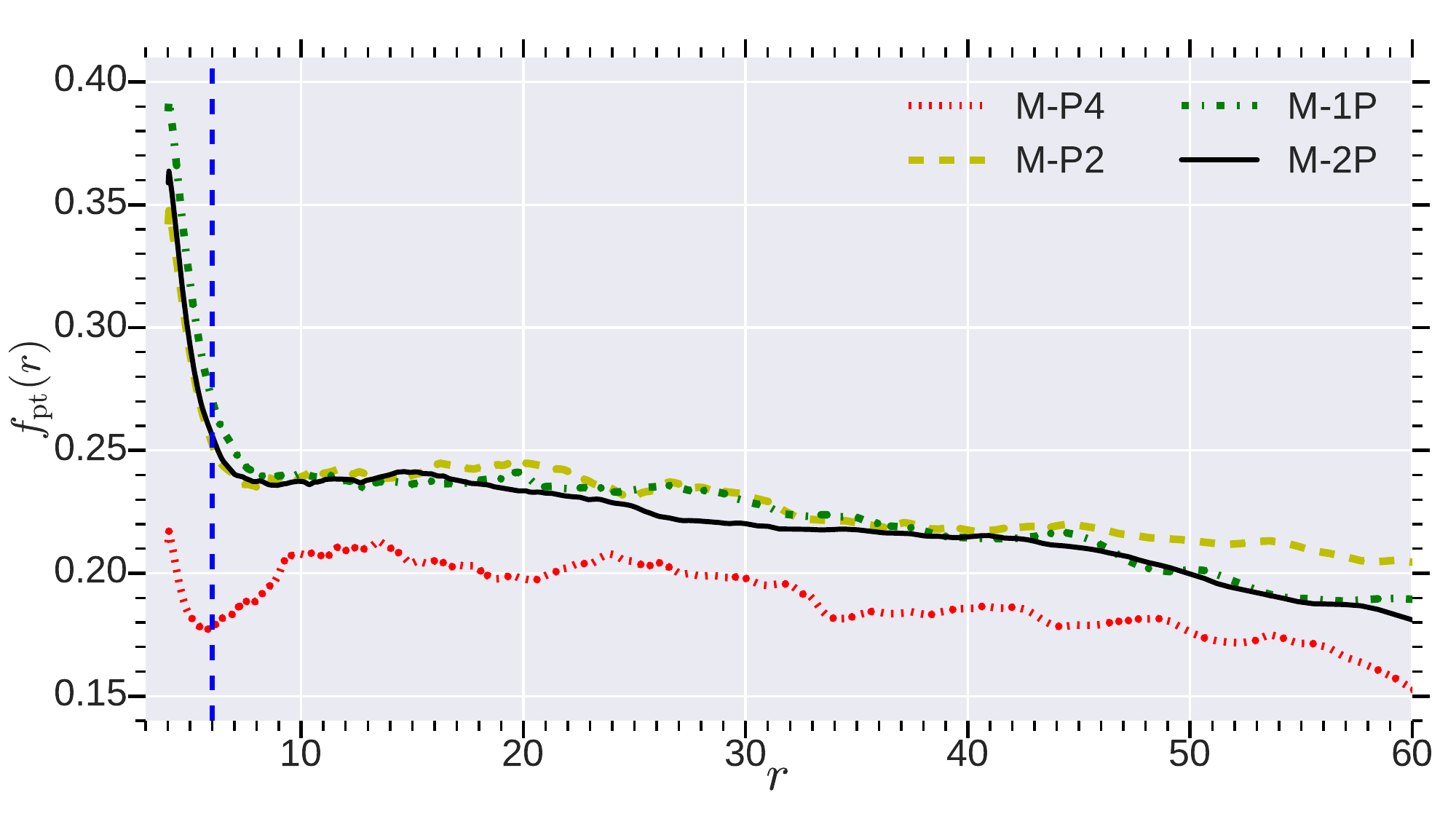}
    \caption{Radial variation of $f_{\rm pt}$ (see equation \ref{eq:fD}) for the four runs with different azimuthal domain sizes. Time average is done between $N_{\rm ISC0}=200$ and $N_{\rm ISC0}=600$.}
   \label{fig:med_fd}
 \end{figure}
 
 To look at the radial structure of magnetic fields in a concise way, we define a poloidal to toroidal magnetic energy ratio as
 \be 
 \label{eq:fD}
 f_{\rm pt}(r) = \frac{\la \la B^2_r(r) \ra \ra  + \la \la B^2_{\theta}(r) \ra \ra}{\la \la B^2_{\phi}(r) \ra \ra},
\ee
which describes the radial variation of the poloidal to toroidal magnetic energy ratio. It is also a measure of the efficiency of production of poloidal fields (predominantly radial field) out of the toroidal fields. Fig. \ref{fig:med_fd} shows the radial profiles of $f_{\rm pt}(r)$ for four runs M-P4, M-P2, M-1P and M-2P. As we approach ISCO from the outer radii, the poloidal to toroidal field ratio increases very slowly; just outside the ISCO, we see a rapid increase in 
$f_{\rm pt}$. While outside the ISCO, the total magnetic field is dominated by the toroidal component ($\gtrsim 80$ percent of total field energy), inside it the importance of poloidal field (mainly the radial field) increases very rapidly. This result along with those shown in Fig. \ref{fig:velocity_r} suggest that the dynamics of the magnetic field inside ISCO is controlled by flux-freezing and almost radial free fall rather than turbulence that controls the evolution of magnetic field outside ISCO (for a more detail discussion see section 5.2 of \cite{Beckwith2011}).     
 
 Although the above mentioned trend in magnetic field configuration is followed for all runs, run M-P4 stands out with a lower efficiency of conversion of toroidal field to poloidal field. This implies, like the velocity field, the steady state nature of the magnetic field matches for the runs with azimuthal extent $\Phi_0 \geq \pi/2$. 
 
 \section{Dynamo}
 \label{sect:dynamo}
 Previous local (e.g. \citealt{Brandenburg1995, Gressel2010, Bodo2012}) and global (e.g \citealt{Oneill2011, Flock2012, Parkin2013}) studies of MRI 
 driven accretion flow identify the oscillating mean fields with a dynamo process.
 All these works simulate thin disks ($H/R \ll 1$) in a local or global frame-work, in which the scale separation is self-evident. In this work,  we study the dynamo 
 process in a geometrically thick disk in which $H/R \sim 0.5$. It is interesting to see in Fig. \ref{fig:med_butterfly} that we don't see any periodic 
 oscillations of the mean toroidal fields ($\bar{B}_{\phi}$) for any of the runs, rather we see an intermittent sign reversal of ($\bar{B}_{\phi}$) in both 
 the hemispheres. In this section we analyze the runs with the full $2\pi$ azimuthal extent (M-2P).
  
 \subsection{Symmetry of mean fields}
\label{sect:mean_symmetry}
  \begin{figure}
    \includegraphics[scale=0.37]{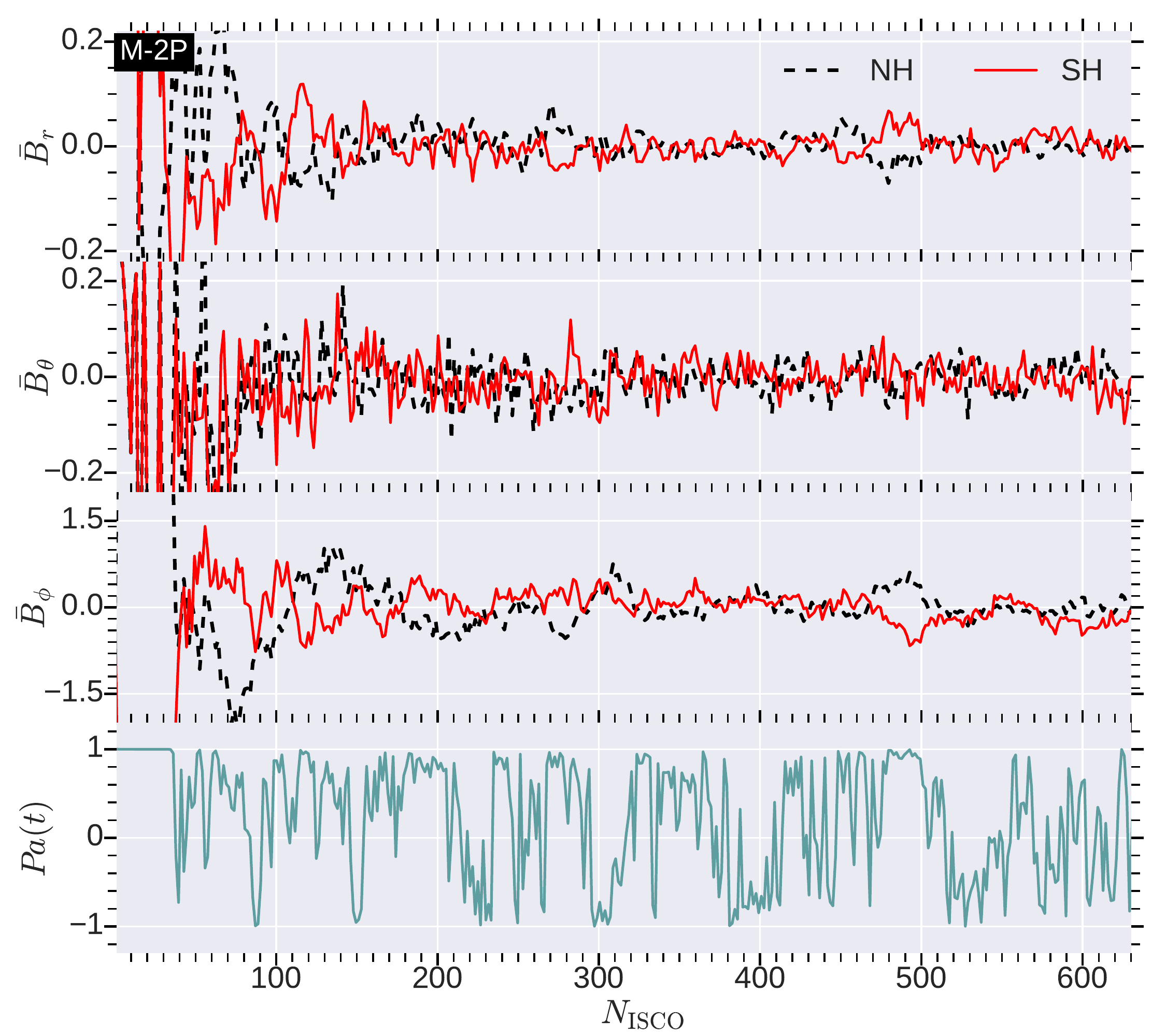}
    \caption{Time variation of average mean fields in both the hemispheres for the most realistic run M-2P. The average is done over the $\theta$ in 
    the northern ($\theta_H < \theta < 90^{\circ} $; NH) and southern hemispheres ($90^{\circ} < \theta < 90^{\circ} + \theta_H$; SH) and calculated 
    at $r=20$. Top three panels show the variations of mean radial ($\la \bar{B}_r (R_0,t) \ra$), meridional ($\la \bar{B}_{\theta} (R_0,t) \ra$) and 
    toroidal ($\la \bar{B}_{\phi} (R_0,t) \ra$) components of the magnetic field respectively. Bottom panel show the variation of mean field parity 
    (Eq. \ref{eq:parity}) with time. On average, dipole symmetry ($Pa>0$) prevails over quadrupole symmetry ($Pa<0$). Note that erratic parity at late
    times coincides with irregular dynamo cycles seen in the bottom right panel of Fig. \ref{fig:med_butterfly}.} 
   \label{fig:med_B_mean_nh_sh}
 \end{figure}
 The geometry of mean fields is described by its symmetry about the mid-plane ($\theta=90^{\circ}$). If radial and azimuthal fields are of different signs in SH and NH, the magnetic field configuration is dipole dominated (or more precisely, dominated by the even $l$ modes). On the hand, quadrupole dominated field configuration demands the same signs of $\bar{B}_r$ and $\bar{B}_{\phi}$ in the two hemispheres. The symmetry of the mean fields {\em within one scale-height} is not very clear from Fig. \ref{fig:med_butterfly}. To examine it more carefully, we look at the time variation of $\theta$-averaged mean fields in northern ($\theta_H < \theta < 90^{\circ} $; NH) and southern ($90^{\circ} < \theta < 90^{\circ} + \theta_H$; SH) hemispheres calculated at $r=R_0=20$  for the most realistic run M-2P. Top three panels of Fig. \ref{fig:med_B_mean_nh_sh} show the variations of mean radial ($\la \bar{B}_r (R_0,t) \ra$), meridional ($\la \bar{B}_{\theta} (R_0,t) \ra$) and toroidal ($\la \bar{B}_{\phi} (R_0,t) \ra$) components of the magnetic field respectively. While $\la \bar{B}_{\theta} (R_0,t) \ra$ in SH and NH does not show any correlation, $\la \bar{B}_r (R_0,t) \ra$ and $\la \bar{B}_{\phi} (R_0,t) \ra$ show oscillations between dipole and quadrupole symmetry.  To describe the symmetry concisely, we look at the parity of the mean fields ( following \citealt{Flock2012}) defined as
\be
\label{eq:parity}
Pa = \frac{B_{\rm dip} - B_{\rm quad}}{B_{\rm dip} + B_{\rm quad}},
\ee
where $B_{\rm dip} = (B^{AS}_{r})^2 + (B^{S}_{\theta})^2 + (B^{AS}_{\phi})^2 $ and  $B_{\rm quad} = (B^{S}_{r})^2 + (B^{AS}_{\theta})^2 
+ (B^{S}_{\phi})^2 $. Here, symmetric and anti-symmetric parts of the field are defined as $B^{S}_{i} = (\bar{B}^{NH}_{i} + \bar{B}^{SH}_{i})/2$ 
and $B^{AS}_{i} = (\bar{B}^{NH}_{i} - \bar{B}^{SH}_{i})/2$; $i=r,\theta,\phi$.
Bottom panel of Fig. \ref{fig:med_B_mean_nh_sh} shows the variation of parity with time. As we initialize with a dipolar field, 
in the beginning, $Pa=+1$. In the quasi-steady state, parity oscillates between dipole ($Pa>0$) and quadrupole ($Pa<0$) 
dominated configurations. In the quasi-steady state ($N_{\rm ISCO}=200-600$), the mean field symmetry is dominated by dipole with time average 
parity $Pa_{\rm av}= 0.1 \pm 0.7$ (see Table \ref{tab:mean_vs_turb} to see the average parity for other runs).

 \subsection{$\alpha-\Omega$ dynamo and $\alpha$ quenching}
 \begin{figure}
    \includegraphics[scale=0.4]{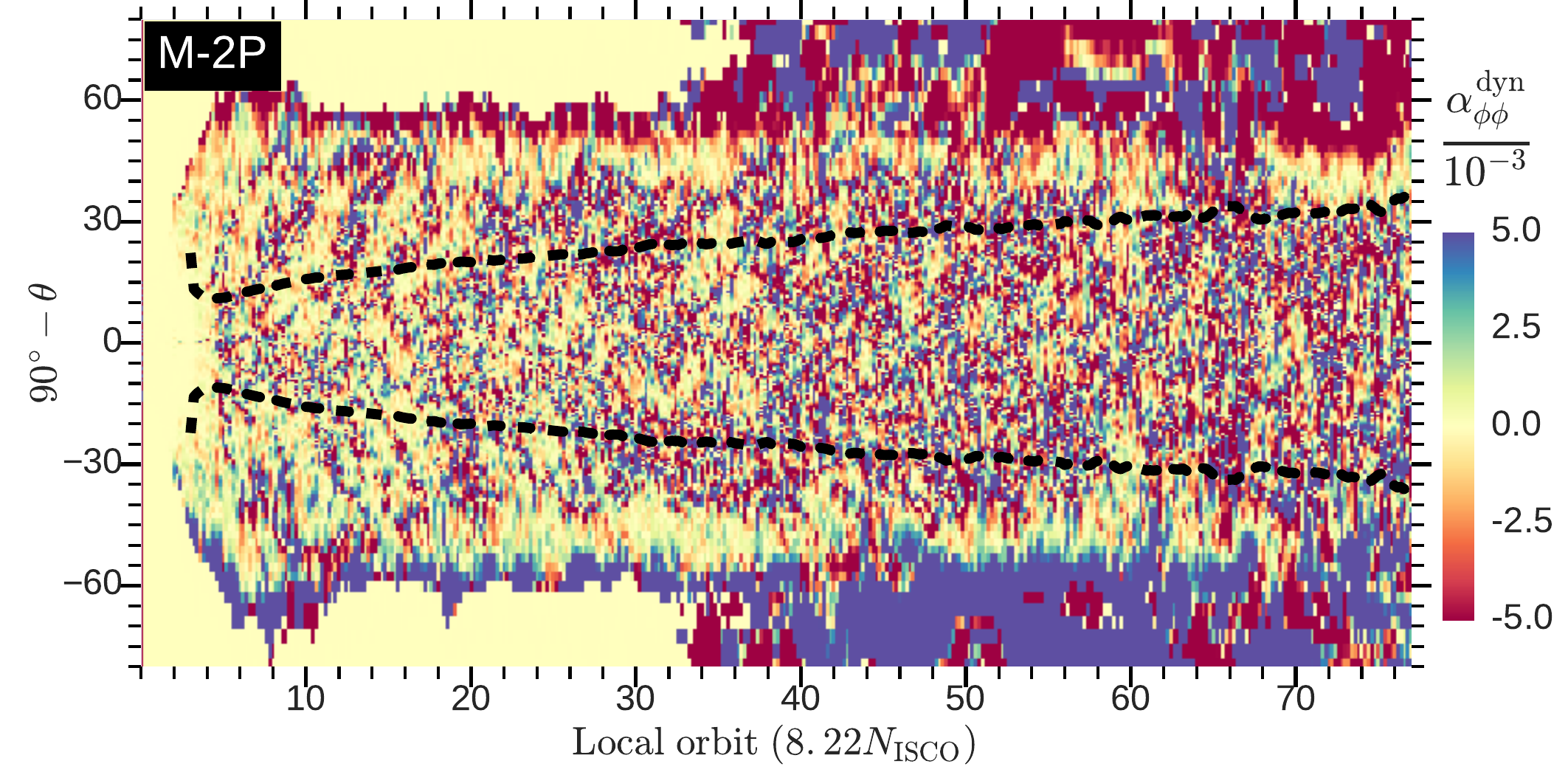}
    \caption{Variation of $\alpha^{\rm dyn}_{\phi \phi}=\bar{\mathcal{E}}_{\phi}/\bar{B}_{\phi}$ with $\theta$ and time at $r=R_0$. Time is expressed in units of local orbit at $r=R_0$. $\alpha^{\rm dyn}_{\phi \phi}$ does not show any coherent pattern as seen in previous local and global simulations (see the text); instead we see a patchy distribution over $\theta$ and time. This may be a result of $\alpha$-quenching. }
   \label{fig:alpha_dyn}
 \end{figure}
 
 \begin{figure}
    \includegraphics[scale=0.4]{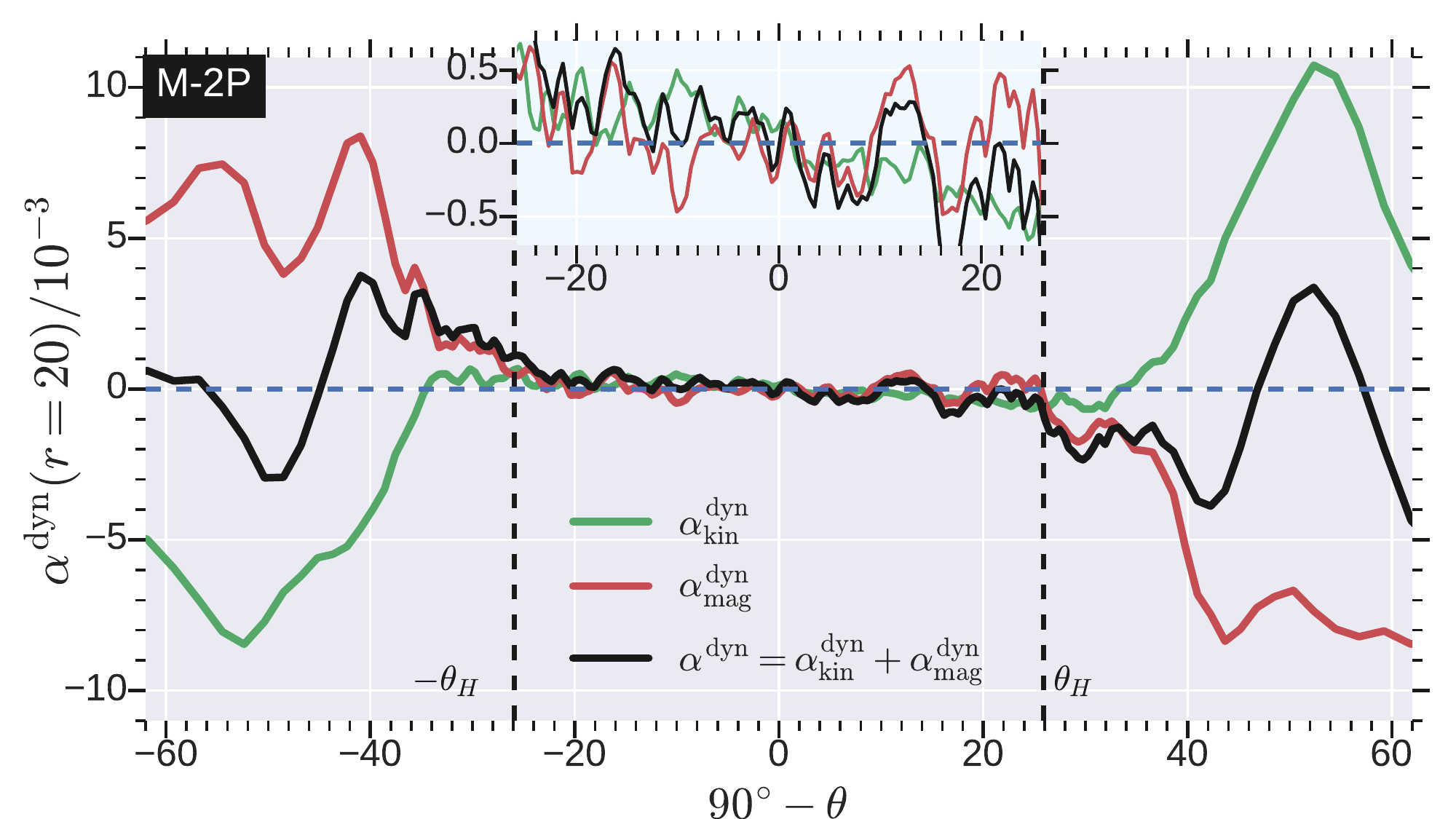}
    \includegraphics[scale=0.4]{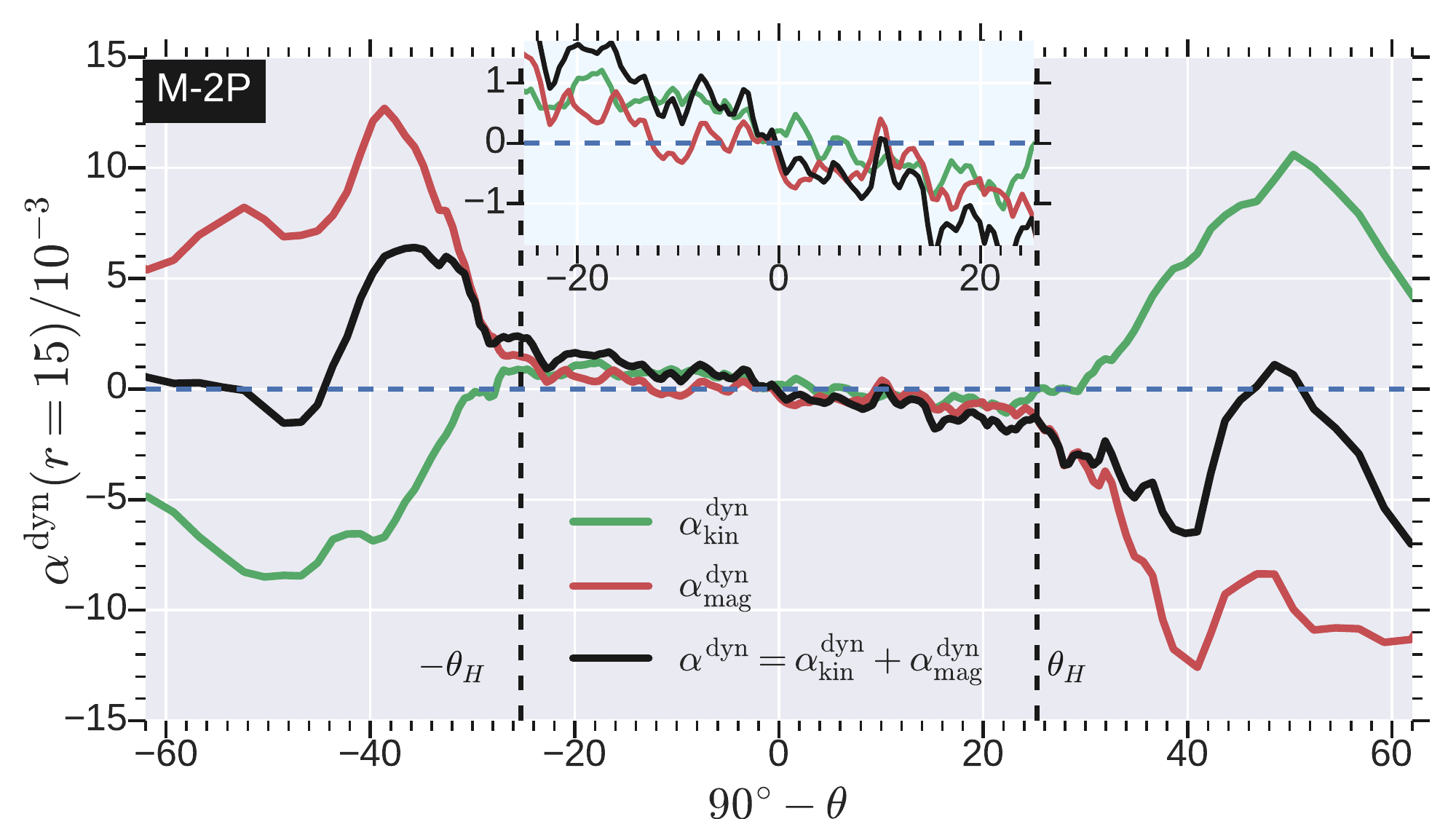}
    \caption{The meridional variation of average (over time) kinetic ($\alpha^{\rm dyn}_{\rm kin}$) and magnetic ($\alpha^{\rm dyn}_{\rm mag}$) helicities at $r=20$ (top panel) and $r=15$ (bottom panel). The sum of  the  $\alpha^{\rm dyn}_{\rm kin}$ and the $\alpha^{\rm dyn}_{\rm mag}$ is shown by black line ($\alpha^{\rm dyn}=\alpha^{\rm dyn}_{\rm kin}+\alpha^{\rm dyn}_{\rm mag}$). }
   \label{fig:alpha_kin_mag}
 \end{figure}
 In addition to intermittent sign flipping of mean fields, the saturation of magnetic energy in our global stratified simulation indicates a 
 dynamo process. MRI is compatible with both direct (non-helical turbulent dynamo) and inverse (helical mean-field dynamo) 
 dynamo processes (\citealt{Blackman_tan2004}). Direct dynamos can work both in unstratified and stratified MRI simulations, in which the 
 field grows by random stretching and twisting (\citealt{Kazantsev1968, Parker1979}). In a stratified medium, turbulence can be 
 helical due to two effects (\citealt{Blackman_tan2004}, \citealt{Gressel2010})- i) traditional Parker-type $\alpha-\Omega$ 
 dynamo mechanism (\citealt{Parker1979, ARC1998}), ii) due to magnetic buoyancy (\citealt{Tout1992, Brandenburg1998, Ruediger2000}). 
 The generation of large-scale field by turbulent fields can be modeled using the non-linear mean field theory.
 
 The mean field evolution equations are given by,
 \be
 \frac{\partial \bar{\bf{B}}}{\partial t} = \nabla \times \left [ \bar{\bf{v}} \times \bar{\bf{B}} + \overline{\bf{v}^{\prime} \times \bf{B}^{\prime}} - \lambda \nabla \times \bar{\bf{B}} \right]
 \ee
 where $\lambda$ is the microscopic diffusivity. The first term within the bracket on the right-hand side describes the the effects of mean 
 fields and flows on its evolution. The effect of turbulence on the mean field evolution is captured through mean emf 
 $\bar{\mathcal{E}} = \overline{\bf{v}^{\prime} \times \bf{B}^{\prime}}$. While shear in the mean  velocity field dominates the generation of mean 
 toroidal field $\bar{B}_{\phi}$, mean poloidal field (both $\bar{B}_{r}$ and $\bar{B}_{\theta}$) generation is  mainly attributed to the poloidal 
 gradient of the mean emf, $\partial \bar{\mathcal{E}}_{\phi}/ \partial \theta$ and $\partial \bar{\mathcal{E}}_{\phi}/ \partial r$. 
 The key idea of mean field dynamo is to express mean emf in terms 
 of mean velocity and magnetic fields and statistical properties of turbulent velocity fields. If we use a closure (e.g see \citealt{ARC1998})
 \be
 \bar{\mathcal{E}}_{i} = \alpha^{\rm dyn}_{ij} \bar{B}_{j} + \lambda^{\prime}_{ijk}\partial_{j} \bar{B}_{k}
 \label{eq:closure}
 \ee
 we obtain the classical $\alpha-\Omega$ dynamo. Here, $\lambda^{\prime}$ is the turbulent diffusion and $\alpha^{\rm dyn}$ is the tensor 
 which is crucial for the mean field amplification. If we assume isotropic turbulence, $\alpha^{\rm dyn}_{ij}$ only contains the diagonal terms. 
 Then the term 
 \be 
 \label{eq:alpha_pp}
 \alpha^{\rm dyn}_{\phi \phi} \approx \bar{\mathcal{E}}_{\phi}/\bar{B}_{\phi}
 \ee
 (neglecting the off-diagonal terms of $\alpha^{\rm dyn}_{ij}$ and second term in equation \ref{eq:closure}) captures toroidal to 
 poloidal field conversion. From symmetry arguments, we expect different sign of $\alpha^{\rm dyn}_{\phi \phi}$ in the two hemispheres. 
 On one hand local simulations provide negative (positive) sign of $\alpha^{\rm dyn}_{\phi \phi}$ in NH (SH) 
 (\citealt{Brandenburg1995, Brandenburg1997, Davis2010}, in \citealt{Gressel2010} for $z < H$), on the other hand global simulations show 
 the opposite sign (\citealt{Arlt2001, Flock2012}). Although recently, \cite{Hogg2018} observed a very weak $\alpha^{\rm dyn}_{\phi \phi}$ of negative sign (in NH) in global simulations. A summary of previous results is given in Table \ref{tab:list_prev_results}.
 
 \begin{table*}
\centering
\begin{tabular}{ |p{3.3cm}||p{0.8cm}||p{0.8cm}||p{0.6cm}|p{1.8cm}||p{0.7cm}||p{0.7cm}||p{1.65cm}||p{1.65cm}|p{1.8cm}}
 \hline
 \multicolumn{10}{|c|}{Details of previous studies} \\
 \hline
 Work  & Model & Strati- \newline fication  & $\Phi_0$ & EOS & $\gamma$ &  $H/R$ & $\alpha^{\rm dyn}_{\phi \phi}$ \newline (NH) & $\alpha^{\rm dyn}_{\rm kin}$ \newline (NH) & $\alpha^{\rm dyn}_{\rm mag}$ \newline (NH) \\
 \hline
\cite{Brandenburg1995}   & Local  & Yes   & --  & Ideal, cooling    &  $5/3$  & --  & $-ve$   & $+ve$  & $+ve$ \\
\cite{Davis2010}        & Local   & Yes  & -- & Isothermal   & --  &  -- & $-ve$ &  -- & -- \\
\cite{Gressel2010}       & Local   & Yes  & -- & Isothermal   & --  &  -- & $-ve$ $(z<H)$ \newline $+ve$ $(z>H)$ &  $-ve$ &  $-ve$ $(z<H)$ \newline $+ve$ $(z>H)$\\
\cite{Oishi2011}     & Local   & Yes  & -- & Isothermal   & --  &  -- & -- & -- &  $-ve$ $(z<2H)$ \newline $+ve$ $(z >2H)$\\
\cite{Arlt2001}      & Global  & Yes  & $2 \pi$  & Isothermal  & -- & 0.07  & $+ve$  & -- & -- \\
\cite{Flock2012}    & Global  & Yes  & $2 \pi$  & Isothermal  & -- & 0.07  & $+ve$  & -- & -- \\
\cite{Hogg2018} & Global  & Yes  & $\pi/3$  & Ideal, \newline {\em adhoc} cooling  & 5/3 & 0.05  & $-ve$  & -- & -- \\
Current work   & Global  & Yes  & $2 \pi$  & Ideal, \newline adiabatic  & 5/3 & 0.5  & Noisy  & $-ve$ $(z<H)$ \newline $+ve$ $(z>H)$ & $-ve$ \\
 
 \hline
\end{tabular}
\caption{Summary of a few previous local and global studies. `-' implies that either the information is not applicable or not given in the work. \citealt{Hogg2018} used different values of $H/R$ with $x =0.05,~ 0.1,~0.2,~0.4$.}
\label{tab:list_prev_results}
\end{table*}
 
 Fig. \ref{fig:alpha_dyn} shows the variation of $\alpha^{\rm dyn}_{\phi \phi}$ (calculated at $r=R_0$) with $\theta$ and time. Time is 
 expressed in units of local orbit at $r=R_0$. Unlike previous studies, we do not see any coherent spatio-temporal distribution of 
 $\alpha^{\rm dyn}_{\phi \phi}$; instead we see a patchy distribution. This is probably due to $\alpha$-quenching 
 (\citealt{Pouquet1976, Ruediger1993}) which arises due to the back reaction of the Lorentz force on the helical fluid motions. 
 As a result, the effective $\alpha^{\rm dyn}$ 
 turns out to be the combination of kinematic and magnetic contributions; $\alpha^{\rm dyn} = \alpha^{\rm dyn}_{\rm kin} + 
 \alpha^{\rm dyn}_{\rm mag}$ (\citealt{Gressel2010}). While $\alpha^{\rm dyn}_{\rm kin}$ gives the forcing, $\alpha^{\rm dyn}_{\rm mag}$ 
 quenches the kinematic forcing through non-linear response.
 
 Assuming isotropy (which need not  be valid in rotating MHD turbulence), we calculate kinematic and magnetic $\alpha$ in the following way,
 \ba
 \label{eq:kin_alpha}
 && \alpha^{\rm dyn}_{\rm kin} = -\frac{1}{3}\tau_c \overline{\textbf{v}^{\prime}.(\nabla \times \textbf{v}^{\prime})},\\
 \label{eq:mag_alpha}
 && \alpha^{\rm dyn}_{\rm mag} = \frac{1}{3}\tau_c \overline{\frac{\textbf{B}^{\prime}}{\sqrt{\rho}}.\left (\nabla \times \frac{\textbf{B}^{\prime}}{\sqrt{\rho}}\right )},
 \ea
 where the correlation time $\tau_c$ is a free parameter. As we are interested in the relative importance of $^{\rm dyn}_{\rm kin}$ and 
 $\alpha^{\rm dyn}_{\rm mag}$, not on their absolute values, we simply  use $\tau_c= \Omega^{-1}$. Fig. \ref{fig:alpha_kin_mag} shows the 
 meridional variation of average kinetic and magnetic $\alpha$ at $r=20$ (top panel) and $r=15$ (bottom panel). We also plot the sum 
 of the $\alpha$-s to  check the relative strength of the kinematic and magnetic $\alpha$-s. We look at the meridional variation of $\alpha$-s 
 at different radii to check consistency of its trend. The vertical dashed black line describes the angle ($\theta_H$) corresponding to one 
 scale-height. At both radii, $\alpha^{\rm dyn}_{\rm kin}$ shows a definite trend. In the NH, close to the mid-plane, it is negative  and changes 
 sign above one scale height(see insets). In SH,  $\alpha^{\rm dyn}_{\rm kin}$ just follows the opposite trend. On the other hand, 
 $\alpha^{\rm dyn}_{\rm mag}$ is always negative in the NH with little bit of randomness near the mid-plane. Although at $r=15$, due 
 to a better behaved $\alpha^{\rm dyn}_{\rm mag}$, the resultant $\alpha^{\rm dyn}$ shows a 
 regular trend within one-scale height with 
 a negative sign (in NH).  Above one scale-height,  $\alpha^{\rm dyn}_{\rm kin}$ and $\alpha^{\rm dyn}_{\rm mag}$ have different signs with 
 almost equal magnitudes. Their sum have different signs at different $\theta$ depending on the relative strength of kinetic and magnetic 
 $\alpha$. This may be the reason behind the randomness of $\alpha^{\rm dyn}_{\phi \phi}$ seen in Fig. \ref{fig:alpha_dyn}.
 
 \section{Discussion and conclusions}
 \label{sect:discuss_disk}
 \subsection{Convergence}
It is always important to construct a computational model of MRI driven accretion disk turbulence which is independent of the grid-scales. To look at the convergence of turbulence we investigate the temporal behaviour of different quantities as discussed in section \ref{sect:convergence}. 

To minimize computational cost we use simulation set-ups with azimuthal extents $\Phi_0=\pi/4$ for convergence studies. 
Based on the long term evolution of different convergence metrics, we find our medium (M-P4) and high (H-P4) resolutions are converged. 
Like \cite{Parkin2013}, we find accretion stress $\alpha^T$ and plasma-$\beta$ to be good indicators of convergence, unlike 
\cite{Sorathia2012} who found them to be poor indicators. Non-converged low resolution run L-P4 displays reduced stress as well 
as gradual decrease in magnetization (see Figs. \ref{fig:alpha_conv_p4} \& \ref{fig:beta_conv_p4}). Also, the mass accretion rate and 
specific angular momentum at the ISCO are found to be useful to investigate convergence. 

Local isothermal simulations often quote the threshold number of cells in the vertical direction per scale height $H$ for convergence. 
For example, \cite{Davis2010} found that $64-128$ cells$/H$ are necessary for convergence. For our {\em adiabatic} global simulations, 
scale height $H(r)=\la c_s(r,\theta=\pi/2) \ra/\la \Omega(r,\theta=\pi/2) \ra$ increases with time as temperature increases and angular momentum decreases with time. Therefore, the number of vertical cells per 
$H$ is difficult to quantify for our simulations. Although in QSS ($N_{\rm ISCO}> 200$), our medium resolution runs have 
$H/R \Delta \theta \approx 42-64$ compared to non-converged low resolution run (L-P4) with $H/R \Delta \theta \approx 16$. It is worth noting 
that other global studies where $H/R$ is kept constant throughout the run time, the minimum number of cells per H required for convergence 
are - $25/H$ (\citealt{Fromang_nelson2006}), $32-64/H$ (\citealt{Flock2011}), $32/H$ (\citealt{Sorathia2012}), $27/H$ (\citealt{Parkin2013}) 
(note that different papers use different definitions of $H$!).

Like \cite{Hawley2013},  for our {\em adiabatic} simulations, we find quality factors $Q_{\theta}$ and $Q_{\phi}$, which are essentially  
the number of cells per unit wavelength of the fastest growing MRI mode in $\theta$ and $\phi$-directions; to be more useful than 
$H/R \Delta \theta$ related to the thermodynamics. Our converged medium resolution run M-P4 shows average poloidal quality factor 
$\la \la Q_{\theta, {\rm sat}} \ra \ra = 10.5 \pm 0.9$, which is on the lower side according to \cite{Sorathia2012} and \cite{Hawley2013} 
(who mention that $\la \la Q_{\theta} \ra \ra \gtrsim 10-15$ for convergence). While the poloidal quality factor shows the sign of marginal 
resolvability, toroidal quality factor $\la \la Q_{\phi, {\rm sat}} \ra \ra = 33 \pm 3$, is well above the required value of 
$\la \la Q_{\phi} \ra \ra \gtrsim 20$. Comparisons 
among the accretion rates (mass and angular momentum) (Fig. \ref{fig:jnet_conv_p4}), stresses (Fig. \ref{fig:alpha_conv_p4}), 
plasma-$\beta$ (Fig. \ref{fig:beta_conv_p4}) between the run M-P4 and the higher resolution run H-P4 indicates that the former run is also 
well resolved. This justifies the claim that both the quality factors are intertwined, smaller value of one can be compensated by the 
larger value of the other provided $\la \la Q_{\theta, {\rm sat}}  \ra \ra \la \la Q_{\phi, {\rm sat}}  \ra \ra \geq 250$  (\citealt{Hawley2013}).

Magnetic tilt angle, defined in Eq. \ref{eq:tilt_angle}, above a critical value confirms the transition from linear growth of MRI to saturated 
turbulence.The saturated value of quality factors increases with the increasing resolution but the magnetic tilt angles $\theta_B$ for all 
our converged runs remain almost constant (see Table \ref{tab:results_tab}). Even runs with different azimuthal extents $\Phi_0$ which 
produce different accretion stresses, show similar $\la \la \theta_{B,{\rm sat}} \ra \ra = 13^{\circ}-14^{\circ}$. This is the unique property of 
$\theta_B$  among all the convergent metrics. This feature of magnetic tilt angle was first noticed by \cite{Hawley1995}. Later  
\citealt{Blackman2008} provided an empirical value of $\theta_B$ by analyzing several published 3D simulations. 
\cite{Sorathia2012} and \cite{Hawley2013} (there $\alpha_{\rm mag}$) report the saturated value of tilt angle $\approx 12^{\circ}-14^{\circ}$ for the converged runs. Other global simulations also confirm this narrow range of the tilt angle (\citealt{Parkin2013}, \citealt{Hogg2016}).

 \subsection{Dependence on azimuthal domain size}
 \label{sect:domain_size}
  Our 3D global ideal adiabatic MHD simulations of  RIAFs with different azimuthal extents show some similarities and dissimilarities with the 
  previous  (mostly isothermal) studies. We discuss them one by one.
 
 \subsubsection{Higher accretion stress for $\Phi=\pi/2, \pi$}
 While runs with  $\Phi=\pi/2$ and $\pi$ overestimate the accretion stress $\alpha^T$ compared to the natural azimuthal extent of $\Phi_0=2 \pi$, 
 $\alpha^T$ for the run M-P4 ($\Phi_0=\pi/4$) underestimates it (see Fig. \ref{fig:med_alpha_R_M_t}). 
 This result does not match either with  \cite{Hawley2001}, which finds independence of accretion stress on azimuthal domain size (more precisely, 
 about 10 percent less magnetic stress on average for the reduced domain) or with \cite{Flock2012}, which finds a larger stress for smaller $\Phi_0$. 
 Therefore, although we find a dependence of  the accretion stress on azimuthal extent of computational domain, the trend does not follow 
 \cite{Flock2012}. The difference lies in the evolution of mean and turbulent Maxwell stresses in the two studies. \cite{Flock2012} find both higher 
 turbulent and mean  Maxwell stresses for the restricted domain sizes. On the contrary, we find higher mean Maxwell stress for the the runs with 
 smaller azimuthal domains, turbulent stress does not show any tend (see Table \ref{tab:mean_vs_turb}). 
 
 \subsubsection{Higher mean fields for the runs with smaller azimuthal extent}
 
Smaller azimuthal domains produce stronger mean magnetic fields (Table \ref{tab:mean_vs_turb}). This is in accordance with \cite{Flock2012}.  
The reason behind the stronger mean fields for the reduced azimuthal domain is not very clear. \cite{Hawley2001} attributes the larger 
fluctuations of accretion stress seen in the model with reduced azimuthal extent to the existence of the channel solution for vertical fields. 
It is easier for the channel solutions to maintain a spatial coherence for longer time for the smaller domain size. We also observe larger 
fluctuation level for the runs with smaller azimuthal domain (see Table \ref{tab:results_tab}). 
 
 To understand the dominance of mean fields in the restricted domain size, we look at the effects of azimuthal boundary conditions. In 
 Fig. \ref{fig:med_psd} we see that most of turbulent energy is at the largest scales. The periodic boundary conditions used in $\phi$  
 enforces the fields to be the same at the two boundaries. As a result, the natural build up of low$-m$ field is prohibited. For the smaller 
 domain size the probability of cancellation of fields is small as $\bar{B}_i = \sum_{\phi} B_i (\phi)$ and the magnetic fields have tendency 
 to appear as the mean field. This is manifested in stronger mean fields for the restricted azimuthal domains. 
 
 \subsubsection{The appropriate azimuthal extent} 
The steady state properties of the accretion flow  are very similar for the runs with the azimuthal extent $\geq \pi/2$( M-P2, M-1P and M-2P), 
but the run M-P4 with $\Phi_0=\pi/4$ stands out with a lower specific mass accretion rate (see Fig. \ref{fig:med_mdot_jnet}) and less efficient 
toroidal to poloidal magnetic field conversion process (see Figs. \ref{fig:velocity_r} and \ref{fig:med_fd}). Despite many similarities, because 
of the boundary effects, runs with restricted azimuthal domains of $\pi/2$ and $\pi$ give rise to stronger axisymmetric mean fields compared 
to the run with the natural azimuthal extent $2 \pi$. 
 We conclude that the appropriate restricted azimuthal extent of the simulation domain 
 depends on the aim of the study. If one wants to study the steady structure of the flow, the minimum requirement of the azimuthal extent is 
 $\Phi_0=\pi/2$. On the other hand, to study turbulence and dynamo in a geometrically thick disk ($H/R \sim 0.5$), one must use 
 $\Phi_0=2 \pi$ in comparison to the minimum requirement $\phi_0=\pi$ proposed by \cite{Flock2012} for their study of  geometrically 
 thin disk ($H/R = 0.07$). Comparison of these two studies imply that the appropriate domain size essentially depends on the thickness ($H/R$ ratio) of the disc. Because
the number of disc scale-height can be accommodated in the azimuthal direction is different for different disc thickness and mean fields converge only if the simulation domain contains sufficient number of scale-heights in the azimuthal direction. For example, for $H/R\sim 0.07$, azimuthal extent of $\pi$ (which can accommodate $44H$) is sufficient for converged behvaiour of mean fields (\citealt{Flock2012}), whereas for a larger $H/R\sim 0.5$, we need a full $2\pi$ extent.
 
\subsection{Structure of RIAFs}
\label{sect:RIAFs}

 \begin{figure}
    \centering
    \includegraphics[scale=0.42]{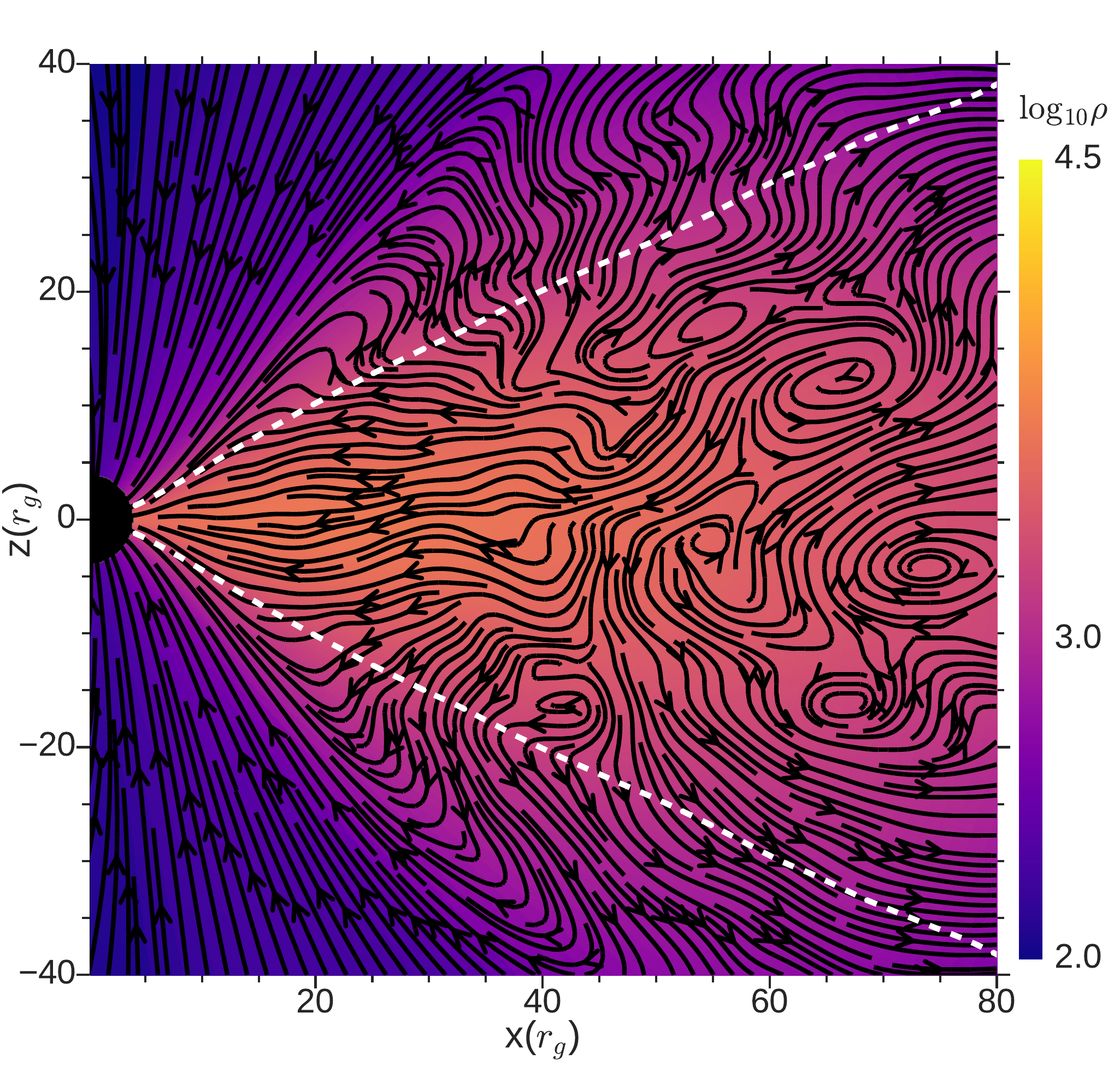}
    \caption{The mean density snapshot (azimuthally and time averaged from 370 to 530 orbits at ISCO) and the superimposed mean 
    poloidal velocity streamlines for the turbulent, steady RIAF.  The dashed white lines mark the average disk scale-height. Note the swirling
    eddies in the poloidal flow at large radii ($\gtrsim 40 r_g$) where averaging does not get rid of temporal variations. Note the mean inflow 
    within the disk mid-plane and the mean outflow across the disk surface.}
   \label{fig:mean_flow}
 \end{figure}

RIAFs are more difficult to describe as compared to the standard thin accretion disks. For the latter in steady state the mass accretion 
rate through the disk is 
constant as a function of radius,  and the vertically-averaged radial structure of the accretion disk is determined by the turbulent `viscosity'. 
The conservation of angular momentum implies that the inward advection of angular momentum at each radius is 
almost balanced by the turbulent flux of outward angular momentum, resulting in a net inward angular momentum flux equal to 
$\dot{M} l_{\rm ISCO}$ ($l_{\rm ISCO}$ is the specific angular momentum corresponding to the ISCO). The energy equation is also simple, with the
gravitational energy release, mediated by viscous dissipation, radiated locally as an optically thick black body.

For RIAFs, $H/R \sim 0.5$ and the radial and vertical dynamics are coupled. Thermal energy produced by accretion is not radiated but 
can be channeled into outflows, or can lead to convection that stifles accretion. Outflows can carry away angular momentum and assist 
accretion in the mid-plane. The standard thin disk accretion models are no longer applicable in this case.

Mean field approach provides a useful framework to describe the essentially 2-D RIAFs. Figure \ref{fig:mean_flow} shows the mean density 
and poloidal flow streamlines in steady state of our fiducial simulation. The steady state is attained  only within $40 r_g$. Once can clearly see
a mean inflow in the equatorial plane and a mean outflow across the disk boundary at radii $\gtrsim 20 r_g$. Looking at the mean flow in the 
poloidal plane provides a more useful way of understanding the flow structure than simply averaging over all $\theta$s (e.g., as in Fig. 7 
of \citealt{Stone2001}). Averaging over all angles cannot distinguish between outflows and circulating motions.

We compare the 1D-profiles obtained from our simulation with that from different 1D RIAF models proposed in the literature. Apart from the flows which are purely advection dominated with
most of the gravitational energy advected on to the black hole (ADAFs; \citealt{Narayan_Yi1994}), there are flows where outflows 
(ADIOS; \citealt{Blandford1999}) and convection (CDAF; \citealt{Narayan2000,Quataert2000}) allow only a small fraction of the available 
mass to be accreted.
All these idealized models consider unmagnetized flows. \cite{Akizuki2006} studied self-similarity in an ADAF with the pure toroidal magnetic fields and 
provide the following scalings,
\be
\label{eq:adaf_scaling}
v_r \sim  -\alpha r^{-0.5}, ~ v_{\phi} \sim r^{-0.5},~ c_{s} \sim r^{-0.5}, ~ v_{A} \sim \beta^{0.5} r^{-0.5};
\ee
where $\alpha$ is the viscosity parameter and $\beta$ is the plasma $\beta$.
The classic unmagnetized ADAF scalings (\citealt{Narayan_Yi1994}) are also very similar. 
On the other hand, the convection dominated accretion flow (CDAF) follows the scalings
\be
\label{eq:cdaf_scaling}
v_r \sim - r^{-1.5}, ~ v_{\phi} \sim r^{-0.5},~ c_{s} \sim r^{-0.5}.
\ee
If wind affects the accretion by carrying away mass, angular momentum and energy, scalings are provided by the advection-dominated inflow-outflow solution (ADIOS) with 
\be
\label{eq:adios_scaling}
v_r \sim - \alpha r^{-0.5}, ~ v_{\phi} \sim r^{-0.5},~ c_{s} \sim r^{-0.5}.
\ee

In Fig. \ref{fig:velocity_r} we show the radial structure of the RIAF from our M-2P run. 
Between $r=10$ and $r=r_{\rm eq}=40$, the characteristic velocities follow the following scalings,
\be
\label{eq:riaf_scaling}
v_r \sim  -r^{-2.5}, ~ v_{\phi} \sim r^{-0.7},~ c_{s} \sim r^{-0.6}, ~ v_{A} \sim r^{-0.9}.
\ee
If we use the measured spatial variation of $\alpha \sim r^{-0.8}$, $v_r /\alpha \sim r^{-1.7}$.
Therefore, the scalings we get for the 
characteristic velocities do not match with any of the standard RIAF models. However, from Figure \ref{fig:mean_flow} it is clear that
outflows play a role in reducing the mass accretion rate on to the black hole. 
It should be noted that a much larger radial extent of the numerical simulation is needed to compare the simulation results with analytic
scalings (\citealt{Narayan2012}). Due to a small dynamical range ($r_{\rm eq}=40$; $r \sim 10 r_g$ is affected by non-self-similarity due to the 
Paczynski-Wiita potential) in our simulations, we are unable to obtain a reliable power-law scaling for all the mean physical quantities.

 \subsection{Dynamo}
 \label{sect:discuss_dynamo}
We observe the saturation of magnetic energy due to a mean field dynamo in our 3D global simulation of a geometrically thick ($H/R \approx 0.5$) radiatively 
inefficient accretion flow. The mean field dynamo give rise to an intermittent sign reversal of the mean toroidal field, unlike previous local and 
global simulations with Keplerian angular velocity profile ($\Omega \propto r^{-q}, q=1.5$) which show a dynamo cycle with definite time 
period. Recently, \cite{Hogg2018} observed similar intermittency in the dynamo cycle of a thick disc with $H/R \approx 0.4$ using the same code PLUTO. Although, the set-ups are very similar,
the main difference is that they use {\em adhoc} cooling which keeps $H/R$ constant throughout the simulation run time. On the other hand, we use an adiabatic assumption with advection as the only source of cooling. As a result, in the quasi-steady state, disc scale-height $H$ has a very weak time dependence in our simulations (e.g. see Fig. \ref{fig:med_butterfly}). Because both the studies (\cite{Hogg2018} and the current work) give similar results, it implies that the adiabatic assumption does not really influence intermittency, which is a generic feature in the dynamo cycle of a geometrically thick accretion flow. 

Recently \cite{Nauman2015} and \cite{Gressel2015} found that the dynamo period becomes shorter 
and less well-defined (e. g. see the butterfly diagram for $q=1.8$ in Fig. 6 in  \cite{Nauman2015}) when the shear parameter $q$ is increased from its 
Keplerian value. Interestingly, in the quasi-steady state of our simulations the angular velocity is sub-Keplerian with 
$\Omega = v_{\phi} r \propto r^{-1.7}$ (see equation \ref{eq:riaf_scaling}). Thus sub-Keplerian 
velocity and intermittent dynamo cycle in our work agree with the observations in previous studies.

The axisymmetric fields produced by the dynamo have admixture of dipolar and quadrupolar symmetry, with the former dominating over the 
latter. Unlike previous global simulations of thin disks (\citealt{Arlt2001, Flock2012}), which found $\alpha^{\rm dyn}_{\phi \phi} \equiv 
\bar{\mathcal{E}}_{\phi}/\bar{B}_{\phi}$ to be positive in NH, we do not find any coherent pattern of $\alpha^{\rm dyn}_{\phi \phi}$. 
Instead we find a random meridional distribution of it due to the occurrence of a direct dynamo (close to the mid-plane), and 
$\alpha$-quenching (away from the mid-plane; \citealt{Blackman_Field2002}) that is due to the interplay between kinetic and 
current helicities. 

Small and random $\alpha^{\rm dyn}$ close to the mid-plane indicates the dominance of a direct dynamo in which the field grows 
due to random stretching of field lines by MRI driven turbulence. Another intriguing feature is the same sign  of 
$\alpha^{\rm dyn}_{\rm kin}$ and $\alpha^{\rm dyn}_{\rm mag}$ within one scale-height for both hemispheres. The same sign indicates 
reinforcing rather than quenching (as expected from Fig. \ref{fig:alpha_dyn}) within one scale-height. This is also observed in 
previous studies (\citealt{Brandenburg1995,Gressel2010}). \cite{Brandenburg1998} proposed the existence of a buoyancy driven 
dynamo and introduced another term ($\alpha^{\rm dyn}_{\rm buo}$) which has the opposite sign to both 
$\alpha^{\rm dyn}_{\rm kin}$ and $\alpha^{\rm dyn}_{\rm mag}$.

Our work probably is the first study which investigates the effects of dynamical quenching on magnetic field saturation in global 
accretion disk simulations. Interestingly, sign of  $\alpha^{\rm dyn}_{\rm mag}$ (negative in NH) is opposite to that found in 
local shearing-box simulations by \cite{Gressel2010} (compare their Fig. 3 with our Fig. \ref{fig:alpha_kin_mag}), 
but matches with that in \cite{Oishi2011}.   \cite{Oishi2011} attributed  the disagreement in the sign of $\alpha^{\rm dyn}_{\rm mag}$ 
to the vertical extent of the box and recovered the negative sign at larger height in the NH by extending the vertical domain size. 
On the other hand, although  the sign of $\alpha^{\rm dyn}_{\rm kin}$ matches with \cite{Gressel2010} within one scale-height, it 
differs at higher latitudes.  A comparison of $\alpha^{\rm dyn}$-s from different local and global studies is listed in Table \ref{tab:list_prev_results}.  

Another interesting difference is in the relative strength between the $\alpha$-s: both \cite{Gressel2010} and \cite{Oishi2011} found 
an order of magnitude stronger $\alpha^{\rm dyn}_{\rm mag}$ compared to  $\alpha^{\rm dyn}_{\rm kin}$; we find $\alpha$-s of 
{\em almost equal strength}. This may be the  reason behind the patchy pattern we obtain for 
$\alpha^{\rm dyn}_{\phi \phi}= \bar{\mathcal{E}}_{\phi}/\bar{B}_{\phi}$. In addition, the presence of the mean flow (see Fig. \ref{fig:mean_flow}) in our global simulation can have significant effects on the dynamo mechanism (\citealt{Choudhuri1995}). Anisotropic nature of turbulence may also be effective 
in our simulations so that the off-diagonal terms in $\alpha$ and $\lambda$ (see equation \ref{eq:closure}) tensors are 
non-negligible. In future we would like to calculate the coefficients of $\alpha$ and $\lambda$ tensors rigorously.
 
 \section{Summary}
 \label{sect:summary}
 In this work we perform 3D ideal MHD simulations of accretion tori with different grid resolutions and azimuthal extents. 
 Our aim is to investigate - i) convergence, ii) effects of azimuthal extent of the simulation domain on accretion flow properties, 
 and iii) saturation mechanism of magnetic energy that is governed by the dynamo process. The key findings of the study are listed below in the order of importance.  
 \begin{itemize}
  \item We see an intermittent dynamo cycle for our geometrically thick ($H/R \sim 0.5$) radiatively inefficient accretion flows (Fig. \ref{fig:med_butterfly}). By looking at the symmetry of mean fields in the northern and southern hemispheres, we find that mean field parity is an admixture of dipole and quadrupole, with the former dominating over the latter (Fig. \ref{fig:med_B_mean_nh_sh} and section \ref{sect:mean_symmetry}). The irregularity found in the spatio-temporal evolution of mean fields in our global simulations (for which shear parameter $q=1.7$) is similar to that found in the butterfly diagrams in previous local studies (\citealt{Nauman2015,Gressel2015}) for $q=1.8$. 
  
  We also find an irregular behaviour of the dynamo-$\alpha$ (Fig. \ref{fig:alpha_dyn}) unlike previous global simulations in \cite{Flock2012}. This is because of  two reasons -- i) dominance of direct dynamo close to the mid-plane, ii) suppression of kinetic $\alpha$ by the magnetic $\alpha$ of a similar magnitude away from the mid-plane (section \ref{sect:discuss_dynamo}). On the contrary, previous local studies found $\alpha_{\rm mag} \sim 10 \alpha_{\rm kin}$ (\citealt{Gressel2010, Oishi2011}). Probably, the effects of $\alpha$-quenching are studied explicitly for the first time in global simulations of accretion flows.
  
   \item We get stronger mean magnetic fields for the runs with smaller azimuthal domains (similar to \cite{Flock2012}) due to the periodic boundary condition used in the azimuthal direction, and  the tendency of magnetic fields to be at the largest scales (Fig. \ref{fig:med_psd}). However, turbulent magnetic fields do not show any trend with  the azimuthal domain size unlike \cite{Flock2012}. For the azimuthal extent $\Phi_0 \geq \pi/2$, the runs with restricted azimuthal domains show stronger accretion stresses compared to the run with the natural domain size $\Phi_0=2 \pi$. For all the runs the stress due to turbulent fields dominates over that due to mean fields (section \ref{sect:domain_size}).
   
   We conclude that the appropriate azimuthal domain size depends on the aim of the study. If one wants to study the structure of the flow, the runs with the azimuthal domain size $\Phi_0 \geq \pi/2$ are able to  produce the results obtained by run with $\Phi_0=2 \pi$. On the other hand, for the study of turbulence and dynamo in a RIAF with $H/R \sim 0.5$, $\Phi_0=2 \pi$ is the necessary azimuthal domain size compared to the minimum requirement $\Phi_0=\pi$ proposed by \cite{Flock2012} for the thin accretion discs with $H/R \sim 0.07$. Essentially the appropriate domain size depends on the $H/R$ ratio of the disc as the number of disc scale-height can be accommodated in the azimuthal direction is different for different disc thickness.
   
   \item Decomposing the flow into a mean and fluctuations provides a useful insight at understanding the structure of RIAFs. Although we have a small radial extent 
   for the steady-state flow ($\lesssim 40$), Fig. \ref{fig:mean_flow} shows the presence of vertical outflows that can carry away substantial mass 
   (energy and angular momentum) away from the disk. This may explain the key feature of RIAFs -- the much smaller accretion rate on to the black hole compared to
   the available mass supply (see section \ref{sect:RIAFs}).
   
  \item We attain convergence for with  $42-64$ cells per scale-height
  in the vertical direction. We use the number of cells in the radial and azimuthal direction such that the three 
  dimensional structure of the cells is close to cubes. Exceeding the minimum requirements for 
  convergence, the quality factors are $\la \la Q_{\theta}\ra \ra=10.5 \pm 0.9$ and $\la \la Q_{\phi} \ra \ra = 33 \pm 3$. 
  The magnetic tilt angle $\theta_B$ turns out to be an excellent indicator of convergence with $\la \la \theta_B \ra\ra = 13^{\circ}- 14^{\circ}$.

 \end{itemize}

\section*{Acknowledgments}
PD thanks Bhupendra Mishra, Mario Flock and Xue-Ning Bai for the discussions regarding the simulation set-up.  PD thanks Gopal Hazra for numerous discussions on dynamo. We thank Arnab Rai Choudhuri, Banibrata Mukhopadhyay, Kandaswamy Subramanian and Ramesh Narayan for  discussions. We thank Oliver Gressel, Piyali  Chatterjee and Sharanya Sur for useful suggestions. We thank KITP for our participation in the program ``Confronting MHD Theories of Accretion Disks with Observations''. This research was supported in part by the National Science Foundation
under Grant No. NSF PHY-1125915 and by an India-Israel joint research grant (6-10/2014[IC]).  This research was supported in part by the International Centre for Theoretical Sciences (ICTS) during a visit for participating in the program - Turbulence from Angstroms to light years (Code: ICTS/Prog-taly2018/01). All the simulations were carried out on Cray XC40-SahasraT cluster at Supercomputing Education and Research Centre (SERC), IISc. Some of the visualizations are done using {\tt VisIt} (\citealt{HPV:VisIt}).

\section*{Additional links}
The movies of the evolution of torus are available in the following links \\
i) \url{https://www.youtube.com/watch?v=zDT5PqxgWMI}, \\
ii) \url{https://www.youtube.com/watch?v=kTc9yYQDDOA}.
\bibliographystyle{mnras}
\bibliography{bibtex_thesis}

\begin{thebibliography}{}
\makeatletter
\relax
\def\mn@urlcharsother{\let\do\@makeother \do\$\do\&\do\#\do\^\do\_\do\%\do\~}
\def\mn@doi{\begingroup\mn@urlcharsother \@ifnextchar [ {\mn@doi@}
  {\mn@doi@[]}}
\def\mn@doi@[#1]#2{\def\@tempa{#1}\ifx\@tempa\@empty \href
  {http://dx.doi.org/#2} {doi:#2}\else \href {http://dx.doi.org/#2} {#1}\fi
  \endgroup}
\def\mn@eprint#1#2{\mn@eprint@#1:#2::\@nil}
\def\mn@eprint@arXiv#1{\href {http://arxiv.org/abs/#1} {{\tt arXiv:#1}}}
\def\mn@eprint@dblp#1{\href {http://dblp.uni-trier.de/rec/bibtex/#1.xml}
  {dblp:#1}}
\def\mn@eprint@#1:#2:#3:#4\@nil{\def\@tempa {#1}\def\@tempb {#2}\def\@tempc
  {#3}\ifx \@tempc \@empty \let \@tempc \@tempb \let \@tempb \@tempa \fi \ifx
  \@tempb \@empty \def\@tempb {arXiv}\fi \@ifundefined
  {mn@eprint@\@tempb}{\@tempb:\@tempc}{\expandafter \expandafter \csname
  mn@eprint@\@tempb\endcsname \expandafter{\@tempc}}}

\bibitem[\protect\citeauthoryear{{Abramowicz}, {Czerny}, {Lasota}  \&
  {Szuszkiewicz}}{{Abramowicz} et~al.}{1988}]{Abramowicz1988}
{Abramowicz} M.~A.,  {Czerny} B.,  {Lasota} J.~P.,   {Szuszkiewicz} E.,  1988,
  \mn@doi [\apj] {10.1086/166683}, \href
  {http://adsabs.harvard.edu/abs/1988ApJ...332..646A} {332, 646}

\bibitem[\protect\citeauthoryear{{Abramowicz}, {Chen}, {Kato}, {Lasota}  \&
  {Regev}}{{Abramowicz} et~al.}{1995}]{Abramowicz1995}
{Abramowicz} M.~A.,  {Chen} X.,  {Kato} S.,  {Lasota} J.-P.,   {Regev} O.,
  1995, \mn@doi [\apjl] {10.1086/187709}, \href
  {http://adsabs.harvard.edu/abs/1995ApJ...438L..37A} {438, L37}

\bibitem[\protect\citeauthoryear{{Akizuki} \& {Fukue}}{{Akizuki} \&
  {Fukue}}{2006}]{Akizuki2006}
{Akizuki} C.,  {Fukue} J.,  2006, \mn@doi [\pasj] {10.1093/pasj/58.2.469},
  \href {http://adsabs.harvard.edu/abs/2006PASJ...58..469A} {58, 469}

\bibitem[\protect\citeauthoryear{{Arlt} \& {R{\"u}diger}}{{Arlt} \&
  {R{\"u}diger}}{2001}]{Arlt2001}
{Arlt} R.,  {R{\"u}diger} G.,  2001, \mn@doi [\aap]
  {10.1051/0004-6361:20010797}, \href
  {http://adsabs.harvard.edu/abs/2001A%26A...374.1035A} {374, 1035}

\bibitem[\protect\citeauthoryear{{Balbus} \& {Hawley}}{{Balbus} \&
  {Hawley}}{1991}]{Balbus_hawley1991}
{Balbus} S.~A.,  {Hawley} J.~F.,  1991, \mn@doi [\apj] {10.1086/170270}, \href
  {http://adsabs.harvard.edu/abs/1991ApJ...376..214B} {376, 214}

\bibitem[\protect\citeauthoryear{{Balbus} \& {Hawley}}{{Balbus} \&
  {Hawley}}{1998}]{Balbus_hawley1998}
{Balbus} S.~A.,  {Hawley} J.~F.,  1998, \mn@doi [Reviews of Modern Physics]
  {10.1103/RevModPhys.70.1}, \href
  {http://adsabs.harvard.edu/abs/1998RvMP...70....1B} {70, 1}

\bibitem[\protect\citeauthoryear{{Beckwith}, {Armitage}  \& {Simon}}{{Beckwith}
  et~al.}{2011}]{Beckwith2011}
{Beckwith} K.,  {Armitage} P.~J.,   {Simon} J.~B.,  2011, \mn@doi [\mnras]
  {10.1111/j.1365-2966.2011.19043.x}, \href
  {http://adsabs.harvard.edu/abs/2011MNRAS.416..361B} {416, 361}

\bibitem[\protect\citeauthoryear{{Blackman} \& {Field}}{{Blackman} \&
  {Field}}{2002}]{Blackman_Field2002}
{Blackman} E.~G.,  {Field} G.~B.,  2002, \mn@doi [Physical Review Letters]
  {10.1103/PhysRevLett.89.265007}, \href
  {http://adsabs.harvard.edu/abs/2002PhRvL..89z5007B} {89, 265007}

\bibitem[\protect\citeauthoryear{{Blackman} \& {Tan}}{{Blackman} \&
  {Tan}}{2004}]{Blackman_tan2004}
{Blackman} E.~G.,  {Tan} J.~C.,  2004, \mn@doi [\apss]
  {10.1023/B:ASTR.0000045043.87692.4a}, \href
  {http://adsabs.harvard.edu/abs/2004Ap%26SS.292..395B} {292, 395}

\bibitem[\protect\citeauthoryear{{Blackman}, {Penna}  \&
  {Varni{\`e}re}}{{Blackman} et~al.}{2008}]{Blackman2008}
{Blackman} E.~G.,  {Penna} R.~F.,   {Varni{\`e}re} P.,  2008, \mn@doi [\na]
  {10.1016/j.newast.2007.10.004}, \href
  {http://adsabs.harvard.edu/abs/2008NewA...13..244B} {13, 244}

\bibitem[\protect\citeauthoryear{{Blandford} \& {Begelman}}{{Blandford} \&
  {Begelman}}{1999}]{Blandford1999}
{Blandford} R.~D.,  {Begelman} M.~C.,  1999, \mn@doi [MNRAS]
  {10.1046/j.1365-8711.1999.02358.x}, \href
  {http://adsabs.harvard.edu/abs/1999MNRAS.303L...1B} {303, L1}

\bibitem[\protect\citeauthoryear{{Blandford} \& {Payne}}{{Blandford} \&
  {Payne}}{1982}]{Blandford_Payne1982}
{Blandford} R.~D.,  {Payne} D.~G.,  1982, \mn@doi [\mnras]
  {10.1093/mnras/199.4.883}, \href
  {http://adsabs.harvard.edu/abs/1982MNRAS.199..883B} {199, 883}

\bibitem[\protect\citeauthoryear{{Blandford} \& {Znajek}}{{Blandford} \&
  {Znajek}}{1977}]{Blandford_Znajek1977}
{Blandford} R.~D.,  {Znajek} R.~L.,  1977, \mn@doi [\mnras]
  {10.1093/mnras/179.3.433}, \href
  {http://adsabs.harvard.edu/abs/1977MNRAS.179..433B} {179, 433}

\bibitem[\protect\citeauthoryear{{Bodo}, {Cattaneo}, {Ferrari}, {Mignone}  \&
  {Rossi}}{{Bodo} et~al.}{2011}]{Bodo2011}
{Bodo} G.,  {Cattaneo} F.,  {Ferrari} A.,  {Mignone} A.,   {Rossi} P.,  2011,
  \mn@doi [\apj] {10.1088/0004-637X/739/2/82}, \href
  {http://adsabs.harvard.edu/abs/2011ApJ...739...82B} {739, 82}

\bibitem[\protect\citeauthoryear{{Bodo}, {Cattaneo}, {Mignone}  \&
  {Rossi}}{{Bodo} et~al.}{2012}]{Bodo2012}
{Bodo} G.,  {Cattaneo} F.,  {Mignone} A.,   {Rossi} P.,  2012, \mn@doi [\apj]
  {10.1088/0004-637X/761/2/116}, \href
  {http://adsabs.harvard.edu/abs/2012ApJ...761..116B} {761, 116}

\bibitem[\protect\citeauthoryear{{Bodo}, {Cattaneo}, {Mignone}  \&
  {Rossi}}{{Bodo} et~al.}{2014}]{Bodo2014}
{Bodo} G.,  {Cattaneo} F.,  {Mignone} A.,   {Rossi} P.,  2014, \mn@doi [\apjl]
  {10.1088/2041-8205/787/1/L13}, \href
  {http://adsabs.harvard.edu/abs/2014ApJ...787L..13B} {787, L13}

\bibitem[\protect\citeauthoryear{{Brandenburg} \& {Donner}}{{Brandenburg} \&
  {Donner}}{1997}]{Brandenburg1997}
{Brandenburg} A.,  {Donner} K.~J.,  1997, \mn@doi [\mnras]
  {10.1093/mnras/288.2.L29}, \href
  {http://adsabs.harvard.edu/abs/1997MNRAS.288L..29B} {288, L29}

\bibitem[\protect\citeauthoryear{{Brandenburg} \& {Schmitt}}{{Brandenburg} \&
  {Schmitt}}{1998}]{Brandenburg1998}
{Brandenburg} A.,  {Schmitt} D.,  1998, \aap, \href
  {http://adsabs.harvard.edu/abs/1998A%26A...338L..55B} {338, L55}

\bibitem[\protect\citeauthoryear{{Brandenburg} \& {Subramanian}}{{Brandenburg}
  \& {Subramanian}}{2005}]{Brandenburg2005}
{Brandenburg} A.,  {Subramanian} K.,  2005, \mn@doi [\physrep]
  {10.1016/j.physrep.2005.06.005}, \href
  {http://adsabs.harvard.edu/abs/2005PhR...417....1B} {417, 1}

\bibitem[\protect\citeauthoryear{{Brandenburg}, {Nordlund}, {Stein}  \&
  {Torkelsson}}{{Brandenburg} et~al.}{1995}]{Brandenburg1995}
{Brandenburg} A.,  {Nordlund} A.,  {Stein} R.~F.,   {Torkelsson} U.,  1995,
  \mn@doi [\apj] {10.1086/175831}, \href
  {http://adsabs.harvard.edu/abs/1995ApJ...446..741B} {446, 741}

\bibitem[\protect\citeauthoryear{Chandrasekhar}{Chandrasekhar}{1960}]{Chandrasekhar1960}
Chandrasekhar S.,  1960, Proceedings of the National Academy of Sciences, 46,
  253

\bibitem[\protect\citeauthoryear{Childs et~al.,}{Childs
  et~al.}{2012}]{HPV:VisIt}
Childs H.,  et~al., 2012, in , {High Performance Visualization--Enabling
  Extreme-Scale Scientific Insight}.
pp 357--372

\bibitem[\protect\citeauthoryear{{Choudhuri}}{{Choudhuri}}{1998}]{ARC1998}
{Choudhuri} A.~R.,  1998, {The physics of fluids and plasmas : an introduction
  for astrophysicists /}

\bibitem[\protect\citeauthoryear{{Choudhuri}, {Schussler}  \&
  {Dikpati}}{{Choudhuri} et~al.}{1995}]{Choudhuri1995}
{Choudhuri} A.~R.,  {Schussler} M.,   {Dikpati} M.,  1995, \aap, \href
  {http://adsabs.harvard.edu/abs/1995A%26A...303L..29C} {303, L29}

\bibitem[\protect\citeauthoryear{{Das} \& {Sharma}}{{Das} \&
  {Sharma}}{2013}]{Das2013}
{Das} U.,  {Sharma} P.,  2013, \mn@doi [MNRAS] {10.1093/mnras/stt1452}, \href
  {http://adsabs.harvard.edu/abs/2013MNRAS.435.2431D} {435, 2431}

\bibitem[\protect\citeauthoryear{{Davis}, {Stone}  \& {Pessah}}{{Davis}
  et~al.}{2010}]{Davis2010}
{Davis} S.~W.,  {Stone} J.~M.,   {Pessah} M.~E.,  2010, \mn@doi [\apj]
  {10.1088/0004-637X/713/1/52}, \href
  {http://adsabs.harvard.edu/abs/2010ApJ...713...52D} {713, 52}

\bibitem[\protect\citeauthoryear{{Evans} \& {Hawley}}{{Evans} \&
  {Hawley}}{1988}]{Evans1988}
{Evans} C.~R.,  {Hawley} J.~F.,  1988, \mn@doi [\apj] {10.1086/166684}, \href
  {http://adsabs.harvard.edu/abs/1988ApJ...332..659E} {332, 659}

\bibitem[\protect\citeauthoryear{{Flock}, {Dzyurkevich}, {Klahr}  \&
  {Mignone}}{{Flock} et~al.}{2010}]{Flock2010}
{Flock} M.,  {Dzyurkevich} N.,  {Klahr} H.,   {Mignone} A.,  2010, \mn@doi
  [\aap] {10.1051/0004-6361/200912443}, \href
  {http://adsabs.harvard.edu/abs/2010A%26A...516A..26F} {516, A26}

\bibitem[\protect\citeauthoryear{{Flock}, {Dzyurkevich}, {Klahr}, {Turner}  \&
  {Henning}}{{Flock} et~al.}{2011}]{Flock2011}
{Flock} M.,  {Dzyurkevich} N.,  {Klahr} H.,  {Turner} N.~J.,   {Henning} T.,
  2011, \mn@doi [\apj] {10.1088/0004-637X/735/2/122}, \href
  {http://adsabs.harvard.edu/abs/2011ApJ...735..122F} {735, 122}

\bibitem[\protect\citeauthoryear{{Flock}, {Dzyurkevich}, {Klahr}, {Turner}  \&
  {Henning}}{{Flock} et~al.}{2012}]{Flock2012}
{Flock} M.,  {Dzyurkevich} N.,  {Klahr} H.,  {Turner} N.,   {Henning} T.,
  2012, \mn@doi [\apj] {10.1088/0004-637X/744/2/144}, \href
  {http://adsabs.harvard.edu/abs/2012ApJ...744..144F} {744, 144}

\bibitem[\protect\citeauthoryear{{Fromang} \& {Nelson}}{{Fromang} \&
  {Nelson}}{2006}]{Fromang_nelson2006}
{Fromang} S.,  {Nelson} R.~P.,  2006, \mn@doi [\aap]
  {10.1051/0004-6361:20065643}, \href
  {http://adsabs.harvard.edu/abs/2006A%26A...457..343F} {457, 343}

\bibitem[\protect\citeauthoryear{{Fromang} \& {Papaloizou}}{{Fromang} \&
  {Papaloizou}}{2007}]{Fromang2007_I}
{Fromang} S.,  {Papaloizou} J.,  2007, \mn@doi [\aap]
  {10.1051/0004-6361:20077942}, \href
  {http://adsabs.harvard.edu/abs/2007A%26A...476.1113F} {476, 1113}

\bibitem[\protect\citeauthoryear{{Gardiner} \& {Stone}}{{Gardiner} \&
  {Stone}}{2005}]{Gardiner_stone2005}
{Gardiner} T.~A.,  {Stone} J.~M.,  2005, \mn@doi [Journal of Computational
  Physics] {10.1016/j.jcp.2004.11.016}, \href
  {http://adsabs.harvard.edu/abs/2005JCoPh.205..509G} {205, 509}

\bibitem[\protect\citeauthoryear{{Goodman} \& {Xu}}{{Goodman} \&
  {Xu}}{1994}]{Goodman1994}
{Goodman} J.,  {Xu} G.,  1994, \mn@doi [\apj] {10.1086/174562}, \href
  {http://adsabs.harvard.edu/abs/1994ApJ...432..213G} {432, 213}

\bibitem[\protect\citeauthoryear{{Gressel}}{{Gressel}}{2010}]{Gressel2010}
{Gressel} O.,  2010, \mn@doi [\mnras] {10.1111/j.1365-2966.2010.16440.x}, \href
  {http://adsabs.harvard.edu/abs/2010MNRAS.405...41G} {405, 41}

\bibitem[\protect\citeauthoryear{{Gressel} \& {Pessah}}{{Gressel} \&
  {Pessah}}{2015}]{Gressel2015}
{Gressel} O.,  {Pessah} M.~E.,  2015, \mn@doi [\apj]
  {10.1088/0004-637X/810/1/59}, \href
  {http://adsabs.harvard.edu/abs/2015ApJ...810...59G} {810, 59}

\bibitem[\protect\citeauthoryear{{Guan}, {Gammie}, {Simon}  \&
  {Johnson}}{{Guan} et~al.}{2009}]{Guan2009}
{Guan} X.,  {Gammie} C.~F.,  {Simon} J.~B.,   {Johnson} B.~M.,  2009, \mn@doi
  [\apj] {10.1088/0004-637X/694/2/1010}, \href
  {http://adsabs.harvard.edu/abs/2009ApJ...694.1010G} {694, 1010}

\bibitem[\protect\citeauthoryear{{Guilet} \& {Ogilvie}}{{Guilet} \&
  {Ogilvie}}{2012}]{Guilet_Ogilvie2012}
{Guilet} J.,  {Ogilvie} G.~I.,  2012, \mn@doi [\mnras]
  {10.1111/j.1365-2966.2012.21361.x}, \href
  {http://adsabs.harvard.edu/abs/2012MNRAS.424.2097G} {424, 2097}

\bibitem[\protect\citeauthoryear{{Hawley}}{{Hawley}}{2000}]{Hawley2000}
{Hawley} J.~F.,  2000, \mn@doi [\apj] {10.1086/308180}, \href
  {http://adsabs.harvard.edu/abs/2000ApJ...528..462H} {528, 462}

\bibitem[\protect\citeauthoryear{{Hawley}}{{Hawley}}{2001}]{Hawley2001}
{Hawley} J.~F.,  2001, \mn@doi [\apj] {10.1086/321348}, \href
  {http://adsabs.harvard.edu/abs/2001ApJ...554..534H} {554, 534}

\bibitem[\protect\citeauthoryear{{Hawley}, {Gammie}  \& {Balbus}}{{Hawley}
  et~al.}{1995}]{Hawley1995}
{Hawley} J.~F.,  {Gammie} C.~F.,   {Balbus} S.~A.,  1995, \mn@doi [\apj]
  {10.1086/175311}, \href {http://adsabs.harvard.edu/abs/1995ApJ...440..742H}
  {440, 742}

\bibitem[\protect\citeauthoryear{{Hawley}, {Gammie}  \& {Balbus}}{{Hawley}
  et~al.}{1996}]{Hawley1996}
{Hawley} J.~F.,  {Gammie} C.~F.,   {Balbus} S.~A.,  1996, \mn@doi [\apj]
  {10.1086/177356}, \href {http://adsabs.harvard.edu/abs/1996ApJ...464..690H}
  {464, 690}

\bibitem[\protect\citeauthoryear{{Hawley}, {Guan}  \& {Krolik}}{{Hawley}
  et~al.}{2011}]{Hawley2011}
{Hawley} J.~F.,  {Guan} X.,   {Krolik} J.~H.,  2011, \mn@doi [\apj]
  {10.1088/0004-637X/738/1/84}, \href
  {http://adsabs.harvard.edu/abs/2011ApJ...738...84H} {738, 84}

\bibitem[\protect\citeauthoryear{{Hawley}, {Richers}, {Guan}  \&
  {Krolik}}{{Hawley} et~al.}{2013}]{Hawley2013}
{Hawley} J.~F.,  {Richers} S.~A.,  {Guan} X.,   {Krolik} J.~H.,  2013, \mn@doi
  [\apj] {10.1088/0004-637X/772/2/102}, \href
  {http://adsabs.harvard.edu/abs/2013ApJ...772..102H} {772, 102}

\bibitem[\protect\citeauthoryear{{Hogg} \& {Reynolds}}{{Hogg} \&
  {Reynolds}}{2016}]{Hogg2016}
{Hogg} J.~D.,  {Reynolds} C.~S.,  2016, \mn@doi [\apj]
  {10.3847/0004-637X/826/1/40}, \href
  {http://adsabs.harvard.edu/abs/2016ApJ...826...40H} {826, 40}

\bibitem[\protect\citeauthoryear{{Hogg} \& {Reynolds}}{{Hogg} \&
  {Reynolds}}{2018}]{Hogg2018}
{Hogg} J.~D.,  {Reynolds} C.~S.,  2018, \mn@doi [\apj]
  {10.3847/1538-4357/aac439}, \href
  {http://adsabs.harvard.edu/abs/2018ApJ...861...24H} {861, 24}

\bibitem[\protect\citeauthoryear{{Ichimaru}}{{Ichimaru}}{1977}]{Ichimaru1977}
{Ichimaru} S.,  1977, \mn@doi [\apj] {10.1086/155314}, \href
  {http://adsabs.harvard.edu/abs/1977ApJ...214..840I} {214, 840}

\bibitem[\protect\citeauthoryear{{Johansen}, {Youdin}  \& {Klahr}}{{Johansen}
  et~al.}{2009}]{Johansen2009}
{Johansen} A.,  {Youdin} A.,   {Klahr} H.,  2009, \mn@doi [\apj]
  {10.1088/0004-637X/697/2/1269}, \href
  {http://adsabs.harvard.edu/abs/2009ApJ...697.1269J} {697, 1269}

\bibitem[\protect\citeauthoryear{{Kazantsev}}{{Kazantsev}}{1968}]{Kazantsev1968}
{Kazantsev} A.~P.,  1968, Soviet Journal of Experimental and Theoretical
  Physics, \href {http://adsabs.harvard.edu/abs/1968JETP...26.1031K} {26, 1031}

\bibitem[\protect\citeauthoryear{{Koratkar} \& {Blaes}}{{Koratkar} \&
  {Blaes}}{1999}]{Koratkar1999}
{Koratkar} A.,  {Blaes} O.,  1999, \mn@doi [\pasp] {10.1086/316294}, \href
  {http://adsabs.harvard.edu/abs/1999PASP..111....1K} {111, 1}

\bibitem[\protect\citeauthoryear{{Lesur} \& {Ogilvie}}{{Lesur} \&
  {Ogilvie}}{2008}]{Lesur2008}
{Lesur} G.,  {Ogilvie} G.~I.,  2008, \mn@doi [\aap]
  {10.1051/0004-6361:200810152}, \href
  {http://adsabs.harvard.edu/abs/2008A%26A...488..451L} {488, 451}

\bibitem[\protect\citeauthoryear{{Lovelace}, {Rothstein}  \&
  {Bisnovatyi-Kogan}}{{Lovelace} et~al.}{2009}]{Lovelace2009}
{Lovelace} R.~V.~E.,  {Rothstein} D.~M.,   {Bisnovatyi-Kogan} G.~S.,  2009,
  \mn@doi [\apj] {10.1088/0004-637X/701/2/885}, \href
  {http://adsabs.harvard.edu/abs/2009ApJ...701..885L} {701, 885}

\bibitem[\protect\citeauthoryear{{Lubow}, {Papaloizou}  \& {Pringle}}{{Lubow}
  et~al.}{1994}]{Lubow1994}
{Lubow} S.~H.,  {Papaloizou} J.~C.~B.,   {Pringle} J.~E.,  1994, \mn@doi
  [\mnras] {10.1093/mnras/267.2.235}, \href
  {http://adsabs.harvard.edu/abs/1994MNRAS.267..235L} {267, 235}

\bibitem[\protect\citeauthoryear{{McKinney} \& {Gammie}}{{McKinney} \&
  {Gammie}}{2002}]{McKinney2002}
{McKinney} J.~C.,  {Gammie} C.~F.,  2002, \mn@doi [\apj] {10.1086/340761},
  \href {http://adsabs.harvard.edu/abs/2002ApJ...573..728M} {573, 728}

\bibitem[\protect\citeauthoryear{{Mignone}, {Bodo}, {Massaglia}, {Matsakos},
  {Tesileanu}, {Zanni}  \& {Ferrari}}{{Mignone} et~al.}{2007}]{Mignone2007}
{Mignone} A.,  {Bodo} G.,  {Massaglia} S.,  {Matsakos} T.,  {Tesileanu} O.,
  {Zanni} C.,   {Ferrari} A.,  2007, \mn@doi [ApJS] {10.1086/513316}, \href
  {http://adsabs.harvard.edu/abs/2007ApJS..170..228M} {170, 228}

\bibitem[\protect\citeauthoryear{{Miller} \& {Stone}}{{Miller} \&
  {Stone}}{2000}]{Miller2000}
{Miller} K.~A.,  {Stone} J.~M.,  2000, \mn@doi [\apj] {10.1086/308736}, \href
  {http://adsabs.harvard.edu/abs/2000ApJ...534..398M} {534, 398}

\bibitem[\protect\citeauthoryear{{Miyoshi} \& {Kusano}}{{Miyoshi} \&
  {Kusano}}{2005}]{Miyoshi2005}
{Miyoshi} T.,  {Kusano} K.,  2005, \mn@doi [Journal of Computational Physics]
  {10.1016/j.jcp.2005.02.017}, \href
  {http://adsabs.harvard.edu/abs/2005JCoPh.208..315M} {208, 315}

\bibitem[\protect\citeauthoryear{{Narayan} \& {Yi}}{{Narayan} \&
  {Yi}}{1994}]{Narayan_Yi1994}
{Narayan} R.,  {Yi} I.,  1994, \mn@doi [\apjl] {10.1086/187381}, \href
  {http://adsabs.harvard.edu/abs/1994ApJ...428L..13N} {428, L13}

\bibitem[\protect\citeauthoryear{{Narayan} \& {Yi}}{{Narayan} \&
  {Yi}}{1995}]{Narayan_Yi1995}
{Narayan} R.,  {Yi} I.,  1995, \mn@doi [\apj] {10.1086/176343}, \href
  {http://adsabs.harvard.edu/abs/1995ApJ...452..710N} {452, 710}

\bibitem[\protect\citeauthoryear{{Narayan}, {Igumenshchev}  \&
  {Abramowicz}}{{Narayan} et~al.}{2000}]{Narayan2000}
{Narayan} R.,  {Igumenshchev} I.~V.,   {Abramowicz} M.~A.,  2000, \mn@doi
  [\apj] {10.1086/309268}, \href
  {http://adsabs.harvard.edu/abs/2000ApJ...539..798N} {539, 798}

\bibitem[\protect\citeauthoryear{{Narayan}, {S{\"A} dowski}, {Penna}  \&
  {Kulkarni}}{{Narayan} et~al.}{2012}]{Narayan2012}
{Narayan} R.,  {S{\"A} dowski} A.,  {Penna} R.~F.,   {Kulkarni} A.~K.,  2012,
  \mn@doi [\mnras] {10.1111/j.1365-2966.2012.22002.x}, \href
  {http://adsabs.harvard.edu/abs/2012MNRAS.426.3241N} {426, 3241}

\bibitem[\protect\citeauthoryear{{Nauman} \& {Blackman}}{{Nauman} \&
  {Blackman}}{2015}]{Nauman2015}
{Nauman} F.,  {Blackman} E.~G.,  2015, \mn@doi [\mnras]
  {10.1093/mnras/stu2226}, \href
  {http://adsabs.harvard.edu/abs/2015MNRAS.446.2102N} {446, 2102}

\bibitem[\protect\citeauthoryear{{Noble}, {Krolik}  \& {Hawley}}{{Noble}
  et~al.}{2010}]{Noble2010}
{Noble} S.~C.,  {Krolik} J.~H.,   {Hawley} J.~F.,  2010, \mn@doi [\apj]
  {10.1088/0004-637X/711/2/959}, \href
  {http://adsabs.harvard.edu/abs/2010ApJ...711..959N} {711, 959}

\bibitem[\protect\citeauthoryear{{Novikov} \& {Thorne}}{{Novikov} \&
  {Thorne}}{1973}]{Novikov_thorne1973}
{Novikov} I.~D.,  {Thorne} K.~S.,  1973, in {Dewitt} C.,  {Dewitt} B.~S.,  eds,
  Black Holes (Les Astres Occlus). pp 343--450

\bibitem[\protect\citeauthoryear{{O'Neill}, {Reynolds}, {Miller}  \&
  {Sorathia}}{{O'Neill} et~al.}{2011}]{Oneill2011}
{O'Neill} S.~M.,  {Reynolds} C.~S.,  {Miller} M.~C.,   {Sorathia} K.~A.,  2011,
  \mn@doi [\apj] {10.1088/0004-637X/736/2/107}, \href
  {http://adsabs.harvard.edu/abs/2011ApJ...736..107O} {736, 107}

\bibitem[\protect\citeauthoryear{Oishi \& Mac~Low}{Oishi \&
  Mac~Low}{2011}]{Oishi2011}
Oishi J.~S.,  Mac~Low M.-M.,  2011, The Astrophysical Journal, 740, 18

\bibitem[\protect\citeauthoryear{{Paczy{\'n}sky} \& {Wiita}}{{Paczy{\'n}sky} \&
  {Wiita}}{1980}]{Paczynsky_wiita1980}
{Paczy{\'n}sky} B.,  {Wiita} P.~J.,  1980, \aap, \href
  {http://adsabs.harvard.edu/abs/1980A%26A....88...23P} {88, 23}

\bibitem[\protect\citeauthoryear{{Papaloizou} \& {Pringle}}{{Papaloizou} \&
  {Pringle}}{1984}]{Papaloizou_pringle1984}
{Papaloizou} J.~C.~B.,  {Pringle} J.~E.,  1984, \mn@doi [\mnras]
  {10.1093/mnras/208.4.721}, \href
  {http://adsabs.harvard.edu/abs/1984MNRAS.208..721P} {208, 721}

\bibitem[\protect\citeauthoryear{{Parker}}{{Parker}}{1979}]{Parker1979}
{Parker} E.~N.,  1979, {Cosmical magnetic fields: Their origin and their
  activity}

\bibitem[\protect\citeauthoryear{{Parkin}}{{Parkin}}{2014}]{Parkin2014}
{Parkin} E.~R.,  2014, \mn@doi [\mnras] {10.1093/mnras/stt2379}, \href
  {http://adsabs.harvard.edu/abs/2014MNRAS.438.2513P} {438, 2513}

\bibitem[\protect\citeauthoryear{{Parkin} \& {Bicknell}}{{Parkin} \&
  {Bicknell}}{2013}]{Parkin2013}
{Parkin} E.~R.,  {Bicknell} G.~V.,  2013, \mn@doi [\mnras]
  {10.1093/mnras/stt1450}, \href
  {http://adsabs.harvard.edu/abs/2013MNRAS.435.2281P} {435, 2281}

\bibitem[\protect\citeauthoryear{{Pessah}}{{Pessah}}{2010}]{Pessah2010}
{Pessah} M.~E.,  2010, \mn@doi [\apj] {10.1088/0004-637X/716/2/1012}, \href
  {http://adsabs.harvard.edu/abs/2010ApJ...716.1012P} {716, 1012}

\bibitem[\protect\citeauthoryear{{Pessah} \& {Goodman}}{{Pessah} \&
  {Goodman}}{2009}]{Pessah_goodman2009}
{Pessah} M.~E.,  {Goodman} J.,  2009, \mn@doi [\apjl]
  {10.1088/0004-637X/698/1/L72}, \href
  {http://adsabs.harvard.edu/abs/2009ApJ...698L..72P} {698, L72}

\bibitem[\protect\citeauthoryear{{Pessah}, {Chan}  \& {Psaltis}}{{Pessah}
  et~al.}{2007}]{Pessah2007}
{Pessah} M.~E.,  {Chan} C.-k.,   {Psaltis} D.,  2007, \mn@doi [\apjl]
  {10.1086/522585}, \href {http://adsabs.harvard.edu/abs/2007ApJ...668L..51P}
  {668, L51}

\bibitem[\protect\citeauthoryear{{Pouquet}, {Frisch}  \& {Leorat}}{{Pouquet}
  et~al.}{1976}]{Pouquet1976}
{Pouquet} A.,  {Frisch} U.,   {Leorat} J.,  1976, \mn@doi [Journal of Fluid
  Mechanics] {10.1017/S0022112076002140}, \href
  {http://adsabs.harvard.edu/abs/1976JFM....77..321P} {77, 321}

\bibitem[\protect\citeauthoryear{{Quataert} \& {Gruzinov}}{{Quataert} \&
  {Gruzinov}}{2000}]{Quataert2000}
{Quataert} E.,  {Gruzinov} A.,  2000, \mn@doi [\apj] {10.1086/309267}, \href
  {http://adsabs.harvard.edu/abs/2000ApJ...539..809Q} {539, 809}

\bibitem[\protect\citeauthoryear{{Regev} \& {Umurhan}}{{Regev} \&
  {Umurhan}}{2008}]{Regev2008}
{Regev} O.,  {Umurhan} O.~M.,  2008, \mn@doi [\aap]
  {10.1051/0004-6361:20078413}, \href
  {http://adsabs.harvard.edu/abs/2008A%26A...481...21R} {481, 21}

\bibitem[\protect\citeauthoryear{{Remillard} \& {McClintock}}{{Remillard} \&
  {McClintock}}{2006}]{Remillard2006}
{Remillard} R.~A.,  {McClintock} J.~E.,  2006, \mn@doi [\araa]
  {10.1146/annurev.astro.44.051905.092532}, \href
  {http://adsabs.harvard.edu/abs/2006ARA%26A..44...49R} {44, 49}

\bibitem[\protect\citeauthoryear{{R{\"u}diger} \& {Pipin}}{{R{\"u}diger} \&
  {Pipin}}{2000}]{Ruediger2000}
{R{\"u}diger} G.,  {Pipin} V.~V.,  2000, \aap, \href
  {http://adsabs.harvard.edu/abs/2000A%26A...362..756R} {362, 756}

\bibitem[\protect\citeauthoryear{{Ruediger} \& {Kichatinov}}{{Ruediger} \&
  {Kichatinov}}{1993}]{Ruediger1993}
{Ruediger} G.,  {Kichatinov} L.~L.,  1993, \aap, \href
  {http://adsabs.harvard.edu/abs/1993A%26A...269..581R} {269, 581}

\bibitem[\protect\citeauthoryear{{Ryan}, {Gammie}, {Fromang}  \&
  {Kestener}}{{Ryan} et~al.}{2017}]{Ryan2017}
{Ryan} B.~R.,  {Gammie} C.~F.,  {Fromang} S.,   {Kestener} P.,  2017, \mn@doi
  [\apj] {10.3847/1538-4357/aa6a52}, \href
  {http://adsabs.harvard.edu/abs/2017ApJ...840....6R} {840, 6}

\bibitem[\protect\citeauthoryear{{Sano}, {Inutsuka}, {Turner}  \&
  {Stone}}{{Sano} et~al.}{2004}]{Sano2004}
{Sano} T.,  {Inutsuka} S.-i.,  {Turner} N.~J.,   {Stone} J.~M.,  2004, \mn@doi
  [\apj] {10.1086/382184}, \href
  {http://adsabs.harvard.edu/abs/2004ApJ...605..321S} {605, 321}

\bibitem[\protect\citeauthoryear{{Shakura} \& {Sunyaev}}{{Shakura} \&
  {Sunyaev}}{1973}]{Shakura_sunyaev1973}
{Shakura} N.~I.,  {Sunyaev} R.~A.,  1973, \aap, \href
  {http://adsabs.harvard.edu/abs/1973A%26A....24..337S} {24, 337}

\bibitem[\protect\citeauthoryear{{Shi}, {Krolik}  \& {Hirose}}{{Shi}
  et~al.}{2010}]{Shi2010}
{Shi} J.,  {Krolik} J.~H.,   {Hirose} S.,  2010, \mn@doi [\apj]
  {10.1088/0004-637X/708/2/1716}, \href
  {http://adsabs.harvard.edu/abs/2010ApJ...708.1716S} {708, 1716}

\bibitem[\protect\citeauthoryear{{Shi}, {Stone}  \& {Huang}}{{Shi}
  et~al.}{2016}]{Shi2016}
{Shi} J.-M.,  {Stone} J.~M.,   {Huang} C.~X.,  2016, \mn@doi [\mnras]
  {10.1093/mnras/stv2815}, \href
  {http://adsabs.harvard.edu/abs/2016MNRAS.456.2273S} {456, 2273}

\bibitem[\protect\citeauthoryear{{Shiokawa}, {Dolence}, {Gammie}  \&
  {Noble}}{{Shiokawa} et~al.}{2012}]{Shiokawa2012}
{Shiokawa} H.,  {Dolence} J.~C.,  {Gammie} C.~F.,   {Noble} S.~C.,  2012,
  \mn@doi [\apj] {10.1088/0004-637X/744/2/187}, \href
  {http://adsabs.harvard.edu/abs/2012ApJ...744..187S} {744, 187}

\bibitem[\protect\citeauthoryear{{Simon}, {Beckwith}  \& {Armitage}}{{Simon}
  et~al.}{2012}]{Simon2012}
{Simon} J.~B.,  {Beckwith} K.,   {Armitage} P.~J.,  2012, \mn@doi [\mnras]
  {10.1111/j.1365-2966.2012.20835.x}, \href
  {http://adsabs.harvard.edu/abs/2012MNRAS.422.2685S} {422, 2685}

\bibitem[\protect\citeauthoryear{{Sorathia}, {Reynolds}, {Stone}  \&
  {Beckwith}}{{Sorathia} et~al.}{2012}]{Sorathia2012}
{Sorathia} K.~A.,  {Reynolds} C.~S.,  {Stone} J.~M.,   {Beckwith} K.,  2012,
  \mn@doi [\apj] {10.1088/0004-637X/749/2/189}, \href
  {http://adsabs.harvard.edu/abs/2012ApJ...749..189S} {749, 189}

\bibitem[\protect\citeauthoryear{{Stone} \& {Pringle}}{{Stone} \&
  {Pringle}}{2001}]{Stone2001}
{Stone} J.~M.,  {Pringle} J.~E.,  2001, \mn@doi [\mnras]
  {10.1046/j.1365-8711.2001.04138.x}, \href
  {http://adsabs.harvard.edu/abs/2001MNRAS.322..461S} {322, 461}

\bibitem[\protect\citeauthoryear{{Stone}, {Hawley}, {Gammie}  \&
  {Balbus}}{{Stone} et~al.}{1996}]{Stone1996}
{Stone} J.~M.,  {Hawley} J.~F.,  {Gammie} C.~F.,   {Balbus} S.~A.,  1996,
  \mn@doi [\apj] {10.1086/177280}, \href
  {http://adsabs.harvard.edu/abs/1996ApJ...463..656S} {463, 656}

\bibitem[\protect\citeauthoryear{{Suzuki} \& {Inutsuka}}{{Suzuki} \&
  {Inutsuka}}{2014}]{Suzuki2014}
{Suzuki} T.~K.,  {Inutsuka} S.-i.,  2014, \mn@doi [\apj]
  {10.1088/0004-637X/784/2/121}, \href
  {http://adsabs.harvard.edu/abs/2014ApJ...784..121S} {784, 121}

\bibitem[\protect\citeauthoryear{{Tout} \& {Pringle}}{{Tout} \&
  {Pringle}}{1992}]{Tout1992}
{Tout} C.~A.,  {Pringle} J.~E.,  1992, \mn@doi [\mnras]
  {10.1093/mnras/259.4.604}, \href
  {http://adsabs.harvard.edu/abs/1992MNRAS.259..604T} {259, 604}

\bibitem[\protect\citeauthoryear{Velikhov}{Velikhov}{1959}]{Velikhov1959}
Velikhov E.,  1959, Sov. Phys. JETP, 36, 995

\bibitem[\protect\citeauthoryear{{Yuan} \& {Narayan}}{{Yuan} \&
  {Narayan}}{2014}]{Yuan2014}
{Yuan} F.,  {Narayan} R.,  2014, \mn@doi [\araa]
  {10.1146/annurev-astro-082812-141003}, \href
  {http://adsabs.harvard.edu/abs/2014ARA%26A..52..529Y} {52, 529}

\makeatother
\end{thebibliography}
\label{lastpage}

\end{document}